\documentstyle[12pt]{article}

\newcommand{\noi}{\noindent}
\newcommand{\hsp}{\hspace{5.7mm}}
\newcommand{\ssp}{\hspace*{4mm}}
\newcommand{\tsp}{\hspace{2.7mm}}
\newcommand{\fsp}{\hspace*{1.8mm}}
\newcommand{\fssp}{\hspace*{5mm}}
\newcommand{\sspp}{\hspace*{11.3mm}}
\newcommand{\mini}{\hspace{.3mm}}    
\newcommand{\fmini}{\hspace{.5mm}}
\newcommand{\mmini}{\hspace{-.4mm}}   
\newcommand{\mimini}{\hspace{-.3mm}}
\newcommand{\mE}{\hspace{-1.7mm}}
\newcommand{\mA}{\hspace{-1mm}}

\let\ssection=\section
\renewcommand{\section}{\setcounter{equation}{0}\ssection}

\newfont{\BBB}{msbm10 scaled\magstephalf}
\newcommand{\BB}[1]{\mbox{\BBB #1}}
\newcommand{\BBn}[2]{\mbox{{\BBB #1}$^{#2}$}}

\newcommand{\lsim}{\,\raisebox{.9mm}{$<$} \hspace{-3.3mm}
   \raisebox{-.9mm}{$\sim$}\,}
\newcommand{\rank}{\mbox{rank\,}}
\newcommand{\ranks}{\mbox{\scriptsize rank\mini}}
\newcommand{\lcm}{\mbox{lcm}}
\newcommand{\emb}{\mbox{\raisebox{.8mm}{$\hspace{1.8mm}\scriptstyle
   \subset\mA$}\raisebox{-.4mm}{$\longrightarrow\hspace{1.8mm}$}}}
\newcommand{\bfdelta}{\mbox{$\delta \hspace{-2.1mm}
   \delta \hspace{-2.1mm} \delta \hspace{-2.1mm} \delta$}}

\newcommand{\be}{\begin{equation}}
\newcommand{\ee}{\end{equation} \vspace{0mm}}
\newcommand{\ben}{$$}       
\newcommand{\een}{$$ \vspace{0mm}}
\newcommand{\bea}{\begin{eqnarray}}
\newcommand{\eea}{\end{eqnarray} \vspace{0mm}}
\newcommand{\ba}{\begin{array}{c}}
\newcommand{\ea}{\end{array}}
\newcommand{\barr}{\begin{array}{r}}
\newcommand{\earr}{\end{array}}
\newcommand{\bal}{\begin{array}{l}}
\newcommand{\eal}{\end{array}}

\newcommand{\maxsym}{
  \begin{picture}(0,0)(.025,0)\circle{.31}\end{picture}}
\newcommand{\lowdim}{
  \begin{picture}(0,0)(.12,.1)\framebox(.19,.19){}\end{picture}}
\newcommand{\chacon}{
  \begin{picture}(0,0)(.185,.16)\dashbox{.04}(.32,.32){}\end{picture}}
\newcommand{\noLmin}{\raisebox{-.6mm}
  {\hspace*{-.1mm}$\scriptstyle\times$}}
\newcommand{\nolodiLmin}{\raisebox{-.6mm}
  {\hspace*{-.7mm}$\scriptstyle(\!\times\!)$}}
\newcommand{\dimms}[1]{\makebox(0.16,-0.21)[t]{$\scriptstyle #1$}}
\newcommand{\dimg}[1]{\makebox(0.16,0.35)[t]{$\scriptstyle #1$}}

\newcommand{\hah}{\hsp\mbox{and}\hsp}
\newcommand{\ha}{\hsp\mbox{and}}
\newcommand{\hfa}{\hsp\mbox{for all }\,}
\newcommand{\hfs}{\hsp\mbox{for some }\,}
\newcommand{\hwh}{\hsp\mbox{with}\hsp}

\newcommand{\fh}{\mbox{for}\hsp}
\newcommand{\vav}{\vspace{-4.5mm}\noi and \vspace{-3mm}}
\newcommand{\vwv}{\vspace{-4.5mm}\noi where \vspace{-3mm}}

\newcommand{\SUs}{$SU(2)_{spin}$}
\newcommand{\ui}{$u(1)$}

\newcommand{\sH}{\sigma_H}
\newcommand{\nH}{n_H}
\newcommand{\dH}{d_H}
\newcommand{\ndH}{{{n_H\rule[-.8mm]{0mm}{2mm}} \over d_H}}
\newcommand{\eh}{{e^2\over h}}
\newcommand{\eht}{e^2\!/\mmini h}
\newcommand{\Oth}{\Omega^{(\theta)}}
\newcommand{\vx}{\vec{x}}
\newcommand{\vxi}{\vec{\xi}}
\newcommand{\Bc}{{\bf B}_c}
\newcommand{\Bcpara}{{\bf B}_c^{\mini\parallel}}
\newcommand{\Bcperp}{B_{\!c}^\perp}
\newcommand{\txi}{(\tau,\vec{\xi}\,)}

\newcommand{\qel}{q_{el}}
\newcommand{\Q}{{\bf Q}}
\newcommand{\Qe}{{\bf Q}_{e}}
\newcommand{\Qh}{{\bf Q}_{h}}
\newcommand{\G}{\Gamma}
\newcommand{\Gp}{\Gamma_{\!phys}}
\newcommand{\Gs}{\Gamma^\ast}
\newcommand{\Gw}{\Gamma_W}
\newcommand{\Gwi}[1]{\Gamma_{W_#1}}
\newcommand{\Gws}{\Gamma_W^\ast}
\newcommand{\Ge}{\Gamma_{\!e}}
\newcommand{\Gh}{\Gamma_{\!h}}
\newcommand{\QHLe}{\mbox{$(\Ge,\Qe)$}}
\newcommand{\QHLh}{\mbox{$(\Gh,\Qh)$}}
\newcommand{\QHL}{\mbox{$(\G,\Q)$}}
\newcommand{\veb}{{\bf e}}
\newcommand{\ve}[1]{{\bf e}_#1}
\newcommand{\veed}{\mbox{\boldmath $\varepsilon$}}
\newcommand{\ved}[1]{\mbox{\boldmath $\varepsilon$}^#1}
\newcommand{\vom}{\mbox{\boldmath $\omega$}}
\newcommand{\omv}{\mbox{\raisebox{-2.2mm}{$\stackrel{\displaystyle
   \omega}{\scriptstyle \rightarrow}$}}}
\newcommand{\omvi}[1]{\mbox{\raisebox{-2.2mm}{$\stackrel{\displaystyle
   \omega}{\scriptstyle \rightarrow}$}
   \raisebox{-1.2mm}{$\scriptstyle\hspace{-1.5mm}#1$}}}
\newcommand{\hw}{h_{\vom}}
\newcommand{\hws}{{h_{\mbox{$\scriptstyle\omega\hspace{-1.9mm}
   \omega\hspace{-1.9mm}\omega\hspace{-1.9mm}\omega$}}}}
\newcommand{\vq}{{\bf q}}
\newcommand{\vv}{{\bf v}}
\newcommand{\vn}{{\bf n}}
\newcommand{\Ki}{K^{-1}}
\newcommand{\D}{\Delta}
\newcommand{\Qv}{\mbox{\raisebox{-2.2mm}{$\stackrel{\displaystyle
   Q}{\scriptstyle \rightarrow}$}}}
\newcommand{\Qvp}{\mbox{\raisebox{-2.2mm}{$\stackrel{\displaystyle
   Q^\prime}{\scriptstyle \rightarrow}$}}}
\newcommand{\lb}{\lambda}

\newcommand{\subs}[1]{\raisebox{-2mm}{$\scriptstyle #1$}}
\newcommand{\sub}[1]{\raisebox{-1.6mm}{$\scriptstyle #1$}}
\newcommand{\subi}[1]{\raisebox{-1.2mm}{$\scriptstyle #1$}}
\newcommand{\thesymbol}{\raisebox{-3.6mm}{$\scriptstyle N$}
   {\left( \ndH \right)}^g_\lambda\;[\mini\lm,\lM\mini]}
\newcommand{\thesymbolpr}{\raisebox{-3.6mm}{$\scriptstyle
   {N^\prime}$}{\left( \ndH \right)}^{g^\prime}_{\lambda^\prime}\;
   [\mini\lm^{\mini\prime},\lM^{\mini\prime}\mini]}
\newcommand{\QHLsymbol}[4]{\raisebox{-3.6mm}{$\scriptstyle #1$}
   {\left( #2 \right)}^{#3}\hspace{-1,6mm}
   \raisebox{-3.6mm}{$\scriptstyle #4$}}

\newcommand{\LM}{L_{max}}
\newcommand{\Lm}{L_{min}}
\newcommand{\Lii}{\ell_2}
\newcommand{\Ls}{\ell_\ast}
\newcommand{\lM}{\ell_{max}}
\newcommand{\lm}{\ell_{min}}
\newcommand{\BQ}{{\cal B}_{\!\Q}}
\newcommand{\BQs}{\raisebox{-.5mm}{\mbox{${\textstyle\cal
   B}_{\scriptscriptstyle\bf Q}$}}}
\newcommand{\oBQ}{o{\cal B}_{\mA\Q}}
\newcommand{\oBQs}{\raisebox{-.5mm}{\mbox{${\textstyle o{\cal
   B}}_{\!\scriptscriptstyle\bf Q}$}}}
\newcommand{\Sp}{{\cal S}_p}
\newcommand{\Si}{{\cal S}_1}
\newcommand{\Sigp}{\Sigma_p}
\newcommand{\Sigi}{\Sigma_1}
\newcommand{\Lmini}{$L\mini$-minimal}
\newcommand{\Hp}{{\cal H}_p}
\newcommand{\Hi}{{\cal H}_1}

\newcommand{\bulf}[1]{$\bullet\hspace{.7mm}{#1}$}
\newcommand{\pbulf}[2]{{\scriptsize\em #1}$\:
   \bullet\hspace{.7mm}{#2}$}
\newcommand{\cirf}[1]{$\circ\hspace{.7mm}{#1}$}
\newcommand{\tcirf}[1]{{\scriptsize\em (2)}$\,
   \circ\hspace{.7mm}{#1}$}
\newcommand{\dof}[1]{$\hspace*{.4mm}\cdot\hspace{.8mm}{#1}$}
\newcommand{\nof}[1]{$\hspace*{2.1mm}{#1}$}
\newcommand{\tbulf}[1]{{\scriptsize\em (2)}$\,
   \bullet\hspace{.7mm}{#1}$}
\newcommand{\ptbulf}[2]{{\scriptsize\em #1\,(2)}$\:
   \bullet\hspace{.7mm}{#2}$}

\newcommand{\OO}{{\cal O}\mini}
\newcommand{\GG}{{\cal G}\mini}
\newcommand{\GGh}{\hat{\cal G}}

\newcommand{\PRL}{Phys.\ Rev.\ Lett.\ }
\newcommand{\PRB}{Phys.\ Rev.\ B }
\newcommand{\PRD}{Phys.\ Rev.\ D }

\newcommand{\PL}{Phys.\ Lett.\ }
\newcommand{\NPB}{Nucl.\ Phys.\ B }
\newcommand{\AP}{Ann.\ Phys.\ (N.Y.) }
\newcommand{\CMP}{Commun.\ Math.\ Phys.\ }
\newcommand{\SurS}{Surf.\ Sci.\ }
\newcommand{\SSC}{Solid State Comm.\ }
\newcommand{\HPA}{Helv.\ Phys.\ Acta }
\newcommand{\IJMP}{Int.\ J.\ Mod.\ Phys.\ }

\begin{document}


\thispagestyle{empty}

\begin{flushright}
{\footnotesize\tt preprint KUL-TF-94/35\\February 1995}
\end{flushright}

\vspace{20mm}

\begin{center}
{\LARGE\bf A CLASSIFICATION OF\rule[-4mm]{0mm}{6mm} \\
QUANTUM HALL FLUIDS}

\vspace{15mm}

J\"urg Fr\"ohlich,$^1$ Urban M.~Studer,$^2$\footnote{Present
address: Institut f\"ur Theoretische Physik, ETH-H\"onggerberg,
8093 Z\"urich, Switzerland} and Emmanuel Thiran$^1$

\vspace{10mm}

$^1\,${\small Institut f\"ur Theoretische Physik, ETH-H\"onggerberg,
8093 Z\"urich, Switzerland} \\
$^2\,${\small Instituut voor Theoretische Fysica, Katholieke
Universiteit Leuven, 3001 Leuven, Belgium}
\end{center}

\vspace{25mm}

{\footnotesize\bf Abstract.} In this paper, the key ideas of
characterizing universality classes of dissipation-free
(incompressible) quantum Hall fluids by mathematical objects called
quantum Hall lattices are reviewed. Many general theorems about the
classification of quantum Hall lattices are stated and their
physical implications are discussed. Physically relevant subclasses
of quantum Hall lattices are defined and completely classified. The
results are carefully compared with experimental data and also with
other theoretical schemes (the hierarchy schemes). Several proposals
for new experiments are made which could help to settle interesting
issues in the theory of the (fractional) quantum Hall effect and thus
would lead to a deeper understanding of this remarkable effect.

\newpage
\thispagestyle{empty}
\vspace*{10mm}
{\small
\tableofcontents}


\newpage
\setcounter{page}{10}
\begin{flushleft}
\vspace*{3mm}
\begin{picture}(10,8)(-.4,0)

\put(4.74,0){\makebox(0,0)[l]
  {$\hspace*{.4mm}\cdot\hspace{1.2mm}{\bf 9\over 19}$}}

\put(4.71,.8){\makebox(0,0)[l]
  {$\circ\hspace{.6mm}{\bf 8\over 17}$}}
\put(5.29,.8){\makebox(0,0)[l]
  {$\circ\hspace{.6mm}{\bf 9\over 17}$}}
\put(5.88,.8){\makebox(0,0)[l]
  {$\hspace*{.4mm}\cdot\hspace{1.2mm}{\bf 10\over 17}$}}

\put(2.67,1.6){\makebox(0,0)[l]
  {$\circ\hspace{.7mm}{\bf 4\over 15}$}}
\put(4.67,1.6){\makebox(0,0)[l]
  {$\circ\hspace{.7mm}{\bf 7\over 15}$}}
\put(5.33,1.6){\makebox(0,0)[l]
  {$\circ\hspace{.7mm}{\bf 8\over 15}$}}

\put(2.3,2.4){\makebox(0,0)[l]
  {$\circ\hspace{.7mm}{\bf 3\over 13}$}}
\put(3.07,2.4){\makebox(0,0)[l]
  {$\hspace*{.4mm}\cdot\hspace{1.2mm}{\bf 4\over 13}$}}
\put(4.62,2.4){\makebox(0,0)[l]
  {$\bullet\hspace{.7mm}{\bf 6\over 13}$}}
\put(5.38,2.4){\makebox(0,0)[l]
  {$\bullet\hspace{.7mm}{\bf 7\over 13}$}}
\put(6.15,2.4){\makebox(0,0)[l]
  {$\bullet\hspace{.7mm}{\bf 8\over 13}$}}
\put(6.92,2.4){\makebox(0,0)[l]
  {$\circ\hspace{.7mm}{\bf 9\over 13}$}}

\put(1.82,3.2){\makebox(0,0)[l]
  {$\circ\hspace{.7mm}{\bf 2\over 11}$}}
\put(2.73,3.2){\makebox(0,0)[l]
  {$\bullet\hspace{.7mm}{\bf 3\over 11}$}}
\put(3.64,3.2){\makebox(0,0)[l]
  {$\hspace*{.4mm}\cdot\hspace{1.2mm}{\bf 4\over 11}$}}
\put(4.54,3.2){\makebox(0,0)[l]
  {$\bullet\hspace{.5mm}{\bf 5\over 11}$}}
\put(5.45,3.2){\makebox(0,0)[l]
  {$\bullet\hspace{.7mm}{\bf 6\over 11}$}}
\put(6.36,3.2){\makebox(0,0)[l]
  {$\circ\hspace{.7mm}{\bf 7\over 11}$}}
\put(7.27,3.2){\makebox(0,0)[l]
  {$\bullet\hspace{.7mm}{\bf 8\over 11}$}}

\put(2.22,4){\makebox(0,0)[l]
  {$\bullet\hspace{.7mm}{\bf 2\over 9}$}}
\put(4.44,4){\makebox(0,0)[l]
  {$\bullet\hspace{.7mm}{\bf 4\over 9}$}}
\put(5.56,4){\makebox(0,0)[l]
  {$\bullet\hspace{.7mm}{\bf 5\over 9}$}}

\put(1.43,4.8){\makebox(0,0)[l]
  {$\circ\hspace{.7mm}{\bf 1\over 7}$}}
\put(2.86,4.8){\makebox(0,0)[l]
  {$\bullet\hspace{.7mm}{\bf 2\over 7}$}}
\put(4.29,4.8){\makebox(0,0)[l]
  {$\bullet\hspace{.7mm}{\bf 3\over 7}$}}
\put(5.71,4.8){\makebox(0,0)[l]
  {$\bullet\hspace{.7mm}{\bf 4\over 7}$}}
\put(7.14,4.8){\makebox(0,0)[l]
  {$\bullet\hspace{.7mm}{\bf 5\over 7}\:${\scriptsize\em
  (B-p\hspace{-.2mm}) }}}

\put(2,5.6){\makebox(0,0)[l]
  {$\bullet\hspace{.7mm}{\bf 1\over 5}$}}
\put(4,5.6){\makebox(0,0)[l]
  {$\bullet\hspace{.7mm}{\bf 2\over 5}\:${\scriptsize\em
  (B-p\hspace{-.2mm}) }}}
\put(6,5.6){\makebox(0,0)[l]
  {$\bullet\hspace{.7mm}{\bf 3\over 5}\:${\scriptsize\em B-p}}}
\put(8,5.6){\makebox(0,0)[l]
  {$\bullet\hspace{.7mm}{\bf 4\over 5}$}}

\put(3.33,6.4){\makebox(0,0)[l]
  {$\bullet\hspace{.7mm}{\bf 1\over 3}$}}
\put(6.67,6.4){\makebox(0,0)[l]
  {$\bullet\hspace{.7mm}{\bf 2\over 3}\:${\scriptsize\em B/n-p}}}

\put(10,7.2){\makebox(0,0)[l]
  {$\bullet\hspace{1mm}{\bf 1}$}}

\put(.73,7.2){\makebox(0,0)[r]{\small$\dH=1$}}
\put(.73,6.4){\makebox(0,0)[r]{\small$3$}}
\put(.73,5.6){\makebox(0,0)[r]{\small$5$}}
\put(.73,4.8){\makebox(0,0)[r]{\small$7$}}
\put(.73,4){\makebox(0,0)[r]{\small$9$}}
\put(.73,3.2){\makebox(0,0)[r]{\small$11$}}
\put(.73,2.4){\makebox(0,0)[r]{\small$13$}}
\put(.73,1.6){\makebox(0,0)[r]{\small$15$}}
\put(.73,.8){\makebox(0,0)[r]{\small$17$}}
\put(.73,0){\makebox(0,0)[r]{\small$19$}}

\thinlines
\put(-.5,-.45){\hspace{1mm}\line(1,0){11}}
\put(-.5,7.6){\hspace{1mm}\line(1,0){11}}

\multiput(0,7.6)(0,-0.1){3}{\hspace{1mm}\line(0,-1){0.04}} 
\multiput(0,7)(0,-0.1){74}{\hspace{1mm}\line(0,-1){0.04}}
\put(0,-.4){\hspace{1mm}\line(0,-1){0.1}}

\multiput(1.43,7.6)(0,-0.1){28}{\hspace{1mm}\line(0,-1){0.04}} 
\multiput(1.43,4.7)(0,-0.1){51}{\hspace{1mm}\line(0,-1){0.04}}
\put(1.43,-.4){\hspace{1mm}\line(0,-1){0.1}}

\multiput(1.67,7.6)(0,-0.2){13}{\hspace{1mm}\line(0,-1){0.1}} 
\put(1.67,4.56){\hspace{1mm}\line(0,-1){0.06}}
\multiput(1.67,4.4)(0,-0.2){25}{\hspace{1mm}\line(0,-1){0.1}}

\multiput(2,7.6)(0,-0.1){20}{\hspace{1mm}\line(0,-1){0.04}} 
\multiput(2,5.5)(0,-0.1){21}{\hspace{1mm}\line(0,-1){0.04}}
\multiput(2,2.9)(0,-0.1){33}{\hspace{1mm}\line(0,-1){0.04}}
\put(2,-.4){\hspace{1mm}\line(0,-1){0.1}}

\multiput(2.5,7.6)(0,-0.2){17}{\hspace{1mm}\line(0,-1){0.1}} 
\put(2.5,3.76){\hspace{1mm}\line(0,-1){0.06}}
\multiput(2.5,3.6)(0,-0.2){5}{\hspace{1mm}\line(0,-1){0.1}}
\put(2.5,2.16){\hspace{1mm}\line(0,-1){0.06}}
\multiput(2.5,2)(0,-0.2){13}{\hspace{1mm}\line(0,-1){0.1}}

\multiput(3.33,7.6)(0,-0.1){12}{\hspace{1mm}\line(0,-1){0.04}} 
\multiput(3.33,6.3)(0,-0.1){37}{\hspace{1mm}\line(0,-1){0.04}}
\multiput(3.33,2.1)(0,-0.1){25}{\hspace{1mm}\line(0,-1){0.04}}
\put(3.33,-.4){\hspace{1mm}\line(0,-1){0.1}}

\multiput(5,7.6)(0,-0.2){25}{\hspace{1mm}\line(0,-1){0.1}} 
\put(5,2.16){\hspace{1mm}\line(0,-1){0.06}}
\put(5,2){\hspace{1mm}\line(0,-1){0.1}}
\put(5,1.36){\hspace{1mm}\line(0,-1){0.06}}
\put(5,1.2){\hspace{1mm}\line(0,-1){0.1}}
\put(5,.56){\hspace{1mm}\line(0,-1){0.06}}
\put(5,.4){\hspace{1mm}\line(0,-1){0.1}}
\put(5,-.24){\hspace{1mm}\line(0,-1){0.07}}
\put(5,-.4){\hspace{1mm}\line(0,-1){0.1}}

\multiput(10,7.6)(0,-0.1){4}{\hspace{1mm}\line(0,-1){0.04}} 
\multiput(10,7.1)(0,-0.1){75}{\hspace{1mm}\line(0,-1){0.04}}
\put(10,-.4){\hspace{1mm}\line(0,-1){0.1}}

\thicklines
\put(0,-.6){\hspace{1mm}\makebox(0,0)[t]{$0$}}
\put(.71,-.66){\hspace{1mm}\makebox(0,0)[t]{$\cdots$}}
\put(-.26,-1){\mbox{$\underbrace{\hspace*{19mm}}_{
  \renewcommand{\arraystretch}{.7} \ba \mbox{\small\em Wigner
  crystal} \\ \mbox{\small\em or carrier} \\ \mbox{\small\em
  freeze-out}\ea}$}}
\put(1.43,-.54){\hspace{1mm}\makebox(0,0)[t]{$1\over 7$}}
\put(1.43,-.41){\hspace{1mm}\line(1,0){.24}}
\put(1.67,-.54){\hspace{1mm}\makebox(0,0)[t]{$1\over 6$}}

\put(2,-.54){\hspace{1mm}\makebox(0,0)[t]{$1\over 5$}}
\put(2,-.41){\hspace{1mm}\line(1,0){.5}}
\put(2.5,-.54){\hspace{1mm}\makebox(0,0)[t]{$1\over 4$}}
\put(1.76,-1.5){\mbox{$\renewcommand{\arraystretch}{.7} \ba \uparrow
  \rule[-2mm]{0mm}{5mm} \\ \mbox{\small\em Fermi liquid} \\
  \mbox{\small\em behaviour}\ea$}}

\put(3.33,-.54){\hspace{1mm}\makebox(0,0)[t]{$1\over 3$}}
\put(3.33,-.41){\hspace{1mm}\line(1,0){1.67}}
\put(5,-.54){\hspace{1mm}\makebox(0,0)[t]{$1\over 2$}}
\put(4.26,-1.5){\mbox{$\renewcommand{\arraystretch}{.7} \ba \uparrow
  \rule[-2mm]{0mm}{5mm} \\ \mbox{\small\em Fermi liquid} \\
  \mbox{\small\em behaviour}\ea$}}

\put(10,-.6){\hspace{1mm}\makebox(0,0)[t]{$1$}}
\put(10,-.41){\hspace{1mm}\line(1,0){.5}}
\put(10.4,-.66){\hspace{1mm}\makebox(0,0)[t]{$\sH$}}
\put(8.41,-1){\mbox{$\underbrace{\hspace*{30mm}}_{
  \renewcommand{\arraystretch}{.7} \ba \mbox{\small\em domain of}Ê\\
  \mbox{\small\em attraction} \\ \mbox{\small\em of $\,\sH=1$}\ea}$}}

\put(.71,7.85){\hspace{1mm}\makebox(0,0){$\cdots$}}
\put(1.43,7.56){\hspace{1mm}\line(1,0){.24}}
\put(1.55,7.75){\hspace{.85mm}\makebox(0,0)[b]{$\Sigma_3^+$}}
\put(1.83,7.75){\hspace{2mm}\makebox(0,0)[b]{$\Sigma_3^-$}}

\put(2,7.56){\hspace{1mm}\line(1,0){.5}}
\put(2.25,7.75){\hspace{1.9mm}\makebox(0,0)[b]{$\Sigma_2^+$}}
\put(2.92,7.75){\hspace{1.4mm}\makebox(0,0)[b]{$\Sigma_2^-$}}

\put(3.33,7.56){\hspace{1mm}\line(1,0){1.67}}
\put(4.17,7.75){\hspace{1.2mm}\makebox(0,0)[b]{$\Sigma_1^+$}}
\put(7.5,7.75){\hspace{1.2mm}\makebox(0,0)[b]{$\Sigma_1^-$}}

\put(10,7.56){\hspace{1mm}\line(1,0){.5}}

\end{picture}
\end{flushleft}

\vspace{31mm}
\noi
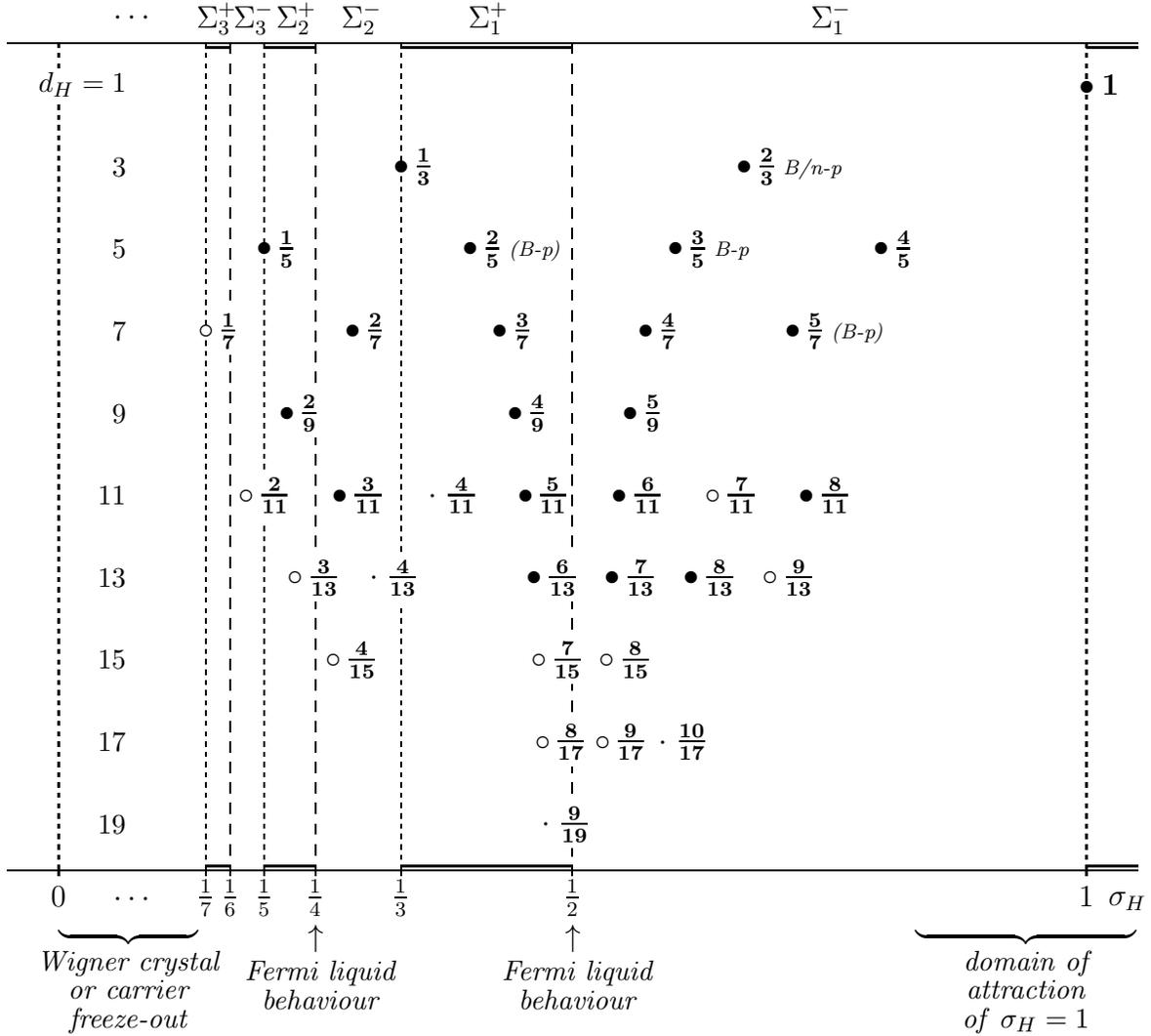
{\bf Figure 1.1. }{\em Observed Hall fractions $\,\sH=\nH/\dH\,$ in
the interval $\,0<\sH\leq 1$, and their experimental status in
single-layer quantum Hall systems.}
\rule[-4mm]{0mm}{5mm}

\noi
{\small Well established Hall fractions are indicated by
``$\,\bullet\,$''. These are fractions for which a
$R_{xx}\mini$-minimum and a plateau in $R_H$ have been clearly
observed, and the quantization accuracy of $\sH=1/R_H$ is typically
better than $0.5$\%. Fractions for which a minimum in $R_{xx}$ and
typically an inflection in $R_H$ (i.e., a minimum in ${d\mini R_H /
d\mini B_{\!c}^\perp}$, but no well developed plateau in $R_H$) have
been observed are indicated by ``$\,\circ\,$''. If there are only
very weak experimental indications or controversial data for a given
Hall fraction, the symbol ``$\,\cdot\,$'' is used. Finally, ``{\em
B/n-p\,}'' is appended to fractions at which a magnetic field
{\em\mimini(B)} and/or density {\em\mimini(n)} driven phase
transition has been observed.}


\newpage
\setcounter{page}{13}
\begin{flushleft}
\begin{picture}(10,8)(1,0)

\put(3.68,0){\noLmin}   
\put(4.21,0){\noLmin}   
\put(4.74,0){\makebox(0,0)[l]{$\hspace*{.4mm}\cdot$}}
  \put(4.74,0){\maxsym}\put(4.74,0){\dimms{9}}  
\put(5.26,0){\chacon}   
\put(5.79,0){\noLmin}   
\put(6.32,0){\maxsym}\put(6.32,0){\dimms{6}}   
\put(6.84,0){\noLmin}   
\put(7.37,0){\lowdim}\put(7.37,0){\dimg{4}}   
\put(7.89,0){\nolodiLmin}\put(7.89,0){\dimms{15}}   
\put(8.42,0){\nolodiLmin}\put(8.42,0){\dimms{19}}   
\put(8.95,0){\nolodiLmin}\put(8.95,0){\dimms{33}}   
\put(9.47,0){\nolodiLmin}\put(9.47,0){\dimms{20}}   

\put(3.53,.8){\noLmin}   
\put(4.12,.8){\noLmin}   
\put(4.71,.8){\makebox(0,0)[l]{$\circ$}}
  \put(4.71,.8){\maxsym}\put(4.71,.8){\dimms{8}}   
\put(5.29,.8){\makebox(0,0)[l]{$\circ$}}
  \put(5.29,.8){\chacon}   
\put(5.88,.8){\makebox(0,0)[l]{$\hspace*{.4mm}\cdot$}}
  \put(5.88,.8){\maxsym}\put(5.88,.8){\dimms{6}}   
\put(6.47,.8){\nolodiLmin}\put(6.47,.8){\dimms{23}}   
\put(7.06,.8){\maxsym}\put(7.06,.8){\dimms{9}}   
\put(7.65,.8){\lowdim}\put(7.65,.8){\dimg{4}}   
\put(8.24,.8){\nolodiLmin}\put(8.24,.8){\dimms{20}}   
\put(8.82,.8){\maxsym}\put(8.82,.8){\dimms{7}}   
\put(9.41,.8){\nolodiLmin}\put(9.41,.8){\dimms{18}}   

\put(4.67,1.6){\makebox(0,0)[l]{$\circ$}}
  \put(4.67,1.6){\maxsym}\put(4.67,1.6){\dimms{7}}   
\put(5.33,1.6){\makebox(0,0)[l]{$\circ$}}
  \put(5.33,1.6){\chacon}   
\put(7.33,1.6){\maxsym}\put(7.33,1.6){\dimms{11}}   
\put(8.67,1.6){\nolodiLmin}\put(8.67,1.6){\dimms{25}}   
\put(9.33,1.6){\maxsym}\put(9.33,1.6){\dimms{8}}   

\put(3.85,2.4){\noLmin}   
\put(4.62,2.4){\makebox(0,0)[l]{$\bullet$}}
  \put(4.62,2.4){\maxsym}\put(4.62,2.4){\dimms{6}}   
\put(5.38,2.4){\makebox(0,0)[l]{$\bullet$}}
  \put(5.38,2.4){\lowdim} \put(5.38,2.4){\chacon}
  \put(5.38,2.4){\dimg{3}}   
\put(6.15,2.4){\makebox(0,0)[l]{$\bullet$}}
  \put(6.15,2.4){\maxsym}\put(6.15,2.4){\dimms{9}}   
\put(6.92,2.4){\makebox(0,0)[l]{$\circ$}}
  \put(6.92,2.4){\maxsym}\put(6.92,2.4){\dimms{9,11}}   
\put(7.69,2.4){\maxsym}\put(7.69,2.4){\dimms{6}}   
\put(8.46,2.4){\lowdim}\put(8.46,2.4){\dimg{4}}   
\put(9.23,2.4){\maxsym}\put(9.23,2.4){\dimms{7,8}}   

\put(3.64,3.2){\makebox(0,0)[l]{$\hspace*{.4mm}\cdot$}}
\put(3.64,3.2){\noLmin}   
\put(4.54,3.2){\makebox(0,0)[l]{$\bullet$}}
  \put(4.54,3.2){\maxsym}\put(4.54,3.2){\dimms{5}}   
\put(5.45,3.2){\makebox(0,0)[l]{$\bullet$}}
  \put(5.45,3.2){\maxsym} \put(5.45,3.2){\chacon}
  \put(5.45,3.2){\dimms{4}}   
\put(6.36,3.2){\makebox(0,0)[l]{$\circ$}}
  \put(6.36,3.2){\maxsym}\put(6.36,3.2){\dimms{7}}   
\put(7.27,3.2){\makebox(0,0)[l]{$\bullet$}}
  \put(7.27,3.2){\maxsym}\put(7.27,3.2){\dimms{11}}
  \put(7.27,3.2){\lowdim}\put(7.27,3.2){\dimg{4}}   
\put(8.18,3.2){\nolodiLmin}\put(8.18,3.2){\dimms{17}}   
\put(9.09,3.2){\lowdim}\put(9.09,3.2){\dimg{4}}   

\put(4.44,4){\makebox(0,0)[l]{$\bullet$}}
  \put(4.44,4){\maxsym}\put(4.44,4){\dimms{4}}   
\put(5.56,4){\makebox(0,0)[l]{$\bullet$}}
  \put(5.56,4){\maxsym}\put(5.56,4){\lowdim}\put(5.56,4){\chacon}
  \put(5.56,4){\dimms{5}}\put(5.56,4){\dimg{4}}   
\put(7.78,4){\maxsym}\put(7.78,4){\dimms{7}}   
\put(8.89,4){\maxsym}\put(8.89,4){\dimms{8,10,11}}   

\put(4.29,4.8){\makebox(0,0)[l]{$\bullet$}}
  \put(4.29,4.8){\maxsym}\put(4.29,4.8){\dimms{3}}   
\put(5.71,4.8){\makebox(0,0)[l]{$\bullet$}}
  \put(5.71,4.8){\maxsym}\put(5.71,4.8){\lowdim}
  \put(5.71,4.8){\chacon}\put(5.71,4.8){\dimms{5,6}}
  \put(5.71,4.8){\dimg{4}}   
\put(7.14,4.8){\makebox(0,0)[l]{$\bullet$}}
  \put(7.14,4.8){\maxsym}
  \put(7.14,4.8){\lowdim \makebox(0,0)[l]{\hspace{1.6mm}
  {\scriptsize\em (B-p\hspace{-.2mm})}}}
  \put(7.14,4.8){\dimms{9,10}}\put(7.14,4.8){\dimg{3,4}}   
\put(8.57,4.8){\maxsym} \put(8.57,4.8){\lowdim}
  \put(8.57,4.8){\dimms{6,7,\ldots}}\put(8.57,4.8){\dimg{4}}  

\put(4,5.6){\makebox(0,0)[l]{$\bullet$}}
  \put(4,5.6){\maxsym \makebox(0,0)[l]{\hspace{1.6mm}
  {\scriptsize\em (B-p\hspace{-.2mm})}}}
  \put(4,5.6){\dimms{2}}   
\put(6,5.6){\makebox(0,0)[l]{$\bullet$}}
  \put(6,5.6){\maxsym} \put(6,5.6){\lowdim}
  \put(6,5.6){\chacon \makebox(0,0)[l]{\hspace{1.6mm}
  {\scriptsize\em B-p}}}
  \put(6,5.6){\dimms{5,6,7}}\put(6,5.6){\dimg{3,4}}   
\put(8,5.6){\makebox(0,0)[l]{$\bullet$}}
  \put(8,5.6){\maxsym} \put(8,5.6){\lowdim}
  \put(8,5.6){\dimms{6,7,\ldots}}\put(8,5.6){\dimg{4}}   

\put(3.33,6.4){\makebox(0,0)[l]{$\bullet$}}
  \put(3.33,6.4){\maxsym}\put(3.33,6.4){\dimms{1}}   
\put(6.67,6.4){\makebox(0,0)[l]{$\bullet$}}
  \put(6.67,6.4){\maxsym}\put(6.67,6.4){\lowdim}
  \put(6.67,6.4){\chacon \makebox(0,0)[l]{\hspace{1.6mm}
  {\scriptsize\em B/n-p}}}\put(6.67,6.4){\dimms{4,5,\ldots}}
  \put(6.67,6.4){\dimms{4,5,\ldots}}\put(6.67,6.4){\dimg{3,4}}  

\put(10,7.2){\makebox(0,0)[l]{$\bullet$}}
  \put(10,7.2){\maxsym}\put(10,7.2){\dimms{1}}   

\put(2.72,7.2){\makebox(0,0)[r]{\small$\dH=1$}}
\put(2.72,6.4){\makebox(0,0)[r]{\small$3$}}
\put(2.72,5.6){\makebox(0,0)[r]{\small$5$}}
\put(2.72,4.8){\makebox(0,0)[r]{\small$7$}}
\put(2.72,4){\makebox(0,0)[r]{\small$9$}}
\put(2.72,3.2){\makebox(0,0)[r]{\small$11$}}
\put(2.72,2.4){\makebox(0,0)[r]{\small$13$}}
\put(2.72,1.6){\makebox(0,0)[r]{\small$15$}}
\put(2.72,.8){\makebox(0,0)[r]{\small$17$}}
\put(2.72,0){\makebox(0,0)[r]{\small$19$}}

\thinlines
\put(2.83,-.45){\hspace{1mm}\line(1,0){7.67}}
\put(2.83,7.6){\hspace{1mm}\line(1,0){7.67}}

\multiput(3.33,7.6)(0,-0.1){11}{\hspace{1mm}\line(0,-1){0.04}}
\multiput(3.33,6)(0,-0.1){64}{\hspace{1mm}\line(0,-1){0.04}}
\put(3.33,-.4){\hspace{1mm}\line(0,-1){0.1}}    
\multiput(5,7.6)(0,-0.2){40}{\hspace{1mm}\line(0,-1){0.1}}
\put(5,-.4){\hspace{1mm}\line(0,-1){0.1}}       
\multiput(10,7.6)(0,-0.1){3}{\hspace{1mm}\line(0,-1){0.04}}
\multiput(10,6.8)(0,-0.1){72}{\hspace{1mm}\line(0,-1){0.04}}
\put(10,-.4){\hspace{1mm}\line(0,-1){0.1}}      

\thicklines
\put(3.33,-.54){\hspace{1mm}\makebox(0,0)[t]{$1\over 3$}}
\put(3.33,-.41){\hspace{1mm}\line(1,0){1.67}}
\put(5,-.54){\hspace{1mm}\makebox(0,0)[t]{$1\over 2$}}
\put(4.26,-1.5){\mbox{$\renewcommand{\arraystretch}{.7} \ba
\uparrow
  \rule[-2mm]{0mm}{5mm} \\ \mbox{\small\em Fermi liquid} \\
  \mbox{\small\em behaviour}\ea$}}
\put(10,-.6){\hspace{1mm}\makebox(0,0)[t]{$1$}}
\put(10,-.41){\hspace{1mm}\line(1,0){.5}}
\put(10.4,-.66){\hspace{1mm}\makebox(0,0)[t]{$\sH$}}
\put(8.41,-1){\mbox{$\underbrace{\hspace*{30mm}}_{
  \renewcommand{\arraystretch}{.7} \ba \mbox{\small\em domain of}Ê\\
  \mbox{\small\em attraction} \\
  \mbox{\small\em of $\,\sH=1$}\ea}$}}

\put(3.33,7.56){\hspace{1mm}\line(1,0){1.67}}
\put(4.17,7.75){\hspace{1.2mm}\makebox(0,0)[b]{$\Sigma_1^+$}}
\put(7.5,7.75){\hspace{1.2mm}\makebox(0,0)[b]{$\Sigma_1^-$}}
\put(10,7.56){\hspace{1mm}\line(1,0){.5}}

\end{picture}
\end{flushleft}

\vspace{31mm}
\noi
{\bf Figure 1.2. }{\em Compilation of \Lmini\ $\,(\mini\lM\!=
\!3\mini)$ chiral quantum Hall lattices (CQHLs) with
odd-denominator Hall fractions $\mini\sH$ in the interval
$\,1/3\leq\sH\leq 1\mini$.}\rule[-4mm]{0mm}{5mm}

\noi
{\small The experimental status of the Hall fractions displayed
here is indicated, for single-layer systems, by
``$\,\bullet\,,\circ\,$'', and ``$\,\cdot\,$'', as in Fig.\,1.1.
Superposed on the interval  $\,1/3\leq\sH\leq 1$ of that figure
is a list of different \Lmini\ CQHLs:
``\raisebox{1.5mm}{\hspace{2.3mm}\maxsym\hspace{3mm}}'' indicates
maximally symmetric, \Lmini\ CQHLs of dimension $\mini
N\!\leq\!11\mini$ (where the corresponding dimensions are given
below the symbols);
``\raisebox{1.4mm}{\hspace{1.6mm}\lowdim\hspace{2.5mm}}''
indicates generic, indecomposable, \Lmini\ CQHLs of low
dimension, $N\!\leq\!4\mini$ (the respective dimensions are given
above the symbols). For fractions decorated with
``\raisebox{1.3mm}{\hspace{1.5mm}\nolodiLmin\,}'', there are no
low-dimensional ($N\!\leq\!4\mini$), \Lmini\ CQHLs. However, there
are maximally symmetric ones in ``high'' dimensions (with the
lowest such dimension indicated below the symbols). At fractions
with ``\raisebox{1.3mm}{\,\mini\noLmin\,}'', there are neither
low-dimensional, \mbox{\Lmini} CQHLs, nor maximally symmetric ones
in ``higher'' dimensions. In addition,
``\raisebox{1.5mm}{\hspace{2.5mm}\chacon\hspace{3.1mm}}'' stands
for non-chiral QH lattices that are ``charge-conjugated'' to the
maximally symmetric, \Lmini\ CQHLs in $\Sigi^+$.}


\newpage
\setcounter{page}{1}
\begin{flushleft}
\section{Introduction: Experimental Facts and Theoretical Ideas}
\label{sI}
\end{flushleft}

In this paper, we describe a classification of (universality classes
of) dissipation-free (incompressible) quantum Hall fluids in terms
of arithmetical invariants connected to integral lattices. The key
insight will be that the theory of certain classes of integral
lattices organizes experimental data in an efficient and accurate
way. We emphasize that the appearance of integral lattices in the
theory of the quantum Hall (QH) effect is not the consequence of
queer mathematical fantasizing devoid of physical insight, but is
the consequence of some fundamental physical principles and
properties, such as the absence of dissipation in an incompressible
QH fluid, electromagnetic gauge invariance, parity and time-reversal
breaking of the quantum mechanics of charged particles in an external
magnetic field, and the Fermi statistics of electrons. It is our aim
to show that integral lattices are fundamental to the theory of the
QH effect. It will therefore be impossible to spare the reader a
certain amount of mathematical reasoning involving lattice theory.

The integer QH effect has been discovered by von Klitzing and
collaborators, fifteen years ago, the fractional effect by Tsui and
collaborators, in $1982\mini$; see~\cite{QHE}. Since then this
remarkable effect of non-relativistic many-body physics has posed
numerous and diverse challenges to experimentalists and
theoreticians. As theorists, we should sadly confess that we have
anticipated few of the real surprises.
\rule[-4mm]{0mm}{5mm}

Experimentally, the QH effect is observed in two-dimensional systems
of electrons and/or holes confined to a planar region $\Omega$ and
under the influence of a strong, uniform magnetic field $\mini\Bc$
transversal to $\Omega$. Such systems can be realized as inversion
layers forming at the interface between an insulator and a
semiconductor when an electric field (gate voltage) perpendicular to
the interface is applied. Imagine that the sample is rectangular,
with $\Omega$ contained in the $(x,y)$-plane. By tuning the total
electric current $\,I=(I_x,I_y)\,$ to some value and measuring the
voltage drops, $V_x$ and $V_y$, in the $x$- and $y$-directions
of the plane of the system, we may determine the resistances
$R_{xx}$, $R_{yy}$, and $R_H$ from the equations

\bea
V_x &=& R_{xx}\, I_x - R_H\, I_y\ , \hah \nonumber \\
V_y &=& R_H\, I_x + R_{yy}\, I_y\ .
\label{URI}
\eea

One finds that, at temperatures $T$ very close to $\mini 0\,K$, $R_H$
is independent of $I\mini$; it only depends on a dimensionless
quantity $\mini\nu$, called {\em filling factor} and defined by

\be
\nu \:=\: {n\over (e \Bcperp/hc)}\ ,
\ee

\noi
where $n$ is the difference between the density of electrons and
the density of holes in the sample, $\Bcperp$ is the component of
the external magnetic field $\mini\Bc$ perpendicular to the plane of
the sample, and $hc/\mmini e$ is the quantum of magnetic flux.
Treating electrons and holes as classical point particles, one finds
by equating electrostatic- and Lorentz force that, in a stationary
state,

\be
{1\over R_H} \:=\: \nu\, {e^2\over h}\ ,
\label{classic}
\ee

\noi
the constant of proportionality, $\eht$, being a universal
constant of nature. Since, experimentally, $n$ can be varied (by
varying the gate voltage) and $\Bcperp$ can be varied, the classical
prediction~(\ref{classic}) can be tested. Experiments at very low
temperatures, with rather pure samples, yield surprising data: The
experimental curve for $R_H^{-1}$ as a function of $\mini\nu\mini$
shows plateaux, i.e., small intervals of values of $\mini\nu$, where
$R_H^{-1}$ is constant. Whenever $(\nu, R_H^{-1})$ belongs to a
plateau then

(i) $R_{xx}$ and $R_{yy}$ very nearly vanish;

(ii) $R_H^{-1}$ is a {\em rational\,} multiple of $\eht$. The
plateaux, where $\,R_H^{-1}=\nH\mini\eht$, for some integer $\,\nH=
1,2,3,\ldots$ (not too large), occur with an astounding precision of
one part in $10^8$. The plateau-height quantization is insensitive
to sample preparation (e.g., to impurities) and -geometry, for all
practical purposes.

(iii) Only a limited (experimentally, a finite) set of rational
numbers appear as plateau-heights of $R_H^{-1}h/\mmini e^2$. The
behaviour of $R_H^{-1}$ as a function of $\mini\nu\mini$ between
neighbouring plateaux appears to exhibit universal features. In such
transition regions, $R_{xx}$ and $R_{yy}$ are non-zero.
\rule[-4mm]{0mm}{5mm}

These (and other) experimental findings pose fascinating problems
to the theorist:

{\bf (1)} Applying non-relativistic many-body theory to a
two-dimensional system of interacting electrons in an external
magnetic field, can one predict the values of $\mini\nu\mini$ at
which $R_{xx}$ and $R_{yy}$ vanish?

{\bf (2)} If $R_{xx}$ and $R_{yy}$ vanish, can one predict the
possible values of $R_H$? Writing

\be
R_H^{-1} \:=\: \sH\,\eh\ ,\hwh \sH \:=\: \ndH\ ,
\ee

\noi
where $\mini\nH\mini$ and $\mini\dH\mini$ are two integers without
common divisor, we would like to understand which set of rational
numbers, $\nH/\dH$, corresponds to plateau-heights of the {\em
dimensionless Hall conductivity (or Hall fraction)} $\mini\sH\mini$
in real samples. Do only special types of integers appear as
numerators, $\nH$, or denominators, $\dH$, of $\mini\sH$;
(``odd-denominator rule'')? Conversely, can we predict which
rational numbers will ``never'' appear as plateau-heights of
$\mini\sH$? How does the set of observed plateau-heights depend
on properties of the sample, e.g., on the number of interacting
layers, the width of the quantum well corresponding to a layer, the
in-plane component, $\Bcpara$, of the applied magnetic field, etc.?
Given an observed plateau-height of $\mini\sH$, can we say something
about the stability of the corresponding state of the system?

{\bf (3)} What is the structure of the quantum-mechanical state of
the system when $(\nu,\sH)$ lies in between two plateaux; e.g., when
$\,\nu=1/4\,$ or $\,\nu=1/2$, in a single-layer sample?
Experimentally, the transitions between plateaux do not appear to
exhibit any hysteresis phenomena. Does this mean that these
transitions are continuous and pass through a critical point where
one should observe critical phenomena? If this is the case what kind
of theories describe the critical points? Can we predict the
(relative) widths of plateaux and of transition regions?
\rule[-4mm]{0mm}{5mm}

During the past five years, we have been involved in theoretical
work on many of these questions. While we feel that theorists have
gained a lot of fairly convincing heuristic insight in the direction
of answering these questions, it is only the questions described
under point {\bf (2)}, above, to which we have what we would like to
think are fairly definitive and mathematically precise answers. The
description and mathematical derivation of some of these answers
form the main contents of this paper. (We hope to present some of
our insights into questions posed in points {\bf (1)} and {\bf (3)}
in future communications.)

The ground work for our approach to the problems described under
point {\bf (2)}, above, has been carried out in Refs.~\cite{FK}
through~\cite{FKST}. It owes much inspiration to work of
Halperin~\cite{Ha} and Read~\cite{Read} and overlaps with work by
Wen and others~\cite{edgetheo}; (see also the books quoted
in~\cite{QHE}, and~\cite{edge,edgeexp}).

Next, we recapitulate the key theoretical facts underlying our
analysis. In this work, we use units where the electron's charge,
$-e$, and Planck's constant, $h$, equal unity. A two-dimensional
system of electrons and/or holes in a transversal, external magnetic
field exhibiting the Hall effect ($R_H\neq 0$) is called a {\em QH
system\mini}. If $R_{xx}$ and $R_{yy}$ vanish it is called an {\em
incompressible QH fluid\,} or, for short, a {\em QH fluid\mini}.

Our purpose, in this paper, is to explain or predict {\em
universal\,} properties of QH fluids at temperatures $\,T\approx
0\,K$. It is therefore reasonable to look for a description of such
systems in the {\em scaling limit\mini}. Thus we consider a family,
parametrized by a scale parameter $\theta$, with $\,1\leq\theta
<\infty$, of ever larger samples confined to regions $\,\Oth:=\{
\,\vx\: |\:Ê\vx/\theta=:\vxi\in\!\Omega \}\,$ in the $(x,y)$-plane.
We describe the system in $\mini\Oth$ in terms of rescaled space- and
time coordinates $(\tau,\vxi\,)$, where $\,\vxi=\vx/\theta,\
\vx\in\!\Oth,\ \tau=t/\theta$, and $\,t\in\!\BB{R}\,$ denotes time.
The property that $R_{xx}$ and $R_{yy}$ vanish in QH fluids can be
interpreted as indicating that the ground state energy of such a
quantum fluid confined to the region $\mini\Oth$ is separated from
the rest of its spectrum of energies of (extended) states by a {\em
mobility gap\,} $\Delta^{(\theta)}$, with

\be
\Delta^{(\theta)} \:\geq\: \Delta_\ast \:>\: 0\ ,
\label{mob}
\ee

\noi
for all $\mini\theta$. {}From assumption~(\ref{mob}) it follows that
the universal physics of QH fluids in the scaling limit,
$\theta\rightarrow \infty$, is described by a {\em topological
field theory\mini}. For the purpose of predicting the values of
$\mini\sH$, or of other electric transport properties, it is
sufficient to determine the Green functions of conserved current
densities, in particular of the electric current density, in the
scaling limit. Thus, let $\,j_1,\ldots,j_N\,$ be a list of {\em
all\,} current densities of a QH fluid which, in the scaling
limit, are {\em independently conserved\mini}. We write

\be
j_k\txi \:=\: (\mini j_k^0\txi,\mini\vec{j}_k\txi\mini)\ ,
\ee

\noi
where $\mini j_k^0\mini$ is the charge density and $\mini\vec{j}_k$
the vector current density associated with $\mini j_k\mini$,
$k=1,\ldots,N$. Saying that $\mini j_k\mini$ is {\em conserved\,}
means that it satisfies the {\em continuity equation}

\be
{1\over c}{\partial\over \partial\tau}\,
j_k^0-\vec{\bigtriangledown}\cdot\vec{j}_k \:=\: 0\ .
\label{continu}
\ee

The {\em total electric\mini} current density, $j_{el}\mini$, must
always be among the conserved current densities of a QH fluid. Thus
there are real numbers $\,Q_1,\ldots,Q_N\,$ such that

\be
j_{el} \:=\: \sum_{k=1}^N Q_k\,j_k\ .
\label{jel}
\ee

Let $\,<\ldots>^{(\theta)}$ denote the quantum-mechanical
expectation in the ground state of a QH fluid confined to
$\mini\Oth$. Let $\,\xi:=(\xi^0,\xi^1,\xi^2)=(c\tau,\vxi\,),\
\vxi\in\!\Omega$, and $\,\partial_\mu:=\partial/\partial_\mu$.
We define the ``vacuum polarization tensor'', $\Pi$, in the scaling
limit by

\be
\Pi_{kl}^{\mu\nu}(\xi,\eta) \,:=\: \lim_{\theta\rightarrow \infty}
\theta^4 <\! T[\,j_k^\mu(\theta\xi)\,j_l^\nu(\theta\eta)\mini]
\!>^{(\theta)}\ ,
\label{pol}
\ee

\noi
for $\,\mu,\nu=0,1,2$, and $\,k,l=1,\ldots,N$. In~(\ref{pol}), we
are using that a conserved current density of a two-dimensional
system scales like the square of an inverse length; (conserved
current densities {\em cannot\,} have anomalous scaling
dimensions). It follows from the continuity
equations~(\ref{continu}) that

\be
\partial_\mu\Pi_{kl}^{\mu\nu} \:=\: \partial_\nu\Pi_{kl}^{\mu\nu}
\:=\: 0\ ,\hfa k,l=1,\ldots,N\ .
\label{divpol}
\ee

\noi
{}From~(\ref{divpol}) and the fact that the current densities
$\mini j_k$ have scaling dimension $2$, it follows that, for
$\mini\vxi\,$ and $\mini\vec{\eta}\,$ in the interior of
$\mini\Omega$,

\be
\Pi_{kl}^{\mu\nu}(\xi,\eta) \:=\: i\,S^{kl}\,
\varepsilon^{\mu\nu\rho} \partial_\rho\mini \delta^{(3)}(\xi-\eta)
\fsp\,(+\,\cdots\,)\ ,
\label{polCS}
\ee

\noi
where the coefficients $S^{kl}$ are the matrix elements of a {\em
symmetric} $N\!\times\!N$ matrix $S$ and are {\em
dimensionless\mini} (in our units, where $h\!=\!-e\!=\!1$). The terms
$(+\cdots\,)$ omitted on the r.h.s.\ of~(\ref{polCS}) involve second
or higher derivatives of $\delta$-functions and have {\em
dimensionful\mini} coefficients, (with dimensions of a first or
higher power of length). They are of subleading order in the scaling
limit. Let $N_+,\ N_-$, and $N_0$ denote the number of positive,
negative, and zero eigenvalues of $S$, respectively. By rescaling
the current densities $\mini j_k\mini$ and introducing suitable
linear combinations thereof, we can always achieve that

\be
S^{kl} \:=\: s_k\,\delta^{kl}\ ,
\label{Sdia}
\ee

\noi
with $\,s_k=1$, for $\,1\leq k\leq N_+,\ s_k=-1$, for
$\,N_+\!+\!1\leq k\leq N_+\!+\!N_-$, and $s_k=0$, otherwise. We may
henceforth assume that the current densities $\mini j_k\mini$ have
been chosen in such a way that~(\ref{Sdia}) holds. In discussing
electric transport properties {\em in the scaling limit\mini} and
predicting the {\em possible values} of $\mini\sH$, current
densities $\mini j_k\mini$ corresponding to $\,s_k=0\,$ are
irrelevant, and we may therefore assume that $\,N_0=0,\
N=N_+\!+\!N_-$.

Note that, for $\,S\neq 0$, the tensor $\mini\Pi\mini$ violates
parity and time-reversal invariance. Thus, the ground state of a QH
fluid is {\em not\,} invariant under parity and time-reversal,
unless $\,N_+=N_-=0$. This is to be expected of a system of charged
particles in an external magnetic field.

It follows from~(\ref{polCS}) and~(\ref{jel}) that

\bea
\Pi_{el}^{\mu\nu}(\xi,\eta) & :=& \lim_{\theta\rightarrow \infty}
\theta^4 <\! T[\,j_{el}^\mu(\theta\xi)\,j_{el}^\nu(\theta\eta)\mini]
\!>^{(\theta)} \nonumber \\
&=& i <\!\Q\,,\Q\!>\,\varepsilon^{\mu\nu\rho}\mini \partial_\rho
\mini\delta^{(3)}(\xi-\eta)\ ,
\label{polel}
\eea

\noi
where $\Q$, with components $\,Q_1,\ldots,Q_N$, introduced
in~(\ref{jel}), is called ``{\em charge vector\,}'', and

\be
<\!\Q\,,\Q\!> \;=\: \sum_{k,l=1}^N Q_k\,S^{kl}\,Q_lÊ\:=\:
\sum_{k=1}^N s_k\,Q_k^2\ ,
\ee

\noi
where the second equality holds if the ``normalization
conditions''~(\ref{Sdia}) are imposed.

{}From the basic equations of the electrodynamics of QH fluids
(see~\cite{FK} and~\cite{FS2}) we know that the coefficient,
$<\!\Q\,,\Q\!>$, on the r.h.s.\ of~(\ref{polel}) is nothing but the
{\em dimensionless Hall conductivity} $\mini\sH$, i.e.,

\be
\sH \:=\; <\!\Q\,,\Q\!>\ .
\label{sHQQ}
\ee

Since the theory describing a QH fluid in the scaling limit is a
{\em topological field theory} ($\Delta_\ast\!>\!0\mini$!), as
remarked above, all excitations above the ground state of a QH fluid
of finite energy and localized in compact regions contained in the
bulk of the system (``{\em quasi-particles\,}'') can be described, in
the scaling limit, as {\em pointlike, static} sources of the
topological field theory (located at points in the interior of
$\mini\Omega$). One can show~\cite{FK,FGM} that one can assign $N$
charges, $q^1,\ldots,q^N$, to every such source. The charge
$\mini q^k\mini$ is an eigenvalue of the conserved total charge
operator corresponding to the conserved current density
$\mini j_k\mini$; this charge operator is normalized in such a way
that the ground state of the system has charge zero. By~(\ref{jel}),
the total electric charge of a source described by a vector
$\mini\vq\mini$ of charges, $q^1,\ldots,q^N\mini$, is given by

\be
\qel(\vq) \:=\: \sum_{k=1}^N Q_k\,q^k\ .
\ee

If a source with a vector $\mini\vq_1\mini$ of charges is transported
(adiabatically) around a source with a vector $\mini\vq_2\mini$ of
charges along a counter-clockwise oriented loop not enclosing other
sources a corresponding quantum-mechanical state vector is
multiplied by an ``Aharonov-Bohm phase factor''

\be
\exp\,(2\mini\pi\mini i\!<\!\vq_1\mini,\vq_2\!>)\ ,
\label{mono}
\ee

\vwv

\be
<\!\vq_1\mini,\vq_2\!> \;=\: \sum_{k,l=1}^N
q_1^k\,(S^{-1})_{kl}\,q_2^l\ .
\ee

\noi
If two identical sources labelled by vectors $\mini\vq_1,\:
\vq_2\mini$ of charges, with $\,\vq_1=\vq_2=\vq$, are
(adiabatically) {\em exchanged\,} along counter-clockwise oriented
paths not enclosing other sources then a corresponding
quantum-mechanical state vector is multiplied by the phase factor

\be
\exp(\mini\pi\mini i\!<\!\vq\,,\vq\!>)\ .
\label{halfmono}
\ee

\noi
These are properties of physical state vectors of the topological
field theory, an abelian Chern-Simons theory of $N$ gauge fields,
that reproduces the current Green functions given in~(\ref{polCS}).
They have been derived and discussed in great detail in previous
papers; see~\cite{FK,FZ,FS2,FT}.

The conventional connection between electric charge and quantum
statistics in a quantum-mechanical gas of non-relativistic
electrons says that whenever the total electric charge, $\qel(\vq)$,
of a localized excitation labelled by a vector $\mini\vq\mini$ of
charges is an {\em even (odd)\,} integer (in units where
$\,e\!=\!-1$), i.e., the excitation is composed of an {\em even
(odd)\,} number of electrons and/or holes, then the excitation
obeys {\em Bose-Einstein (Fermi-Dirac)\,} statistics. This
{\em charge-statistics connection}, together with~(\ref{halfmono}),
implies that every vector $\mini\vq\mini$ corresponding to an
integer electric charge $\mini\qel(\vq)$ satisfies the constraint

\be
\qel(\vq) \:\equiv\; <\!\vq\,,\vq\!>\; \bmod\ 2\ .
\label{charstat}
\ee

\noi
Moreover, it follows from the charge-statistics connection
and~(\ref{mono}) that if $\mini\vq_1\mini$ and $\mini\vq_2\mini$ both
correspond to {\em integer} electric charges, $\qel(\vq_1),\
\qel(\vq_2) \in\!\BB{Z}\,$, then \mbox{$\,<\!\vq_1\mini,\vq_2\!>\,$}
is an {\em integer}. Finally, the vectors $\mini\vq\mini$ for which
$\mini\qel(\vq)$ is an integer form an {\em additive group\mini};
addition corresponding to the composition of two excitations, and
the operation $\,\vq \rightarrow-\vq\,$ corresponds to ``charge
conjugation'' (electron-hole exchange).

A detailed account of the arguments just sketched can be found
in~\cite{FT}. The key result that they imply is that the vectors
$\mini\vq\mini$ of charges belonging to the set

\bea
\G & := & \{\, \vq\in\BBn{R}{N}\,|\ \qel(\vq)\in\BB{Z}\mini,\
\qel(\vq) \:\equiv\; <\!\vq\,,\vq\!>\; \bmod\ 2\, \}
\eea

\noi
form an {\em integral lattice\mini}. In other words, $\G\mini$ is an
additive group (a ``free $\BB{Z}\,$-module''), and, for any pair,
$\vq_1,\,\vq_2\mini$, of vectors in $\G$, $<\!\vq_1\mini,\vq_2\!>$
is an integer. We define the lattice {\em dual\,} to $\G\mini$ by

\bea
\Gs & := & \{\, \vn\in\BBn{R}{N}\, | <\!\vn\,,\vq\!>\:
\in\BB{Z}\mini,\ \mbox{for all }\, \vq \in\!\G \mini\}\ .
\eea

\noi
Since the charge vector $\Q$ introduced in~(\ref{jel})
and~(\ref{polel}) has the property that

\be
<\!\Q\,,\vq\!> \;=\: \qel(\vq)\in\BB{Z}\ ,\hfa \vq\in\!\G\ ,
\ee

\noi
it follows that $\,\Q\in\!\Gs$. This implies that $<\!\Q\,,\Q\!>$
is a rational number, and hence, by~(\ref{sHQQ}), the Hall fraction
$\,\sH=\nH/\dH=\:<\!\Q\,,\Q\!>\,$ is {\em rational\,}!

An electron and a hole are among the localizable, physical
excitations of a QH fluid. Thus there must exist some vector
$\,\vq\in\!\G\,$ with the property that

\be
\qel(\vq) \:=\; <\!\Q\,,\vq\!> \;=\: 1\ .
\label{qone}
\ee

\noi
Then~(\ref{charstat}) implies that $<\!\vq\,,\vq\!>$ is an {\em
odd\,} integer; hence $\G$ is what is called an {\em odd\,} integral
lattice, and, by~(\ref{qone}), $\Q$ is a so-called {\em primitive}
(or {\em visible}) vector of $\mini\Gs$. Moreover, by reading the
charge-statistics connection~(\ref{charstat}) (which holds for
all $\,\vq\in\!\G$) as a constraint on $\Q$, we say that $\Q$ is an
{\em odd\,} vector of $\mini\Gs$.

It is a basic fact of non-relativistic quantum theory that state
vectors are single-valued in the positions of electrons and holes.
Let $\mini\vn\mini$ be a vector of charges of an arbitrary,
localizable physical excitation of a QH fluid, and let
$\,\vq\in\!\G$. Then, by~(\ref{mono}), and since state vectors are
single-valued in the positions of electrons and holes,
$<\!\vn\,,\vq\!>$ must be an integer, and hence

\be
\vn \:\in\, \Gs\ .
\ee

\noi
Thus, the vectors of charges of localizable physical excitations
form a lattice $\mini\Gp$ contained in or equal to $\mini\Gs$.

The conclusion reached, so far, is that: {\em In the scaling limit,
an (incompressible) QH fluid with $N$ conserved current densities,
$j_1,\ldots,j_N$ (we shall speak of $\mini N$ ``channels''),
$N=1,2,\ldots\,$, can be characterized by the data

(i) an $N$-dimensional, odd, integral lattice $\G$;

(ii) an odd, primitive vector $\,\Q\in\!\Gs$, with $<\!\Q\,,
\Q\!>\:=\sH\mini$; and

(iii) a lattice $\Gp$, with $\,\G\subseteq\Gp\subseteq\Gs$.}
\rule[-4mm]{0mm}{5mm}

A pair \QHL\, is called a {\em quantum Hall lattice}. If the
integral quadratic form (or metric) $<\,,>$ defined on
$\mini\G\mini$ is either {\em positive-} or {\em negative-definite},
we say that \QHL\, is a {\em chiral\,} QH lattice (CQHL), for
reasons connected to the chirality of edge currents; see
Sect.\,\ref{sUC} and also~\cite{FS2,FT}. It is a plausible idea
about the physics of QH fluids that if $<\,,>$ is {\em not\,}
positive- or negative-definite then $\mini\G\mini$ can be {\em
decomposed\,} into an (orthogonal) direct sum,

\be
\G \:=\: \Ge\oplus\Gh\ ,
\label{decomp}
\ee

\noi
with the property that $\mini\Ge\ (\Gh)$ is an odd, integral
sublattice of $\mini\G$ on which $<,>$ is positive- (negative-)
definite. Decomposition~(\ref{decomp}) may not hold in general, but
it will serve as a fairly safe ``working hypothesis'' throughout
much of our paper. The physical basis of this working hypothesis
(decomposition of QH fluids into electron- and hole-rich subfluids)
will be discussed in Sect.\,\ref{sUC} and Appendix~E; see
also~\cite{FT}. (In Sect.\,\ref{sUC}, we summarize the basic
physical assumptions of our approach and provide the mathematical
notions connected to (chiral) QH lattices.)
\rule[-4mm]{0mm}{5mm}

Our aim in this paper, is to present a {\em partial classification}
of QH lattices. In view of our working hypothesis~(\ref{decomp}),
our main effort will concern the classification of {\em chiral\,}
QH lattices; (but see Appendix~E). We shall carefully compare our
results with experimental data on QH fluids, focussing our
attention primarily on data for single-layer QH fluids with
$\mini\sH\mini$ in the interval $\,0<\sH\leq 1\mini$. Our job
involves a characterization of QH lattices \QHL\, in terms of
numerical invariants; see Sect.\,\ref{sBI}\mini. Among these
invariants, the following ones play a key role:

(i) The dimension, $N$, of $\mini\G\mini$;

(ii) the discriminant of $\mini\G$, i.e., the order of the abelian
group $\mini\Gs\mA/\G$, where $\mini\Gs\mA/\G$ denotes the family
of cosets of $\mini\Gs$ mod $\mini\G$, (as well as more sophisticated
invariants involving $\mini\Gs\mA/\G$, e.g., the genus of $\mini\G$);

(iii) an invariant, denoted $\mini\lM\mini$, interpreted,
physically, as the smallest relative angular momentum of a certain
pair of two identical excitations of electric charge $1$
(electrons)\,--\,$\lM$ is an {\em odd\,} integer (see
Sect.\,\ref{sBI}); and, of course,

(iv) the dimensionless Hall conductivity (or Hall fraction),
$\sH=\:<\!\Q\,,\Q\!>\mini$.
\rule[-4mm]{0mm}{5mm}

For CQHLs, the invariants $\mini\lM$ and $\mini\sH\mini$ are related
by

\be
\lM \:\geq\: {1\over \sH}\ ,
\label{lMgeq}
\ee

\noi
which is a consequence of the Cauchy-Schwarz inequality; see
Sect.\,\ref{sGT}.

In our comparison between theory and experiment, we shall appeal to a
heuristic (analytically plausible, but mathematically unproven)
{\em stability principle\mini} which says that a QH fluid described
by a QH lattice \QHL\, is the more {\em stable}, the {\em
smaller\mini} the value of the invariant $\mini\lM$ and, given the
value of $\mini\lM$, the {\em smaller\mini} the dimension $N$ (and
the discriminant) of $\mini\G$; see Sects.\,\ref{sGT},~\ref{sLD},
and~\ref{sDis}\mini. A measure for the stability of a QH fluid is,
for example, the width of the plateau of $\mini\sH\mini$ (as a
function of $\mini\nu$) corresponding to that QH fluid.

In view of~(\ref{lMgeq}), it is useful to decompose the interval
$(0\mini,1]$ of values of $\mini\sH\mini$ into subintervals
(``windows'') $\,\Sigp=\Sigp^+\cup\Sigp^-$, where

\be
\Sigp^+ \::=\: [\mini{1\over 2\mini p+1}\mini,\mini {1\over 2\mini
p}\mini)\ , \hah
\Sigp^- \::=\: [\mini{1\over 2\mini p}\mini,\mini {1\over 2\mini
p-1}\mini)\ ,\hsp p=1,2,\ldots\ .
\label{sigpm}
\ee

\noi
The invariant $\mini\lM$ of a CQHL \QHL\, with $\,\sH\in\!\Sigp\,$ is
bounded below by $\mini 2\mini p\!+\!1\mini$. We define
$\mini\Hp^\pm$ to be the class of all CQHLs, \QHL, with
$\,\sH\in\!\Sigp^\pm\,$ and $\,\lM=2\mini p\!+\!1$, (and which
are, to be technically precise, ``primitive'', as specified in
Sect.\,\ref{sUC}). We shall see that, for all $\mini p$, {\em all\,}
CQHLs in $\mini\Hp^+$ can be enumerated explicitly, and that, for
$\,p\!\leq\!3\,$ and sufficiently small values of their dimension
(stability principle\mini!), they correspond to experimentally well
verified plateaux of $\mini\sH$.

There are heuristic analytical and numerical arguments, as well
as convincing phenomenological evidence, indicating that the most
stable state of a QH system with $\,\nu<1/7\,$ is one where the
electrons form a {\em Wigner lattice}. But a Wigner lattice is
incompatible with a positive mobility gap $\D_\ast$, i.e., with
incompressibility. By~(\ref{lMgeq}), this implies that the
invariant $\mini\lM$ of a {\em chiral\,} QH lattice corresponding to
an experimentally realizable QH fluid is bounded above by

\be
\lM \:\leq\: 7\ .
\label{WCsI}
\ee

There is reasonable, analytical evidence~\cite{HLRAIM} that {\em
single-layer\mini} QH systems with filling factors $\,\nu=1/2,\
1/4\,$, and various other even-denominator fractions are described
by gapless (possibly marginal) Fermi liquids. Thus, e.g.,
$\,\sH=1/2\,$ and $\,\sH=1/4\,$ should {\em not\,} correspond to
plateaux of {\em single-layer\mini} QH fluids.

The CQHL $(\G\!=\!\BB{Z}\,,\Q\!=\!1)$ has invariants $\,N=1,\
|\Gs\mA/\G|=1,\ \lM=1$, and $\,\sH=1$. It describes the {\em by far
most stable\mini} QH fluid with a Hall conductivity
$\,\sH\in\!(0\mini,1]$. Thus the plateaux at $\,\sH=1\,$
should have {\em by far the broadest width\mini} among {\em
all\,} plateaux at values of $\mini\sH\mini$ in the interval
$(0\mini,1]$. QH fluids described by QH lattices of dimension
$\,N\!>\!1$, discriminant $>\!1$, and with $\mini\sH\mini$ close to
$1\mini$ (e.g., $\,6/7\,\lsim\,\sH<1$) are expected to be very
unstable against transitions to the QH fluid at $\,\sH=1\,$
described by $\,(\G,\Q)= (\BB{Z}\,,1)\,$ and are therefore likely to
be {\em invisible\mini} experimentally.

In Fig.\,\ref{sI}.1, we display experimentally observed
plateau-values of $\mini\sH\mini$ in the interval $\,0<\sH\leq 1\,$
and indicate the quality of their experimental verification. (For
general experimental reviews of the (fractional) QH effect, see,
e.g., \cite{nus,Saj} and references therein. Recent data on QH
fluids with Hall fractions belonging to the ``two main series'',
$\sH=N/(2\mini N\!\pm\! 1),\ N=1,\ldots,9$, can be found
in~\cite{main,Kang}. For the status of a QH fluid with
$\,\sH=10/17$, see~\cite{GS}. We recall that the signals observed
at $\,\sH=4/11\,$ and $\,\sH=4/13\,$ appear to be very weak;
see~\cite{Hu}. Magnetic field and density driven phase transitions
have been reported at $\,\sH=2/3\,$ in~\cite{Cl90,Eis90,Eng}. A
magnetic field driven phase transition at $\,\sH=3/5\,$ has been
established in~\cite{Eng}, and a possible phase transition at
$\,\sH=5/7\,$ has been discussed in~\cite{Saj}.) In
Fig.\,\ref{sI}.1, we write $\,\sH=\nH/\dH\,$ and display the data
in a ``$\dH\mini$ versus $\mini\sH\mini$ plot''. We subdivide the
interval $(0\mini, 1]$ into the windows $\mini\Sigp^\pm$ introduced
above, for $\,p=1,2$, and $3\mini$. \rule[-4mm]{0mm}{5mm}


\addtocounter{page}{1}
It may happen that there are {\em several\,} QH lattices with the
{\em same\mini} Hall fraction $\mini\sH$. At such values of
$\mini\sH$, we predict {\em phase transitions between ``structurally
different'' QH fluids}, as, e.g., the in-plane component, $\Bcpara$,
of the external magnetic field (and thus the magnitude of Zeeman
energies associated with the magnetic moment of electrons), or the
density of electrons (at fixed filling factor), or the width of the
layer to which the electrons (or holes) are confined are varied. A
theory of such phase transitions is developed in Sect.\,\ref{sDis}
and the results are summarized in Appendix~D. The most likely Hall
fractions $\mini\sH\mini$ at which they may occur are $\,2/3,\ 3/5,\
4/7,\ 5/7,\ 5/9$, and $\,1/2\mini$!

We shall find (see Sect.\,\ref{sMS} and Appendix~B) a nice, simple
CQHL \QHL\, with $\,N=3,\ \lM=3$, and $\,\sH=1/2\mini$. However, in
{\em single-layer\mini} QH systems, there is {\em no\mini} plateau
at $\,\sH=1/2$, and we just said that there is analytical evidence
for the idea that the ground state of a QH system at $\,\nu=1/2\,$
is a gapless Fermi liquid. So, is there a problem with our theory?
In order to understand what is going on at $\,\sH=1/2\,$ (and at
various other values of $\,\sH\in\! (0\mini,1]$), it is useful to
consider yet one further invariant of integral lattices, the
so-called {\em Witt sublattice\mini}. Given an integral lattice
$\mini\G$, its Witt sublattice, $\Gw$, is defined to be the
sublattice generated by all vectors $\,\vq\in\!\G$, with
$\,<\!\vq\,,\vq\!>\:=1\,$ or $2\mini$. It turns out that, for {\em
(indecomposable) chiral\,} QH lattices \QHL, the Witt sublattice
$\mini\Gw$ of $\mini\G$ is always the {\em root lattice\mini} of a
semi-simple Lie algebra $\mini\GG$; more precisely, $\Gw$ is an
orthogonal direct sum of $\mini A$-, $D\mini$-, and
$E_{6,7}\mini$-root lattices. (These notions are explained in
Appendix~A.) Furthermore, the Lie group $\mini G$ corresponding to
the Lie algebra $\mini\GG$ whose root lattice is given by $\mini\Gw$
is a {\em symmetry group\mini} of the topological quantum theory
describing the scaling limit of the QH fluid corresponding to \QHL,
in a sense that has been made precise in~\cite{FZ,FS2,FT} and is
briefly reviewed in Sect.\,\ref{sMS}\mini. Standard physics often
permits us to determine at least {\em some\mini} of the symmetries
of QH fluids (in the scaling limit).

For example, if the {\em effective\mini} gyromagnetic factor of an
electron in a QH fluid is {\em small}, so that Zeeman energies can
be essentially neglected, then the scaling limit of a QH fluid in an
only moderately large magnetic field is expected to exhibit an
\SUs\  global symmetry (spin-flip); see~\cite{HaHPA,FS2}. In this
case, the Witt sublattice, $\Gw$, of the QH lattice describing the
QH fluid must contain the root lattice, $\sqrt{2}\,\BB{Z}\mini$, of
$su(2)$. Furthermore, if we consider a {\em double-layer\mini} QH
fluid which, in the scaling limit, exhibits an $SU(2)_{layer}$
symmetry (coherent superposition of modes in the two layers with
$SU(2)$ symmetry) then $\mini\Gw$ must contain an $su(2)$-root lattice.
One can easily imagine that there are double-layer QH fluids
exhibiting (in the scaling limit) {\em both\mini} symmetries, an
\SUs\ and an $SU(2)_{layer}$ symmetry. Then $\mini\Gw$ must contain the
direct sum of {\em two\mini} $su(2)$-root lattices. It so happens
that there is a {\em three-dimensional CQHL \QHL, with} $\,\lM=3,\
\Gw=\sqrt{2}\,\BB{Z}\oplus\!\sqrt{2}\,\BB{Z}\mini$ (direct sum of two
$su(2)$-root lattices), {\em and\,} $\sH=1/2\mini$. This matches the
recent experimental observation of a plateau at $\,\sH=1/2\,$ in
double-layer (or two-component) QH systems~\cite{1/2,Su}.

Incidentally, ``layer'' could also stand for ``filled Landau level'',
and this remark suggests a theoretical explanation of the observed
plateau at $\,\sH=5/2\mini$~\cite{Wil5/2,5/2}.

There is also a {\em two-dimensional\,} CQHL \QHL, with $\,\lM=3,\
\Gw=\emptyset$, and $\,\sH=1/2\mini$. It might describe an
incompressible QH fluid consisting of {\em two\mini} interacting
layers of {\em spin-polarized\,} electrons with a $\BB{Z}_{\mini 2}$
layer permutation symmetry. Since $\BB{Z}_{\mini 2}$ is a discrete
symmetry, it does not contribute to $\Gw$, but constrains the
structure of \QHL.

The moral to be drawn from this discussion is that we are well
advised to search for global symmetries (discrete and, especially,
continuous ones) of the theory that describes the scaling limit of
a QH fluid. The continuous symmetries appear as root lattices
contained in the Witt sublattice of the QH lattice describing the
fluid.

It has been shown in~\cite{FT} that, for CQHLs \QHL\, with
$\,\sH<2$,

\be
<\!\Q\,,\vq\!> \;=\: 0\ ,\hfa \vq\in\!\Gw\ ,
\ee

\noi
i.e., $\Q\mini$ is orthogonal to $\mini\Gw$. Let $\mini\G_0$ denote
the sublattice of $\mini\G$ consisting of {\em all\,} vectors in
$\mini\G$ that are orthogonal to $\Q$. Clearly, for $\,\sH<2$,
$\G_0$ contains $\mini\Gw$, and, obviously, $\dim\G_0\leq\dim\G-1\mini$.

These remarks suggest that an interesting class of QH lattices
consists of those CQHLs \QHL\, for which

\be
\G_0 \:=\: \Gw\ , \hah \dim\Gw \:=\: \dim\G-1\ .
\ee

\noi
We call such lattices ``{\em maximally symmetric\,}'' CQHLs.
Section~\ref{sMS} of this paper is devoted to a classification of
all maximally symmetric CQHLs with $\,0<\sH\leq 1\mini$.

Recall that $\mini\Hp^\pm,\ p=1,2,\ldots\,$, has been defined to be the
class of all (primitive) CQHLs with $\,\sH\in\!\Sigp^\pm$
(see~(\ref{sigpm})) and with $\,\lM=2\mini p\!+\!1\,$ (which,
by~(\ref{lMgeq}), is the {\em minimal\,} value the invariant
$\mini\lM$ can have, for $\,\sH\in\!\Sigp$). Lattices in
$\mini\Hp^\pm$ are said to be {\em\Lmini\mini}. We shall show that
{\em all\,} lattices in $\Hp^+$ are {\em maximally symmetric}, their
Witt sublattice is an $A_{N-1}\mini$- (or $su(N)\mini$-) root
lattice, and their Hall fraction is $\,\sH=N/(2\mini pN\!+\!1),\
N=(1),2,3,\ldots\,$; see Sect.\,\ref{sGT}\mini. This series of CQHLs
is called the ``basic'' $A\mini$- (or $su(N)\mini$-) series in the
window $\mini\Sigp$. We shall find a {\em bijection}, $\Sp$, called
{\em shift map}, mapping the basic $A$-series in the window $\mini
\Sigi^+$ onto the basic $A$-series in the window
$\mini\Sigma_{p+1}^+,\ p=1,2,\ldots\ $. In fact, the shift map $\Sp$
is defined on $\mini{\cal H}_q^\pm\mini$ and is a bijection from
$\mini {\cal H}_q^\pm\mini$ to $\mini{\cal H}_{q+p}^\pm$. On the
sets $\mini{\cal H}_q^\pm$, the action of the map $\Sp$ on the
invariants $\mini\sH\mini$ and $\mini\lM$ is given by

\be
{1\over \sH} \:\rightarrow\: {1\over \sH} + 2\mini p\ ,\hah
\lM \:\rightarrow\: \lM + 2\mini p \ ,\ p=1,2,\ldots\ .
\label{shiftsI}
\ee

If $(\G^\prime,\Q^\prime)$ is the image of a CQHL \QHL\, under
$\Sp,\ p=1,2,\ldots\,$, then, by~(\ref{shiftsI}), and invoking our
stability principle, the QH fluid corresponding to $(\G^\prime,
\Q^\prime)$ is {\em less stable\mini} then the one corresponding to
\QHL. Hence the number of observed plateau-values in a window
$\mini\Sigp$ {\em decreases\mini} with $\mini p$ (reaching $0$ when
$\,p\!>\!3$).

The existence of the shift maps $\Sp$ and the observation just
described allow us to restrict our classification of \Lmini\ CQHLs
to the class $\,\Hi=\Hi^+\cup\Hi^-$.\,--\,This is {\em not\,} true
if we want to classify {\em all\,} QH lattices, not just chiral ones.
However, among QH lattices that are {\em not\mini} chiral, the
``non-euclidean hierarchy lattices'' are well understood (see
Appendix~E) and they are, perhaps, the only physically important
non-chiral QH lattices.\,--\,{\em All\,} CQHLs in $\mini\Hi^+$ are
classified and are maximally symmetric, as remarked above and proven
in Sect.\,\ref{sMS}\mini. They form the basic $A$-series in
$\mini\Sigi^+$. The classification of lattices in $\mini\Hi^-$ is
much more difficult and remains incomplete. But besides the
maximally symmetric ones (Sect.\,\ref{sMS}), we have also classified
{\em all\,} CQHLs in $\mini\Hi^-$ of dimension $\,N\!\leq\!4\mini$.
Our results can be found in Sect.\,\ref{sLD}\mini. (With more
investment in programming and computer time, our results could be
extended to $\,N=5,6\mini$.)

In Fig.\,\ref{sI}.2, results of our theoretical work concerning QH
lattices with odd-denominator Hall fraction are superposed on the
experimental data (displayed in Fig.\,\ref{sI}.1) in the window
$\,\Sigi=\Sigi^+\cup\Sigi^-$. This figure shows a pretty remarkable
agreement between theory and experiment. {\em All\,} experimentally
observed Hall fractions $\mini\sH$ in the window $\mini\Sigi$, with
the only exception of the ``very weak'' fraction $\,\sH=4/11$, can
be realized by an \Lmini\ CQHL or a QH lattice which is
``charge-conjugated'' to an \Lmini\ one. Note that the
corresponding lattices are all of relatively low dimension, namely
$\,N\!\leq\!9\mini$. In Sect.\,\ref{sLD}, we shall see that,
interestingly, the ``simplest'' {\em non\,}-\Lmini\ CQHL is found
at $\,\sH=4/11\mini$. (It coincides with the proposals of the
hierarchy schemes at that fraction.) Fig.\,\ref{sI}.2 also shows
where experimentalists might wish to look for signals of new QH
fluids, or for new phase transitions between structurally different
QH fluids with the same value of $\sH$. \addtocounter{page}{1}


Meditating Fig.\,\ref{sI}.2, it may look disturbing that one seems to
have observed a phase transition at $\,\sH=2/5$, as the in-plane
component, $\Bcpara$, of the external magnetic field is
varied. There is a {\em unique} \Lmini\ CQHL with $\,\sH=2/5\mini$.
It is two-dimensional, with $\,\Gw=\sqrt{2}\,\BB{Z}\,$ (the root
lattice of $su(2)$). So a QH fluid with $\,\sH=2/5\,$ exhibits
a global $SU(2)$ symmetry (in the scaling limit). For ``{\em
small\,}'' values of the external magnetic field $\mini\Bc$, this
symmetry is \SUs, i.e., electron spins may be flipped. But when
$\mini\Bc$ is {\em large}, essentially all electron spins are
oriented in the direction antiparallel to $\mini\Bc$, and the
$SU(2)$ symmetry is an {\em internal\,} symmetry compatible with
the hierarchy pictures of Refs.~\cite{HH,JG}. A rather similar
story can be told about the plateau at $\,\sH=2/3$, (besides the
possibilities of interesting phase transitions between structurally
different QH fluids). All this and more is discussed in
Sect.\,\ref{sDis}\mini.
\rule[-4mm]{0mm}{5mm}

Two concluding remarks may be clarifying:

(i) The term ``incompressible QH fluid'' can be understood literally
in that shape fluctuations of a droplet of an (incompressible) QH
fluid with free boundaries are {\em area-preserving}. The Lie
algebra of area preserving maps has a central extension which is
connected to the $\mini W_{1+\infty}$ algebra. This algebra is
related to the abelian Chern-Simons theory that describes the
scaling limit of an (incompressible) QH fluid, in a natural way
first discussed by Sakita~\cite{Sak}. Its study in connection with
the QH effect has become a ``hot topic'' (see, e.g.,~\cite{Winf}),
but does not appear to lead to results that go beyond those
in~\cite{FS2,FT}, and in this paper.

(ii) The shift map $\,\Si:\,\sH^{-1}\rightarrow \sH^{-1}+2\,$ and the
map $\,T:\,\sH\rightarrow\sH+1$, cor\-re\-sponding to the addition of
a full Landau level, generate a subgroup, $\Gamma_T(2)$, of the
modular group $\mini PSL(2,\BB{Z})$. For fun, one can study the
action of $\mini\Gamma_T(2)$ on the plateaux values of $\mini\sH$.
More daringly, one can study the action of $\mini\Gamma_T(2)$ on the
complex plane of resisitivites $\,\rho:=\rho_{xx}+i\mini\sH^{-1}$,
(where $\,\rho_{xx}:=R_{xx}\mini l_y/l_x\mini$, with $\mini l_x$ and
$\mini l_y$ the length and width, respectively, of a rectangular QH
system). This has been advocated in~\cite{Luet} as a means to
understand a ``global phase diagram'' for the QH effect. However,
the reader who will make it through Sect.\,\ref{sGT} will see that
these are rather misleading speculations which, in the absence of
real understanding of the physics of QH systems, should not be taken
too seriously.
\rule[-4mm]{0mm}{5mm}

As to the contents of this paper, we have already indicated the
contents of Sects.\,\ref{sMS},~\ref{sLD}, and~\ref{sDis}\mini. In
Sect.\,\ref{sUC}, we recall the basic (physical and mathematical)
notions underlying our analysis. In Sect.\,\ref{sBI}, we introduce
and discuss basic invariants for CQHLs and explain their physical
interpretation. In Sect.\,\ref{sGT}, we present {\em general\,}
results on the classification of CQHLs. Sects.\,\ref{sMS}
and~\ref{sLD} concern the complete classification of {\em special
subclasses\mini} of CQHLs. In Sect.\,\ref{sDis}, we apply our
results to understand some of the {\em physics\mini} of the
fractional QH effect. In Appendix~A, we review some group theory
that is important in our analysis. Appendix~B summarizes all our
results on {\em maximally symmetric\mini} CQHLs with
$\,\sH\in\!(0\mini,1]$, Appendix~C those on {\em low-dimensional\,}
($N\!\leq\!4$) CQHLs. In Appendix~D, we summarize the results of the
theory of embeddings (expounded in Sect.\,\ref{sDis}) of \Lmini\
CQHLs into bigger ones, preserving the value of the Hall fraction
$\sH$. This will clarify the classification of the ``difficult''
classes $\Hp^-$. Finally, in Appendix~E, we present the QH lattices
that reproduce the Haldane-Halperin~\cite{HH} and
Jain-Goldman~\cite{JG} hierarchy fluids.


\newpage
\begin{flushleft}
\section{Universality Classes of QH~Fluids and QH~Lattices: Basic
Notions}
\label{sUC}
\end{flushleft}

In this section, we recall the basic physical principles and
assumptions leading to our characterization of (universality
classes of) QH fluids by QH lattices. We introduce the fundamental
mathematical notion of a chiral QH lattice (CQHL). As mentioned
in Sect.\,\ref{sI}, CQHLs are the ``basic building blocks'' of QH
lattices. Basic notions related to CQHLs are summarized. In order
to exemplify our language, we describe the (chiral) QH lattices
corresponding to the integer QH fluids of the non-interacting
electron approximation and the celebrated Laughlin fluids~\cite{L}.

Since the early theoretical work by Laughlin~\cite{L} on the QH
effect, the electromagnetic gauge symmetry of quantum mechanics has
been instrumental in analyzing this effect. This gauge symmetry also
provides the corner-stone of our
approach~\cite{FK,FZ,FS1,FS2,FT}.\,--\,We remark that a general
framework for a systematic discussion of phenomena related to
electron spin in QH fluids has been developed in~\cite{FS1,FS2}. It is
based on the presence of a non-abelian \SUs-gauge symmetry in
non-relativistic quantum many-body systems. Although we will not
review that general framework here, we emphasize that our results
presented in this paper are fully consistent with that general
picture, and, as a matter of fact, the present classification
results provide a basis for a systematic discussion of spin effects
in QH fluids. For further discussion of this point, see the
remarks about phase transitions in Sect.\,\ref{sDis},
and~\cite{FS2,FT}.\,--\,Besides gauge invariance, our approach
requires the following basic physical assumptions characterizing QH
fluids (see also Sect.\,\ref{sI}):
\rule[-4mm]{0mm}{5mm}

{\bf (A1)} The temperature $T$ of the system is close to $0\,K$. For
an (incompressible) QH fluid at $T=0\,K$, the {\em total electric
charge} is a good quantum number to label physical states of the
system describing excitations above the ground state;
see~\cite{FT,FGM}. The charge of the ground state of the system is
normalized to be zero.
\rule[-4mm]{0mm}{5mm}

{\bf (A2)} In the regime of very short wave vectors and low
frequencies, the {\em scaling limit}, the total electric current
density is the sum of $N=1,2,3,\ldots$ {\em separately conserved\,}
\ui-current densities, describing electron and/or hole transport in
$N$ separate ``channels'' distinguished by conserved quantum
numbers. In our analysis, we regard $N$ as a free
parameter.\,--\,Physically, $N$ turns out to depend on the filling
factor $\nu$ and other parameters characterizing the system.
\rule[-4mm]{0mm}{5mm}

{\bf (A3)} In our units where $\,h=-e=1$, the electric charge of an
{\em electron/hole} is $+1/-1$. Any local excitation
(quasi-particle) above the ground state of the system with {\em
integer} total electric charge $\qel$ satisfies {\em Fermi-Dirac
statistics\mini} if  $\qel$ is {\em odd}, and {\em Bose-Einstein
statistics\mini} if $\qel$ is {\em even}. \rule[-4mm]{0mm}{5mm}

{\bf (A4)} The quantum-mechanical state vector describing an {\em
arbitrary\mini} physical state of an (incompressible) QH fluid is
{\em single-valued\,} in the position coordinates of all those
(local) excitations that are composed of {\em electrons\mini}
and/or {\em holes}.
\rule[-4mm]{0mm}{5mm}

In addition to these four basic assumptions, we put forward, as
in~\cite{FS1,FS2,FT,FKST}, a ``working hypothesis'' expressing a
``{\em chiral factorization\mini}'' property of QH fluids.
\rule[-4mm]{0mm}{5mm}

{\bf (A5)} The fundamental charge carriers of a QH fluid are
electrons and/or holes. We assume that, in the scaling limit, the
dynamics of electron-rich subfluids of a QH fluid is {\em
independent} of the dynamics of hole-rich subfluids, and, up to
charge conjugation, the theoretical analysis of an electron-rich
subfluid is identical to that of a hole-rich subfluid.
\rule[-4mm]{0mm}{5mm}

We make a few remarks on these assumptions. For a finite, but
macroscopic system, assumption {\bf (A2)} implies that there are
$N$ distinct, approximately conserved chiral edge currents
circulating along the boundary of the system.\,--\,Strict
conservation of these \ui-current densities holds in the scaling
limit.\,--\,This generalizes to the fractional QH effect Halperin's
edge current picture~\cite{Ha} of the integer QH effect;
see~\cite{FK,FS2} and also~\cite{edge,edgeexp}. Assumption {\bf
(A5)} implies that, for an electron-rich QH fluid, say, the
chirality of {\em all\,} edge currents is the {\em same}. It is
fixed by the direction of the external magnetic field. The
mathematical virtue of the edge current picture is that it allows
for a natural introduction of the tools of current algebra into the
theory of the QH effect; see~\cite{edgetheo,FK,FS2} and the
references therein. A systematic mathematical implementation of
assumptions {\bf (A1--5)} in the edge current picture of the QH
effect has been given in~\cite{FS2,FT}.

Given the close relationship between two-dimensional chiral
conformal field theory and quantum Chern-Simons theory, as
expounded first by Witten~\cite{Wi}, one can establish a
boundary-bulk duality in QH fluids. By this duality,
quasi-particles excited at the edge of a QH fluid have their
precise counterparts in local bulk sources in a quantum
Chern-Simons theory that is expressed in terms of the ``vector
potentials'' of the $N$ separately conserved \ui-current densities
of the system. Details of this bulk picture and, in particular,
the explicit implementation of assumptions \mbox{{\bf (A1--5)}} in
this picture have been given in~\cite{FT,FKST}. Further
considerations elucidating the boundary-bulk duality in QH fluids
can be found in~\cite{FS2,Bala}.

As recapitulated in Sect.\,\ref{sI}, it follows from assumptions
\mbox{{\bf (A1--4)}} that the properties of a QH fluid in the
scaling limit can be described completely in terms of a
mathematical object that we have call {\em quantum Hall lattice}. A
QH lattice \QHL\, consists of an odd, integral lattice $\mini\G$
and an integer-valued linear functional $\Q$ on $\mini\G$; see
Sect.\,\ref{sI} and the definitions below. The number of positive
eigenvalues of the metric on $\mini\G$ corresponds, physically, to the
number of edge currents of one chirality, the number of negative
eigenvalues to the number of edge currents of the opposite
chirality. In the situation envisaged in the working hypothesis
{\bf (A5)}, $\G$ is an orthogonal direct sum of a lattice
$\mini\Ge$ on which the metric is positive-definite and a lattice
$\mini\Gh$ on which it is negative-definite; see~(\ref{decomp}).
The structure of $\mini\G$  can hence be understood if we are able
to enumerate {\em  positive-definite} lattices. In the most general
situation, however, $\G$ could be an {\em indecomposable},
indefinite lattice or contain an indecomposable, indefinite
sublattice. In this case, there would exist local physical
excitations of the system of edge currents with the quantum numbers
of the electron (electric charge $1$ and Fermi-Dirac statistics)
that are composed out of left- {\em and\,} right-moving excitations
which themselves, however, are {\em not\,} physical quasi-particles
of the system. In other words, the left- and right-moving channels
of edge currents are coupled to each other in such a way that
physical states on the algebra of edge currents {\em cannot\,} be
factorized into a product of a physical state on the algebra of
left-moving edge currents and a physical state on the algebra of
right-moving edge currents. We believe that  those indecomposable,
indefinite lattices do not correspond to stable QH fluids.

While we have not found {\em a priori\,} reasons to rule out
indecomposable, indefinite (sub)lattices $\mini\G$, we shall not
consider this situation in the present paper. Rather, it is our
strategy to adopt the chirality assumption~{\bf (A5)} as a
working hypothesis, and, investigating its strongly predictive
consequences, we intend to lay the ground for testing it in
different experimental situations; for the predictions, see
Fig.\,1.2 in Sect.\,\ref{sI} and the discussion in
Sect.\,\ref{sDis}\mini.

In this context, we note that all the Haldane-Halperin~\cite{HH} and
Jain-Goldman~\cite{JG} ``hierarchy fluids'' satisfy our
assumptions~{\bf (A1--4)}, and most of them satisfy assumption~{\bf
(A5)}, too. The exceptions (corresponding to {\em non-euclidean},
composite QH lattices) can be shown to satisfy a slightly weaker
form of~{\bf (A5)}. This slightly weaker form of~{\bf (A5)} is
given in Appendix~E where details about ``hierarchy QH lattices''
can be found.

The stronger assumption~{\bf (A5)} helps in reducing the
classification problem of QH fluids to a tractable one!
Furthermore, it leads, as we wish to show in this paper, to
interesting results typically complementing and sometimes
challenging the commonly accepted hierarchy schemes of the QH
effect.
\rule[-4mm]{0mm}{5mm}

Defining an (incompressible) QH fluid as a two-dimensional
electronic system with vanishing resistances $R_{xx}$ and $R_{yy}$
(see~(\ref{URI})) and satisfying assumptions {\bf (A1--5)}, the
following contention has been advanced in~\cite{FS2,FT,FKST}:
\rule[-4mm]{0mm}{5mm}

{\bf Classification of QH Fluids.} {\em In the scaling limit, the
quantum-mechanical description of an (incompressible) QH fluid
is universal and completely coded into a pair of chiral quantum
Hall lattices (CQHLs), one CQHL, \QHLe, for the electron-rich
subfluids, and one, \QHLh, for the hole-rich subfluids.}
\rule[-4mm]{0mm}{5mm}

In our units where $e^2/h=1$, the Hall conductivity of the entire
QH fluid is given by

\be
\sH \:=\; <\Qe\,,\Qe> - <\Qh\,,\Qh> \;=\: \sH^e - \sH^h\ ,
\label{sHeh}
\ee

\noi
where $<\!\Qe\,,\Qe\!>$ and $<\!\Qh\,,\Qh\!>$ denote the squared
lengths of the charge vectors $\Qe$ and $\Qh$ which are
integer-valued linear functionals on the euclidean lattices $\mini\Ge$
and $\mini\Gh$, respectively. We remark that, by assumption {\bf (A5)},
it suffices to focus our attention on the analysis of, say, the
electron-rich subfluids of a QH fluid and the corresponding CQHL.
In the following, we drop the subscript $\mini e\mini$ from our
notation. \rule[-4mm]{0mm}{5mm}

{\bf Definition.} {\em A chiral quantum Hall lattice (CQHL) is a
pair, \QHL, where $\mini\G$ is an odd, integral, euclidean lattice
and $\mini\Q$ is an odd, primitive vector in $\mini\Gs$, the dual lattice
of $\mini\G$.}
\rule[-4mm]{0mm}{5mm}

We clarify this definition by recalling some technical notions:
\rule[-4mm]{0mm}{5mm}

{\bf (1)} Let $V$ denote a real, $N$-dimensional vector space with
inner product (or metric) $<\ ,\mmini\ >$. We choose an {\em
integral basis}, $\{\ve{1},\ldots, \ve{N}\}$, in $V$. Integrality
means that the (regular, symmetric) matrix of scalar products
$\mini K=(K_{ij})$, the so-called associated {\em Gram matrix}, is
integral, i.e.,

\be
K_{ij} \,:=\; <\ve{i}\mini, \ve{j}>\; \in\BB{Z}\ ,\hfa
i,j=1,\ldots,N\ . \label{Gram}
\ee

\noi
Taking integral linear combinations of these basis vectors, we can
form the {\em integral lattice}

\bea
\Gamma & := & \{\, \vq\in V\,|\ \vq=\sum_{i=1}^N q^i\,\ve{i}\,,\
q^i\in\BB{Z}\mini,\ \mbox{for all }\,i=1,\ldots,N \,\}\ .
\label{lat}
\eea

\noi
A lattice $\mini\G$ is said to be {\em euclidean\mini} if the metric
$<\ ,\mmini\ >$ is positive-definite (i.e., its Gram matrix $K$ is a
positive-definite matrix).

Introducing the {\em dual basis} $\{\ved{1},\ldots,\ved{N}\}$
which is characterized by the property that $<\!\ved{j},\ve{i}\!>\,
= \delta^j_i$, for all $\,i,j=1,\ldots,N$, (i.e., $\ved{j} =
\sum_{i=1}^N (\Ki)^{ji}\,\ve{i},\ j=1,\ldots, N$), the {\em dual
lattice}, $\Gs$, of the lattice $\mini\G$ is given by

\bea
\Gs & := & \{\, \vn\in V\, | <\!\vn\,,\vq\!>\: \in\BB{Z}\mini,\
\mbox{for all }\, \vq \in\!\G \,\}
\nonumber  \\
& \:= & \{\, \vn\in V\, |\ \vn=\sum_{j=1}^N n_j\,\ved{j},\
n_j\in\BB{Z}\mini,\ \mbox{for all }\,j=1,\ldots,N \,\}
\:\supseteq\: \Gamma\ .
\label{dulat}
\eea

{\bf (2)} We recall Kramer's rule,

\be
(\Ki)^{ij} \:=\: {1\over \D} \tilde{K}^{ij} \ ,
\label{Kram}
\ee

\noi
where $\tilde{K}$ denotes the cofactor (or adjoint) matrix
of $K$, and $\,\Delta := \det K\,$ denotes the {\em
discriminant\,} of the lattice $\mini\G$. We note that $\D$ is the
order of the abelian group $\mini\Gs\mA/\Gamma$, or, from a geometrical
point of view, it specifies the relative size of the lattice
$\mini\G$ when viewed as a sublattice of the dual lattice
$\mini\Gs$.
\rule[-4mm]{0mm}{5mm}

{\bf (3)} An integral lattice $\mini\G$ is said to be {\em odd\,} if it
contains a vector $\mini\vq\mini$ for which $<\!\vq\,,\vq\!>$ is an
odd integer. Thus $\mini\G$ is odd if and only if $\mini K_{ii}$ is odd
for at least one $\mini i\mini$ in $\,1,\ldots,N$. (Otherwise, $\G$
is said to be even.) \rule[-4mm]{0mm}{5mm}

{\bf (4)} A vector $\,\Q= \sum_{j=1}^N Q_j\,\ved{j} \in\!\Gs\,$ is
called {\em primitive} (or {\em visible}) if the greatest common
divisor (gcd) of its dual components $\mini Q_j$ equals unity, i.e.,

\be
\gcd (Q_1,\ldots,Q_N) \:=\: \gcd (<\!\Q\,,\ve{1}\!>,\ldots,
<\!\Q\,,\ve{N}\!>) \:=\: 1\ .
\label{prim}
\ee

\noi
Geometrically, $\Q\in\!\Gs$ is primitive means that the line segment
from the origin to $\Q$ does not contain any point of $\mini\Gs$
other than $0$ and $\Q$.
\rule[-4mm]{0mm}{5mm}

{\bf (5)} The vector $\,\Q\in\!\Gs\,$ is said to be {\em odd\,} if
the following congruence holds

\be
<\!\Q\,,\vq\!> \;\equiv\; <\!\vq\,,\vq\!>\; \bmod\ 2\ , \hfa
\vq\in\!\G\ .
\label{odd}
\ee

{\bf (6)} A lattice $\mini\G$ is called {\em decomposable} (or {\em
composite}) if it can be written as an orthogonal direct sum of
sublattices,

\be
\G \:=\: \G_1 \oplus \G_2 \oplus \cdots \oplus \G_k\ , \hfs k\geq
2\ ,
\label{deco}
\ee

\noi
i.e., for arbitrary vectors $\,\vq_i\in\!\G_i\,$ and
$\,\vq_j\in\!\G_j$\, we have $\,<\!\vq_i\mini,\vq_j\!>\:=0$, for
all $i\!\neq\! j$. Otherwise, $\G$ is said to be {\em
indecomposable\mini}. If \QHL\, is a composite CQHL with
decomposition~(\ref{deco}) then the dual lattice has the associated
decomposition $\,\Gs \:=\: \Gs_1 \oplus \Gs_2 \oplus \cdots \oplus
\Gs_k\,$, and the corresponding decomposition of the charge vector
reads $\,\Q=\Q_1+\Q_2+\cdots+ \Q_k\,$. The
decomposition~(\ref{deco}) is reflected in the formula

\be
\sH \:=\; <\!\Q\,,\Q\!> \;=\; <\!\Q_1\mini,\Q_1\!> + \cdots +
<\!\Q_k\mini,\Q_k\!> \;=\:  \sH^1 + \cdots + \sH^k \ .
\label{sumsH}
\ee

\noi
{}From a physical point of view, {\em indecomposable\mini} CQHLs can
be considered as describing ``{\em elementary\,}'' QH fluids, and,
for this reason, we mainly focus on indecomposable lattices in the
present work. We note that, as suggested by~(\ref{sHeh}), we can
think of QH fluids with electron- {\em and\,} hole-rich subfluids
as being described by particular composite lattices, namely ones
that are orthogonal direct sums of two CQHLs of {\em opposite}
chirality (i.e., there are currents of both chiralities circulating
at the edge of these fluids).
\rule[-4mm]{0mm}{5mm}

{\bf (7)} We introduce two physically natural restrictions on {\em
chiral\,} QH lattices. First, let \QHL\, be a decomposable CQHL with
decomposition~(\ref{deco}) and~(\ref{sumsH}). Then \QHL\, is said
to be {\em proper\,} if no component $\mini\Q_j,\ j=1,\ldots,k$, of
the charge vector $\mini\Q\mini$ is vanishing. Note that if, say,
$\Q_j=0\,$ then $\,\sH^j=0\,$ in~(\ref{sumsH}), and the subfluid
corresponding to \mbox{$(\G_j,\Q_j)$} does not have any interesting
electric properties, (see also the remark after~(\ref{Sdia})). For
this reason, we neglect improper CQHLs in the present work, and
properness will always be understood to hold in the sequel.

Second, from a physical point of view, it is natural to even
sharpen the notion of properness as follows: Let \QHL\, be a
decomposable CQHL as above. Then \QHL\, is said to be {\em
primitive\mini} if every component $\,\Q_j,\ j=1,\ldots,k$, of the
charge vector $\Q$ is a primitive vector in $\mini\Gs_j\mini$;
see~(\ref{prim}).

Primitive CQHLs are proper, but the contrary is not necessarily
true. We will show in Thm.\,\ref{sGT}.6 in Sect.\,\ref{sGT} that,
for a subclass of proper CQHLs, the contrary can be inferred.
Moreover, note that indecomposable CQHLs are proper and primitive.
{\em The classification of primitive CQHLs is the main objective of
the present paper, and the corresponding results are given in
Sects.\,\ref{sGT}--\ref{sLD}\mini.}

We remark that, as explained in Appendix~E, all {\em chiral\,}
hierarchy fluids correspond to primitive CQHLs. In general,
however, there are (non-chiral) hierarchy fluids which are
associated with {\em non-primitive\mini} CQHLs. For some examples,
see~{\bf (b)} in Appendix~E. We do not find these non-primitive
proposals very attractive and shall provide, at some of the
corresponding Hall fractions, primitive CQHLs in Sects.\,\ref{sMS}
and~\ref{sLD}\mini.

\vspace{5mm}

\noi
{\bf QH Lattice--QH Fluid Dictionary.} We briefly summarize the
basic relationship between the language of QH lattices and the
description of physical properties of the corresponding QH fluids;
see Sect.\,\ref{sI}, and, for a detailed discussion,
see~\cite{FS2,FT,FKST}.

Let \QHL\, denote a CQHL. Then any vector $\mini\vq\mini$ in the
lattice $\mini\G$ labels a {\em multi-electron} or {\em multi-hole
excitation\mini} above the ground state of the corresponding QH
fluid which is localized in some bounded region of the plane of the
system. (Here, ``hole'' means a ``missing electron'' in an
electron-rich fluid.) Next, {\em arbitrary\mini} localized
physical excitations of the QH fluid ({\em quasi-particles}), are
labelled by vectors $\mini\vn\mini$ that form a lattice $\mini\Gp$
which clearly has to contain $\mini\G$ and which itself is contained in,
or is equal to $\mini\Gs$:

\be
\G \:\subseteq\: \Gp \:\subseteq\: \Gs\ .
\ee

\noi
In our units where $\,e\!=\!-1$, the {\em total electric charge},
$\qel(\vn)$, of a physical excitation labelled by $\,\vn\in\!\Gp\,$
is given by the inner product of $\mini\vn\mini$ with the charge
vector $\Q$,

\be
\qel(\vn) \:=\; <\!\Q\,,\vn\!>\ ,
\label{cha}
\ee

\noi
and the {\em statistical phase}, $\vartheta(\vn)$, of the excitation
is determined by the squared length (modulo 2) of $\,\vn$,

\be
\vartheta(\vn) \;\equiv\; <\!\vn\,,\vn\!>\; \bmod\ 2\ .
\label{stat}
\ee

\noi
We note that~(\ref{stat}) corresponds to a normalization of the
statistical phase such that bosons have $\,\vartheta\equiv 0\mE\!
\pmod{2}\,$ while fermions have $\,\vartheta\equiv 1\mE\!
\pmod{2}$. As mentioned in Sect.\,\ref{sI}, moving (adiabatically)
one quasi-particle, labelled by a vector $\mini\vn_1$, around
another one, labelled by a vector $\mini\vn_2$, along a
counter-clockwise oriented loop, the state vector describing the
system changes by a phase factor $\,\exp\mini(2\mini\pi\mini
i\!<\!\vn_1\mini,\vn_2\!>)\mini$; see~(\ref{mono}).

\vspace{5mm}

\noi
{\bf Examples.} We conclude this section by describing the two most
basic examples of QH fluids in the language of QH lattices,
introduced above:
\rule[-4mm]{0mm}{5mm}

{\bf (a)} {\em QH fluids with $\,\sH=N,\ N=1,2,\ldots$, in the
non-interacting electron approximation.} These integer QH
fluids correspond to the (self-dual) unit euclidean lattices in N
dimensions: $\G=\Gp=\Gs=\BBn{Z}{N}= \BB{Z}\,\oplus\cdots
\oplus\,\BB{Z}\mini$. Here, $N$ is the number of separately
conserved edge currents~\cite{Ha} or filled Landau levels. Denoting
by $\mini\ve{i}$ the generator of the $i$th summand, $i=1,\ldots,N$,
we have $\,K_{ij}=\:<\!\ve{i},\ve{j}\!>\: =\delta_{ij}$. By the
primitivity condition on the charge vector $\Q$ (see point {\bf
(7)} above), we have $\,\Q= \ve{1}+\cdots + \ve{N}$, and $\,\sH
=\:<\!\Q\,,\Q\!>\:=1+\cdots+1=N$. Finally, we note that, by the
self-duality of $\BBn{Z}{N}$, there are no fractionally charged
excitations with fractional statistics (``anyons'') in these
fluids.
\rule[-4mm]{0mm}{5mm}

{\bf (b)} {\em The Laughlin fluids\,}~\cite{L} {\em at
$\,\sH=1/m\,$ where $\,m=2\mini p+1,\  p=(0),1,2,\ldots\,$.} Here,
$\G=\sqrt{m}\,\BB{Z}\,$ which is the one-dimensional lattice
generated by $\mini\veb\mini$ with squared length
$\,<\!\veb\,,\veb\!>\:=m$. The dual lattice $\,\Gs=(1/\sqrt{m})
\,\BB{Z}\,$ which is generated by \mbox{$\,\veed=\veb/m$}. The
charge vector, being primitive in $\mini\Gs$, takes the form
$\,\Q=\veed$, and thus $\,\sH= \mbox{$<\!\Q\,,\Q\!>$} =1/m$.
The quasi-particles are labelled by $\,\xi\veed\in\!\Gp =\Gs,\
\xi\in\!\BB{Z}\mini$. By~(\ref{cha}), they have fractional
electric charges $\,\qel(\xi)=\mbox{$<\!\Q\,,\xi\veed\!>$}=\xi/m$,
and by the congruence~(\ref{stat}), they have fractional
statistical phases $\,\vartheta(\xi)\equiv
\mbox{$<\!\xi\veed,\xi\veed\!>$}\equiv \mbox{$\xi^2/m$}
\pmod{2}$.

Note that, in this case, the knowledge of the electric charges
$\mini\qel\mini$ of the excitations uniquely determines their
statistical phases $\mini\vartheta$. Such a charge-statistics
relation is a property of many interesting higher-dimensional QH
lattices; see Thm.\,\ref{sGT}.5\mini. However, such a relation does
{\em not\,} hold for arbitrary QH lattices.


\newpage
\begin{flushleft}
\section{Basic Invariants of Chiral QH Lattices (CQHLs) and their
Physical Interpretations}
\label{sBI}
\end{flushleft}

Invariants of CQHLs, most of which seem to be new, provide physically
interesting information about the corresponding chiral (i.e.,
electron- {\em or\mini} hole-rich) QH (sub)fluids. Most of the
invariants summarized below have been introduced in~\cite{FT} where
more details can be found. {}From the classification results presented
in Sects.\,\ref{sMS} and~\ref{sLD} and from the discussion in
Sect.\,\ref{sDis}, it follows that a microscopic understanding and a
corresponding determination of the values of these invariants pose
interesting open problems in the theory of the QH effect.

The invariants of a (proper) CQHL, \QHL, capture {\em intrinsic}
properties of \QHL, i.e., properties that do not depend on the
explicit choice of a basis in $\mini\G$ and on the ``reshuffling''
of electric charge assignments in the lattice corresponding to a
transformation of $\,\Q\mini$ by an orthogonal symmetry of
$\,\G$. Choosing a basis, $\{\ve{1},\ldots,\ve{N}\}$, in $\mini\G$,
the CQHL is specified by the (integral) Gram matrix $\,K_{ij}\!
=\:<\!\ve{i},\ve{j}\!>,\ i,j=1,\ldots,N$, and by the row vector
$\,\Qv=(Q_1\mini,\ldots,Q_N)\,$ which specifies the components of
the charge vector $\mini\Q$ in the associated dual basis
$\{\ved{1},\ldots,\ved{N}\}$ of $\,\Gs$, i.e., $\Q=\sum_{j=1}^N
Q_j\,\ved{j}$; see Eqs.~(\ref{Gram}) through~(\ref{dulat}).
Choosing a different basis in $\mini\G$, the resulting pair
$(K^\prime,\Qv^\prime)$ is related to the pair $(K,\Qv)$ by

\be
K^\prime\:=\:S^T\,K\,S\ , \hah \Qv^\prime\:=\Qv\,S\ ,
\label{equiv}
\ee

\noi
where $S$ is an element in $GL(N,\BB{Z}\mini)$, the group of
integral, non-degenerate $N\times N$-matrices. Note that, for
$S^{-1}$ to be an element of the group, the determinant of any
element $\mini S\mini$ has to be $\mini\pm 1\mini$.

Following the proposal in~\cite{FT}, a concise presentation of
the numerical invariants of a CQHL, \QHL, is given by the
associated {\em symbol}

\be
\raisebox{-3.6mm}{$\scriptstyle N$} {\left( \sH = \ndH
\right)}^g_\lambda\;[\mini\lm,\lM] \ ,
\label{symb}
\ee

\noi
where the invariants summarized in the symbol have the following
mathematical definitions and physical interpretations:
\rule[-4mm]{0mm}{5mm}

{\bf (1)} $N:=\dim\G=\mbox{rank}\,\G$; the {\em lattice dimension}
$N$ gives (in the scaling limit) the number of separately
conserved \ui-current densities in the corresponding QH fluid.
Although no rigorous results are known, we expect $\mini N\mini$ to
depend on the filling factor $\mini\nu\mini$ and on the density or
strength of impurities (disorder) in the system. We expect that the
upper bound $\mini N_\ast$ on the dimension $\mini N\mini$ of
physically realizable CQHLs tends to $\mini\infty$, as the density
or strength of impurities tends to $\mini 0\mini$; see~\cite{FKST}.
\rule[-4mm]{0mm}{5mm}

{\bf (2)} By~(\ref{sHeh}) and~(\ref{equiv}), the {\em Hall
conductivity} (or {\em Hall fraction}) $\sH$ is clearly a CQHL
invariant: $\sH =\:<\!\Q\,,\Q\!> \:= \Qv\cdot\Ki \Qv^T$.
By~(\ref{Kram}) and the definition of $\Qv$, it is a positive
{\em rational\,} number.
\rule[-4mm]{0mm}{5mm}

{\bf (3)} Writing $\,\sH=n_H/d_H$, with $\,\gcd(n_H,d_H)=1$, the
important invariant of the lattice given by its {\em discriminant},
$\D$, can be written as

\be
\D\::=\:\det K \:=\: l\,d_H\ ,
\label{disc}
\ee

\noi
where the invariant $\mini l\,$ is called the {\em level\,} of
the CQHL \QHL; see~(\ref{Kram}).
\rule[-4mm]{0mm}{5mm}

{\bf (4)} The level $\mini l\,$ satisfies an interesting
factorization property, namely $\,l=g\mini\lb$, where $\mini
g\mini$ is defined by $\,g:=\gcd(Q^1, \ldots,Q^N)$, with
$\,Q^j:=\D<\!\Q\,,\ved{j}\!>,\ j=1,\ldots,N$, and $\{\ved{1},
\ldots,\ved{N}\}$ {\em any} dual basis of $\mini\Gs$. Thus,
by~(\ref{disc}), the discriminant is given by $\,\D=g\mini\lb\mini
\dH\mini$. The invariant $\lb$ is called the {\em charge
parameter}, and its physical relevance derives from the following
fact: one can prove~\cite{FT} that, in our units where
$\,e\!=\!-1$, the smallest possible (fractional) electric
charge of a quasi-particle excited above the ground state in the
corresponding QH fluid is given by

\be
e^\ast\::=\: \min_{\vn\in\mini\mbox{$\scriptstyle \Gamma
\raisebox{1.3mm}{$\scriptscriptstyle\ast$}$}\mimini,\,
<\Q\mini,\mini\vn>\mini \neq\mini 0} | <\!\Q\,,\vn\!> |
\:=\: {1 \over \lambda\mini d_H} \ .
\label{emin}
\ee

{\bf (5)} Finally, we provide definitions of the {\em
relative-angular-momentum invariants} $\,\lm$ and $\lM\,$. Since
$\Q$ is a primitive vector in $\mini\Gs$ (see~(\ref{prim})) there is a
basis of $\mini\G,\ \{\vq_1,\ldots,\vq_N\}$, such that
$\,<\!\Q\,,\vq_i\!>\:=\!1,\ i=1,\ldots,N$. The set of all such
``symmetric'' bases is denoted by $\BQ$. Then, for any CQHL \QHL,
we define the invariants

\be
\Lm \::=\: \min_{\vq\in\G\mimini,\,<\Q\mini,\mini\vq>\mini =\mini 1}
\mini <\!\vq,\vq\!>\ ,
\label{Lmin}
\ee

\vav

\be
\LM \::=\: \min_{\{\vq_1,\ldots,\vq_N \}\in \BQs}
\mini \left( \max_{1\leq i\leq N} <\!\vq_i,\vq_i\!>\right)\ .
\label{Lmax}
\ee

In the situation where \QHL\, is a {\em primitive decomposable}
CQHL with decomposition $\,(\G,\Q)=\oplus_{j=1}^k(\G_j,\Q_j)\,$
(see~(\ref{deco})) it is natural to refine the
definitions~(\ref{Lmin}) and~(\ref{Lmax}) as follows:

\be
\lm\QHL \::=\: \min_{1\leq j\leq k} \Lm(\G_j,\Q_j) \:\geq\:
\Lm\QHL\ ,
\label{lmin}
\ee

\vav

\be
\lM\QHL \::=\: \max_{1\leq j\leq k} \LM(\G_j,\Q_j) \:\geq\:
\LM\QHL\ .
\label{lmax}
\ee

We note that, by the oddness of $\,\Q\mini$ (see~(\ref{odd})) the
relative-angular-momentum invariants~(\ref{Lmin})
through~(\ref{lmax}) are positive, {\em odd\,} integers which
satisfy

\be
\Lm \:\leq\: \LM \ ,\hah \lm \:\leq\: \lM\ .
\label{mlM}
\ee

Exploiting the Chern-Simons description of QH fluids, it
has been argued in~\cite{FT,FKST} that, physically, for an
{\em elementary\mini} chiral QH fluid corresponding to the {\em
indecomposable} CQHL \QHL, $\,\lm=\Lm\,$ indicates the smallest
possible relative angular momentum of two electrons excited above
the ground state of the fluid. The physical relevance of the
quantity $\mini\lM\mini$ as well as its role in the classification
of CQHLs will be expounded in great detail below, in
Sects.\,\ref{sGT}--\ref{sLD}\mini.

If the values of the quantities $\mini\lm\mini$ and $\mini\lM\mini$
are clear from context they will be droped from the
symbol~(\ref{symb}).

Note that the elementary invariants in points~{\bf (1--\mini 4)} are
clearly well-defined also for (general) QH lattices; see~\cite{FT}.

\vspace{5mm}

\noi
{\bf Examples.} We illustrate the above invariants by considering
some examples. \rule[-4mm]{0mm}{5mm}

{\bf (a)} The integer QH fluids discussed at the end of
Sect.\,\ref{sUC} (non-interacting electron systems) are
characterized by the symbols

\be
\thesymbol \;=\: \sub{N}{(N)}^1\hspace{-1.7mm}\sub{1}\;[\mini 1,1]
\ ,\hsp N=1,2,\ldots  \ .
\ee

\noi
Note that, by the decomposability of the corresponding CQHLs, we
can write $\,\subi{N}{(N)}^1_1 =\subi{1}{(1)}^1_1\oplus\cdots
\oplus\subi{1}{(1)}^1_1\mini$ in accordance with the physical
picture of $N$ independent, filled Landau levels.
\rule[-4mm]{0mm}{5mm}

{\bf (b)} The Laughlin fluids, also discussed at the end of
Sect.\,\ref{sUC}, correspond to CQHLs for which the associated
symbols read
\vspace{-5mm}

\be
\thesymbol \;=\: \QHLsymbol{1}{1\over 2\mini p+1}{1}{1} \;[\mini
2\mini p+1,\mini 2\mini p+1\mini]\ ,\hsp p=1,2,\ldots  \ .
\ee

\noi
For a discussion of the special status of the Laughlin fluids
from a classification point of view, see Thms.\,\ref{sGT}.4 and
\ref{sGT}.8 in Sect.\,\ref{sGT}\mini.
\rule[-4mm]{0mm}{5mm}

{\bf (c)} For each $\,p=1,2,\ldots\,$, there is the series of Hall
fractions $\,\sH=N/(2\mini pN\!+\!1)\,$ with $\,N=1,2,\ldots\ $.
{}From the data presented in Fig.\,\ref{sI}.1, it is clear that many
of the experimentally most prominent Hall fractions belong to these
series (or to the charge-conjugated partner series of the one with
$\,p=1$; see the discussion in Sect.~\ref{sDis}). We note that
these fractions also figure prominently in Jain's
work~\cite{Jain}\,--\,the basis of the Jain-Goldman hierarchy
scheme~\cite{JG}\,--, and we refer to Thm.\,\ref{sGT}.8 in
Sect.\,\ref{sGT} where, from a classification point of view, the
uniqueness of the associated CQHLs is discussed. The above series
of Hall fractions can be obtained by the following series of {\em
indecomposable} CQHLs: the data pairs $(K,\Qv)$ which determine
these CQHLs are given, in some bases that we call ``normal'', by

\be
K \:=\,
\left.\left(\renewcommand{\arraystretch}{.8}
\begin{array}{c|ccccc}
2\mini p+1 & -1 & 0  & \cdot & \cdot & 0\rule[-1.8mm]{0mm}{4mm} \\
\hline
-1 & 2 & -1 & 0 & \cdot & 0\rule[0mm]{0mm}{4.5mm} \\
0 & -1 & 2 & -1 & \cdot & \cdot \\
\cdot & 0 & -1 & \cdot & \cdot & 0 \\
\cdot & \cdot & \cdot & \cdot & \cdot & -1 \\
0 & 0 & \cdot & 0 & -1 & 2
\end{array} \right)\  \right\}\mbox{\scriptsize\em N}
\ ,\hah
\Qv \:=\:
(\mini\underbrace{1,0,\ldots,0}_{\mbox{\scriptsize\em N}}\,) \ ,
\label{promKQ}
\ee

\noi
and the associated symbols read

\be
\thesymbol \;=\: \QHLsymbol{N}{N\over 2\mini pN+1}{1}{1} \;[\mini
2\mini p+1,\mini 2\mini p+1\mini]\ ,\hsp p,N=1,2,\ldots  \ .
\label{promsymb}
\ee

\noi
Note that the $(N\!-\!1)$-dimensional submatrix in the lower right
of $K$ is the Cartan matrix of the simple Lie algebra $\,A_{N-1}
= su(N),\ N=2,3,\ldots\,$; see Appendix~A. For $\,N=1$, we
recognize in~(\ref{promKQ}) and~(\ref{promsymb}) the expressions
corresponding to the Laughlin fluids; see example~{\bf (b)}. In
connection with the QH effect, the matrices in~(\ref{promKQ}) first
appeared in~\cite{Read}. Combining results of~\cite{Read}
and~\cite{FZ} (see Appendix~E) the CQHLs specified
by~(\ref{promKQ}) can be seen to correspond to the ``basic Jain
states''~\cite{Jain} at $\,\sH=N/(2\mini pN\!+\!1)\mini$. Moreover,
it has been shown in~\cite{FZ} that the QH fluids corresponding
to~(\ref{promKQ}) exhibit large symmetries, namely $\mini
su(N)$-current algebras at level 1; see also~\cite{FS2}. In
Sect.\,\ref{sMS}, we show that the above CQHLs belong to an
interesting class of CQHLs with ``large'' symmetries, the so-called
``maximally symmetric'' CQHLs. The classification of ``maximally
symmetric'' CQHLs will be the main objective of
Sect.\,\ref{sMS}\mini.

We note that, by extending definitions~(\ref{promKQ})
and~(\ref{promsymb}) to $\,p=0$, the {\em composite} integer QH
fluids of example~{\bf (a)} can be included as special cases
of~{\bf (c)}.


\newpage
\begin{flushleft}
\section{General Theorems and Classification Results for~CQHLs}
\label{sGT}
\end{flushleft}

The purpose of this section is to review general facts and
classification results for CQHLs, in order to put the more specific
classification results given in Sects.\,\ref{sMS} and~\ref{sLD}
into a broader perspective. We summarize, in the form of eight
theorems, results that have been presented in our previous
works~\cite{FT,FSTL,FKST} where more details can be found. We
indicate those proofs that have not been given previously.
Moreover, we discuss phenomenological implications of our theorems.

The first two theorems are based on {\em CQHL inequalities\mini}
that establish useful relations between some of the numerical
invariants introduced in Sect.\,\ref{sBI}\mini.
\rule[-4mm]{0mm}{5mm}

{\bf Theorem~\ref{sGT}.1.} {\em The set of (proper) CQHLs
\QHL\, with dimensions $\,N\!\leq\! N_\ast\,$ and
relative-angular-momentum invariants $\,\lM\leq \Ls$, where
$\mini N_\ast$ and $\,\Ls$ are two given integers, is\mini} finite.
\rule[-4mm]{0mm}{5mm}

This theorem implies that the set of Hall fractions $\sH$ that can
be realized by CQHLs which satisfy the above bounds on $N$ and
$\mini\lM$ is finite. We remark, however, that the number of
possible fractions is growing superexponentially fast in $N_\ast$
and $\Ls$,\,--\,e.g., for $N_\ast=2$ and $\Ls=3$, there are
$\mini 10\mini$ CQHLs, while for $N_\ast=3$ and $\Ls=5$, one finds
already more than $\mini 250\mini$ CQHLs.\,--\,Fortunately, in the
physically relevant situation where one also has a natural upper
bound, $\sigma_\ast$, on the Hall fractions to be considered, the
number of CQHLs satisfying this bound and the ones in
Thm.\,\ref{sGT}.1 is drastically reduced! This fact is illustrated
by the classification results in Sects.\,\ref{sMS} and~\ref{sLD}.

The basic tools in proving Thm.\,\ref{sGT}.1 are Hadamard's
inequality for positive-definite quadratic forms (see,
e.g.,~\cite{Sieg}), which implies that

\be
\lb\mini g\, \dH \:=\: \D \:=\: \det K \:\leq\: \lM^{\,N}\ ,
\label{Dbound}
\ee

\noi
and the fact that (see~\cite{FKT})

\be
\lb\mini g\, \nH \:\leq	\: C(N)\,\lM^{\,N-1} \ ,
\ee

\noi
where $C(N)$ is a constant depending on the lattice dimension
$N$,\,--\,e.g., for two-dimensional CQHLs, one finds that
$\,C(2)=4\mini$. \rule{2.5mm}{2.5mm}

Physically, $N$ is the number of separately conserved \ui-current
densities in a QH fluid (in the scaling limit). A larger amount of
disorder (an increased density or strength of impurities) in the
system is expected to reduce the quantity $\mini N\mini$ because of
``channel-mixing''  effects. Hence it is natural to impose an upper
bound, $N_\ast$, {\em depending on disorder}, on the dimension $N$
of physically relevant CQHLs. With respect to an upper bound on the
relative-angular-momentum invariant $\mini\lM$, we argue,
physically, that if $\mini\lM\mini$ were too large then the density
of electrons in the ground state of a (pure) system would be so
small that it would be energetically more favourable for the
electrons to form a Wigner crystal, thereby destroying the
incompressibility of the system; see~\cite{WigCry} and, for a
review of recent experiments,~\cite{WigCryexp}. Given this remark
the following basic CQHL inequality is of interest.
\rule[-4mm]{0mm}{5mm}

{\bf Theorem~\ref{sGT}.2.} {\em For a CQHL \,\QHL, the Hall
fraction $\sH$ and the relative-angular-momentum invariants $\Lm,\
\lm$ and $\,\lM$ satisfy}

\be
{1\over \sH} \:\leq\: \Lm \:\leq\: \lm \:\leq\: \lM\ .
\label{sHbound}
\ee

This theorem is a direct consequence of the Cauchy-Schwarz
inequality (in the real vector space $\,V\supset \G\mini$),
$<\!\Q\,,\vq \!>^2\,\leq\:<\!\vq,\vq\!>\,<\!\Q\,,\Q\!>$, and the fact
that, for any vector $\,\vq\in\!\G$, with $\,\qel(\vq)\!=
\:<\!\Q\,,\vq\!>\:\neq 0$, we have $\,<\!\Q\,,\vq\!>^2\,\geq
1$. \rule{2.5mm}{2.5mm}

If we suppose that, physically, chiral QH fluids satisfy a universal
bound $\,\lM \leq \Ls\,$ then~(\ref{sHbound}) tells us that CQHLs
with $\,\sH\!<\!1/\Ls\,$ are physically {\em irrelevant\mini}. Note
that the data in Fig.\,\ref{sI}.1 are consistent with a choice of
$\,\Ls =7$.
\rule[-4mm]{0mm}{5mm}

Given these observations on the quantities $N$ and $\mini\lM$, one
is led to the following heuristic principle:
\rule[-4mm]{0mm}{5mm}

{\bf Stability Principle.} {\em The smaller the values of the
CQHL invariants $N$ and $\,\lM\mini$, the more stable the
corresponding chiral QH fluid.}
\rule[-4mm]{0mm}{5mm}

This heuristic stability principle will recive further support when
comparing our classification results of Sects.\,\ref{sMS}
and~\ref{sLD} with the experimental data of Fig.\,\ref{sI}.1; see
Fig.\,\ref{sI}.2 and the discussion in Sect.\,\ref{sDis}, where an
even sharper version is proposed.
\rule[-4mm]{0mm}{5mm}

{\bf Theorem~\ref{sGT}.3.} {\em Let \,\QHL\, be a CQHL with\,} even
{\em Hall denominator $\dH$. Then the charge parameter $\lb$ has to
be} even, {\em too.}
\rule[-4mm]{0mm}{5mm}

For a proof of this theorem, we first define the vector $\,\vv:=
\lb\mini\dH\mini\Q\in\!\Gs$. Then, for all dual vectors $\,\vn=
\sum_{j=1}^N n_j\, \ved{j} \in\!\Gs$, we find that
\mbox{$\,<\!\vv,\vn\!>\: = \!\lb\mini\dH <\!\Q\,,\vn\!>\: =$}
\mbox{$\sum_{j=1}^N (\D <\!\Q\,,\ved{j} \!>\mmini /\mini g)\, n_j =
\sum_{j=1}^N (Q^j/g)\,n_j \in \!\BB{Z}\mini$}, by using $\,\D=
g\mini\lb\mini\dH\,$ and the definition of $g\mini$; see points
{\bf (3)} and {\bf (4)} in Sect.\,\ref{sBI}\mini. Thus $\vv$ is
actually an element of $\,(\Gs)^\ast \simeq\G\,$ and hence, by the
oddness of $\,\Q\mini$ (see~(\ref{odd})), the congruence
$\,<\!\Q\,,\vv\!> \:\equiv\: <\!\vv,\vv\!>\mE\mE \pmod{2}$ holds.
Now, by the definitions of $\sH$ and $\vv$, it follows that the
l.h.s.\ of the congruence equals $\,\lb\mini \nH\,$ and the r.h.s.
equals $\,\lb^2\mini\dH\mini\nH\mini$, i.e.,
$\,\lb\mini\nH\equiv\lb^2\mini\dH\mini\nH\mE\! \pmod{2}$. Finally,
since for $\dH$ even, $\nH$ is odd (see point {\bf (3)} in
Sect.\,\ref{sBI}), the latter congruence would be false if
$\mini\lb\mini$ were odd. \rule{2.5mm}{2.5mm}

The phenomenologically interesting implication of Thm.\,\ref{sGT}.3
is that, in QH fluids with an even Hall denominator $\dH$, one
predicts the existence of quasi-particle excitations above the
ground state with ``fractional'' fractional charges, i.e., since
$\,\lb=2,4,\ldots\,$,

\be
e^\ast \:=\: {1 \over \lambda\mini d_H} \:\leq\: {1 \over 2\mini
d_H}\ .
\label{even}
\ee

\noi
It would be interesting to test this model-independent prediction
experimentally for even-denominator QH fluids at $\,\sH\!=\!1/2$ and
$5/2$ mentioned in Sect.\,\ref{sI}\mini: We predict that
$\,e^\ast\!\leq\mmini 1/4$ (in units where $\,e=\mimini-1$)\mini!
\rule[-4mm]{0mm}{5mm}

{\bf Theorem~\ref{sGT}.4.} {\em At every Hall fraction $\,\sH=1/m$,
$m$ odd, there is a} unique {\em indecomposable CQHL with the
property that its level $l=\lb\mini g=1$. This CQHL is
one-dimensional and corresponds to the Laughlin fluid at
$\,\sH=1/m$. Moreover, any CQHL with $\,\sH=1/m$, $m$ odd, and
$\,N\geq 2$ has a charge parameter $\,\lb\geq 2$.}
\rule[-4mm]{0mm}{5mm}

A proof of this theorem has been given in~\cite[Subsect.\,7.5]{FT}.

The CQHLs corresponding to the Laughlin fluids have been described
explicitly in example~{\bf (b)} at the end of Sect.\,\ref{sUC}\mini.
We emphasize that, by an argument similar to the one
in~(\ref{even}), the last statement in Thm.\,\ref{sGT}.4 has
implications that are, in principal, observable! In
Sect.\,\ref{sDis}, an example illustrating this point is
discussed when analyzing possible phase transitions at $\,\sH=1$.
\rule[-4mm]{0mm}{5mm}

An interesting subclass of CQHLs is formed by CQHLs with level
$\,l=1$, i.e., their lattice discriminant $\mini\D\mini$ equals the
Hall denominator $\mini\dH\mini$.  Indecomposable CQHLs with level
$\,l=1$, and thus $\,\lb=g=1$, have been classified for $\,\dH\leq
25\,$ and $N$ below relatively high ``critical'' dimensions
$\mini N_c(\sH)$, typically around $10\mini$; see~\cite{FSTL,FT}.

This subclassification has been achieved by combining the recent
lattice-clas\-si\-fi\-ca\-tion results of Conway and
Sloane~\cite{CSS} with a systematic investigation of the possible
charge vectors $\Q$ in the duals of all the classified (odd,
integral, euclidean) lattices. For the latter search, one makes use
of the following fact: from the Cauchy-Schwarz inequality and the
defining relation $\,\sH\!=\:<\!\Q\,,\Q\!> \mini$, one infers that,
for a CQHL \QHL\ the dual components $\mini Q_j$ of the charge
vector $\,\Q=\sum_{j=1}^N Q_j\,\ved{j}\,$ are constrained by

\be
Q_j^2 \:\leq\: \sH\, \lM\ , \hfa j=1,\ldots, N\ .
\label{Qbound}
\ee

\noi
Thus, restricting ones focus to CQHLs with $\,\lM\leq\Ls\,$ and
$\,\sH\leq\sigma_\ast$, Eq.~(\ref{Qbound}) implies that the search
for all possible charge vectors $\Q$ in the dual of a given lattice
$\mini\G$ is a {\em finite} problem. \rule{2.5mm}{2.5mm}

In the next theorem, we recall a few general properties of CQHLs with
level $\,l=1$; for proofs, see~\cite{FT}.
\rule[-4mm]{0mm}{5mm}

{\bf Theorem~\ref{sGT}.5.} {\em Let \,\QHL\, be a (proper) CQHL with
level $\,l=\lb\mini g=1$. Then}

\hspace*{5.3mm}{\bf (i)} {\em $\dH$ is} odd, {\em and} $\,\Gs/\G
\simeq \BB{Z}_{\mini\dH}\,$;

\ssp{\bf (ii)} {\em in order to realize a Hall fraction $\sH$ with
$\nH$ even (odd), $N\mini$ has to be even

\sspp (odd, and $\,N\equiv\nH$ (mod $4$)\mini);}

\hspace*{2.6mm}{\bf (iii)} {\em for quasi-particles labelled by
$\,\vn\in\!\Gs\!$, a charge-statistics relation holds:

\sspp if $\,\qel(\vn)=\varepsilon/\dH\,$ then $\,\vartheta(\vn)
\equiv (\nH)^{-1}\,\varepsilon^2/\dH\!$ (mod $2$).}
\rule[-4mm]{0mm}{5mm}

We note that, in the last statement of this theorem, the number
$(\nH)^{-1}$ is defined as follows: if $\nH$ is odd, then
$\,\nH\mini(\nH)^{-1} \equiv 1$ (mod $2 \dH$), and if $\nH$ is even,
then $\,\nH:=2\mini(2\nH)^{-1}+\dH\mini$, with $\,2\nH\mini
(2\nH)^{-1} \equiv 1$ (mod $\dH$). A proof of this theorem can be
found in~\cite[Sect.\,5]{FT}.

\vspace{5mm}

\noi
{\bf Shift Maps and their Implications.} In the remaining part of
this section, we study ``structurally similar'' chiral QH fluids.
At the level of CQHLs, ``structural'' relationships are realized by
particular maps, called {\em shift maps}. {}From a classification
point of view, shift maps allow\,--\,under suitable
conditions\,--\,to immediately carry over classification results
for CQHLs with Hall fractions in a given interval to corresponding
results for other intervals. Phenomenologically  interesting
implications of structural  relationships are outlined at the
end of this section and in Sect.\,\ref{sDis}\mini.

First, we divide the interval $(0,\infty)$ of possible Hall
fractions $\mini\sH\mini$ into a sequence of suitable subintervals,
``{\em windows\,}'', $\Sigp^\pm$ defined by

\ben
\Sigp^+ \::=\: \{\, \sH\ |\ {1\over 2\mini p+1}\leq \sH <{1\over
2\mini p}\,\}\ ,\hsp p=1,2,\ldots\ ,
\een

\vav

\be
\Sigp^- \::=\: \{\, \sH\ |\ {1\over 2\mini p}\leq \sH <{1\over
2\mini p-1}\,\}\ ,\hsp p=1,2,\ldots\ .
\label{Sig}
\ee

\noi
The ``$+$'' superscripts in the window symbols $\mini\Sigp^+$ are
chosen because these subintervals contain the ``first
main series'' of Hall fractions, $\,\sH=N/(2\mini pN\!+\! 1),\
N=1,2,\ldots\ $. Similarly, the ``$-$'' superscripts for the
``complementary'' windows remind us that these windows contain the
``second main series'' of Hall fractions, $\,\sH=N/(2\mini p N
\!-\!1),\ N=2,3,\ldots\ $. Moreover, we denote by $\Sigma_0^+$ the
interval $\,[\mini 1\mini,\infty)$, and by $\Sigp$ the union of the
two complementary subintervals $\mini\Sigp^+$ and $\mini\Sigp^-$,
i.e., $\Sigp:=\Sigp^+ \cup\Sigp^-,\ p=1,2,\ldots\ $.

Second, we define a class of CQHLs that will figure prominently in
the sequel.
\rule[-4mm]{0mm}{5mm}

{\bf Definition.} {\em A primitive CQHL \QHL\, (see point
{\bf (7)} in Sect.\,\ref{sUC}) with Hall fraction $\,\sH\in\Sigp\,$
is called\,} \Lmini\ {\em if $\,\lM$ takes the smallest possible
value consistent with~(\ref{sHbound}), namely $\,\lM=2\mini p+1,\
p=1,2,\ldots\ $.}
\rule[-4mm]{0mm}{5mm}

\noi
By (\ref{lmin})--(\ref{mlM}), \Lmini\ CQHLs satisfy $\,\Lm=\lm=
\LM=\lM=2\mini p+1$. General, powerful implications that follow from
$L$-minimality are summarized below in
Thms.\,\ref{sGT}.6--\ref{sGT}.8; for proofs, see~\cite{FKST}.
\rule[-4mm]{0mm}{5mm}

{\bf Theorem~\ref{sGT}.6.} {\em For $p=1,2,\ldots\,$, let \,\QHL\,
be a (proper) CQHL with $\,\sH\in\Sigp\,$ and $\,\LM=2\mini p+1$.
Then \,\QHL\, is primitive and\,} \Lmini, {\em i.e., we also have
$\,\Lm=\lM =2\mini p+1$. Moreover, \,\QHL\, is\,}
indecomposable {\em if $\,\sH<2/3\mini$.}
\rule[-4mm]{0mm}{5mm}

We note that the bound $\,\sH<2/3\,$ for indecomposability is
sharp. As a matter of fact, at $\,\sH=2/3\mini$, there is an
\Lmini $\ (\lM\mimini=3)$ {\em composite} CQHL. It is given by the
direct sum of two Laughlin fluids at $\,\sH=1/3\mini$; see
example~{\bf (b)} in Sect.\,\ref{sUC}\mini.

Next, we give a precise definition of ``shift maps''.
\rule[-4mm]{0mm}{5mm}

{\bf Definition.} {\em Shift maps, denoted by $\mini\Sp\mini,\
p=1,2,\ldots\,$, are maps between (proper) CQHLs of equal
dimensions, $\Sp\mini:\, \QHL \mapsto (\G^\prime,\Q^\prime)$.
Starting from an arbitrary basis $\{\ve{1},\ldots,\ve{N}\}$ of
\,\QHL, the image $(\G^\prime, \Q^\prime)$ is uniquely specified by
constructing a basis $\{\ve{1}^\prime,\ldots,\ve{N}^\prime \}$ and
a charge vector $\mini\Q^\prime$ that satisfy the conditions}

\ben
K_{ij}^\prime \:=\; <\!\ve{i}^\prime,\ve{j}^\prime\!> \;=\;
<\!\ve{i},\ve{j}\!> +\, 2\mini p <\!\Q\,,\ve{i}\!>\,
<\!\Q\,,\ve{j}\!> \;=\: K_{ij} + 2\mini p\, \qel(\ve{i}) \mini
\qel(\ve{j})\ ,
\een

\vav

\be
Q_i^\prime \:=\; <\!\Q^\prime,\ve{i}^\prime\!> \;=\;
<\!\Q\,,\ve{i}\!> \;=\: Q_i\ , \hfa i,j=1,\ldots,N\ .
\label{shiftdef}
\ee

\noi
Note that, although the conditions in~(\ref{shiftdef}) are
formulated w.r.t.\ given bases, they specify the image $(\G^\prime,
\Q^\prime)$ uniquely, since different choices of bases and charge
vectors in~(\ref{shiftdef}) simply lead to data pairs $(K^\prime,
\Qvp)$ for $(\G^\prime, \Q^\prime)$ which are all related by the
equivalence transformations~(\ref{equiv}).

Denoting by $\,\G_0\subset\G\,$ the {\em neutral sublattice} of a
CQHL \QHL, i.e.,

\be
\G_0 \::=\: \{\, \vq\in\G\ |\ <\!\Q\,,\vq\!>\:=\qel(\vq)=0 \,\}\ ,
\label{neutraldef}
\ee

\noi
it is straightforward to show that shift maps leave neutral
sublattices invariant,

\be
\G_0^\prime \:=\: \G_0\ .
\label{neutral}
\ee

\noi
As will be explained in more detail in Sect.\,\ref{sMS},
Eq.~(\ref{neutral}) implies that (in the scaling limit) the
corresponding chiral QH fluids exhibit the {\em same symmetries}.
This equation is the mathematical basis for calling two chiral QH
fluids {\em structurally similar}.

What is the action of the shift map $\,\Sp\mini:\, \QHL \mapsto
(\G^\prime,\Q^\prime)$, for $\,p=1,2,\ldots\,$, on the space of
invariants introduced in Sect.\,\ref{sBI}\mini?

{\bf (i)} The discriminant $\D^\prime$ of the (odd, integral,
euclidean) lattice $\mini\G^\prime$ is given by

\be
\D^\prime \:=\: \D\,(1+2\mini p\,\sH)\ .
\ee

{\bf (ii)} The Hall conductivity changes according to

\be
{1\over \sH^\prime} \:=\: {1\over \sH} + 2\mini p\ ,
\label{sHprime}
\ee

\noi
which corresponds to the ``$D$-operation'' in the
Jain-Goldman hierarchy scheme~\cite{JG}; see also~\cite{Jain}
and~\cite{FZ}. Note that Eq.~(\ref{sHprime}) implies that any CQHL
which is the image under a shift map $\mini\Sp\mini,\
p=1,2,\ldots\,$, necessarily has a Hall fraction strictly below
$1/(2\mini p)\mini$.

{\bf (iii)} The level $\mini l$, $g$, and the charge parameter $\lb$
are all invariant under the action of a shift map $\mini\Sp\mini$.

We summarize {\bf (i)--(iii)} by giving a succinct representation of
the action of the shift map $\mini\Sp\mini$ at the level of the CQHL
symbol,

\be
\QHLsymbol{N}{\sH=\ndH}{g}{\lb}
\;\:\stackrel{\Sp}{\longmapsto}\;
\QHLsymbol{N}{\sH^\prime = {{n_H\rule[-.8mm]{0mm}{2mm}} \over
d_H+2\mini p\,n_H}}{g}{\lb}\ ,\ p=1,2,\ldots\ .
\label{shift}
\ee

{\bf (iv)} The name ``shift map'' for $\Sp$ is motivated by the fact
that the relative-angular-momentum invariants $\Lm$ and $\LM$ are
simply shifted by $2\mini p$,

\be
\Lm^\prime \:=\: \Lm + 2\mini p\ , \hah \LM^\prime \:=\: \LM +
2\mini p\ .
\label{LL}
\ee

\noi
Unfortunately, for the physically relevant invariants $\mini\lm$ and
$\mini\lM$ of {\em generic} primitive CQHLs, there does {\em not,
in general}, hold a transformation rule similarly simple
to~(\ref{LL})!\,--\,Note, however, that for {\em indecomposable}
CQHLs the identities $\,\lm = \Lm\,$ and $\,\lM = \LM\,$ hold.

{}From the definitions above, one sees that the shift maps $\Sp$
are invertible on the set of (proper) CQHLs with Hall fractions
$\,\sH\leq 1/(2\mini p),\ p=1,2,\ldots\
$.\,--\,From~(\ref{shiftdef}) it simply follows that $\,\Sp^{-1} =
{\cal S}_{-p}$.\,--\,The preimages of these CQHLs are readily seen
to be (proper) CQHLs. The set of (proper) CQHLs is closed under the
action of the maps $\Sp$ and their inverses.

However, the maps $\Sp$ and their inverses do {\em not\mini}
necessarily preserve the decomposability properties of
CQHLs.\,--\,E.g., composite CQHLs can be mapped into indecomposable
ones, as illustrated in Thm.\,\ref{sGT}.8 below.\,--\,Moreover, the
maps $\Sp$ and their inverses do {\em not}, in general, preserve the
primitivity property we have imposed on physically relevant
composite CQHLs; see point {\bf (7)} in
Sect.\,\ref{sUC}\mini.\,--\,For an example of a primitive CQHL with
a preimage that is non-primitive, see Sect.\,4
in~\cite{FKST}.\,--\,From these remarks and the
definitions~(\ref{lmin}) and~(\ref{lmax}) of the invariants
$\mini\lm$ and $\mini\lM$, it is clear that the transformation
properties of these invariants under shift maps are not as
straightforward as the ones in~(\ref{LL}).

We recall that the main objective of the present work is the
classification of primitive CQHLs. Although this set is not
closed under the action of shift maps and their inverses, it is
remarkable that a subset of the primitive CQHLs, the class of
{\em\Lmini\,} CQHLs (defined after~(\ref{Sig})) {\em is closed\,}
under the action of shift maps and their inverses. This is the key
to powerful classification results that we state presently.

It is convenient to partition the class of \Lmini\ CQHLs into the
following subsets:

\be
\Hp^\pm \::=\: \{\, \QHL\ |\ \sH\in\Sigp^\pm\mini,\mbox{ \Lmini,
i.e., primitive and }\,\lm=\lM=2\mini p+1\,\}\ ,
\ee

\noi
where $\,p=(0),1,2,\ldots\,$, in accordance with the definition of
the windows $\Sigp^\pm$ given in~(\ref{Sig}).

The next two theorems show that, on the one hand, the sets
$\,\Hp:=\Hp^+\cup\Hp^-\,$ are structurally similar for different
$\mini p$'s, while, on the other hand, there is an essential
structural asymmetry between the sets $\Hp^+$ and $\Hp^-\mmini$,
for a given $\mini p$.
\rule[-4mm]{0mm}{5mm}

{\bf Theorem~\ref{sGT}.7.} {\em The sets $\Hp$ of \mini\Lmini\ CQHLs
with $\,\sH\in\!\Sigp\mini$, for $\,p=2,3,\ldots\,$, are in\mini}
one-to-one correspondence {\em with the set $\Hi$. The corresponding
bijections are realized by the shift maps $\,{\cal S}_{p-1} \mini :
\, \Hi \mapsto \Hp$.}
\rule[-4mm]{0mm}{5mm}

The proof of this theorem rests on Thm.\,\ref{sGT}.6 given above,
and it should be emphasized that chirality and $L$-minimality are
crucial for the theorem to hold; see~\cite{FKST}. Thm.\,\ref{sGT}.7
implies that, for the classification of \Lmini\ CQHLs, we can
restrict our analysis to the lattices with Hall fractions $\sH$ in
the ``fundamental domain'' $\,\Sigi=[\mini 1/3\mini, 1)\mini$!
\rule[-4mm]{0mm}{5mm}

In reference~\cite{FKST}, the set ${\cal H}_0^+$ of \Lmini\ CQHLs
with $\,\sH\in\![\mini 1\mini,\infty)\,$ has been constructed.
Applying the shift map ${\cal S}_1$ to it, we obtain the set
$\Hi^+$ of \Lmini\ CQHLs in the window
$\,\Sigi^+=[\mini 1/3\mini,1/2)\subset \Sigi\mini$. Hence, by
Thm.\,\ref{sGT}.7, all the sets $\Hp^+,\ p\geq 1$, are known. In
fact, we have the following result.
\rule[-4mm]{0mm}{5mm}

{\bf Theorem~\ref{sGT}.8.} {\em  For each $\,p=0,1,2,\ldots\,$, the
set $\mini\Hp^+$ of \mini\Lmini\ CQHLs with $\,\sH\in\!\Sigp^+\,$ is
uniquely given by the (infinite) series, $N=1,2,\ldots\,$, of
maximally symmetric CQHLs with $SU(N)$-symmetry of $\mini N$-ality
$1$, meaning that the one-electron states described by these CQHLs
transfrom under the fundamental representations of $SU(N)$. For a
given $p$, the corresponding symbols read}

\be
\QHLsymbol{N}{\sH={N\over 2\mini pN+1}}{1}{1} \;[\mini \lm=\lM=
2\mini p+1\mini]\ ,\hsp N=1,2,\ldots  \ .
\label{symbHp+}
\ee

The maximally symmetric CQHLs of this theorem are $N$-dimensional
and have been described explicitly in example~{\bf (c)} at the end
of Sect.\,\ref{sBI}\mini. In the notation of the next section,
(see~(\ref{dataLoG}) below) the sets $\Hp^+$ are written as

\be
\Hp^+ \:=\: \{ \: (\mini2\mini p+1\,|\,^1\mA A_{N-1})\ |\
N=1,2,\ldots\: \}\ .
\label{Hp+}
\ee

In Thm.~\ref{sGT}.6, it has been stated that all CQHLs
in~(\ref{Hp+}) with $\,p>0\,$ are {\em indecomposable}.
Furthermore, since their level $l$ equals unity, Thm.\,\ref{sGT}.5
states that a {\em charge-statistics relation} holds for the
quasi-particle excitations of the corresponding QH fluids.

\begin{table}[htb]
\vspace{8mm}
\begin{center}
\parbox{145mm}{{\bf Table~\ref{sGT}.1. }{\em Hall fractions
$\,\sH\in\! \Sigp^+$, for $\,p=0,1,2$, and $\mini 3$, that are
uniquely realizable or that cannot be realized by an\,} \Lmini\
{\em CQHL. The symbols ``$\:\bullet\mini$'', ``$\:\circ\mini$'',
and ``$\:\cdot\mini$'' specify the experimental status of the
fractions as explained in Fig.\,\ref{sI}.1. Fractions with
$\,\dH\!>\!21\mini$ are omitted.\rule[-4.8mm]{0mm}{5mm}}}
\renewcommand{\arraystretch}{1.3} \setlength{\tabcolsep}{2.1mm}
\setlength{\doublerulesep}{.7mm}
\begin{tabular}{|r*{11}{l}|}
\hline
\hline
\multicolumn{12}{|l|}{$\Sigma_0^+=[\mini 1\mini,\infty)\,,\ssp
\lm=\lM=1
  \,:$}
\\
{\small realizable}$\,:$ & \,\bulf{1} & \bulf{2} & \bulf{3} &
  \bulf{4} & \fmini\bulf{5} & \fmini\bulf{6} & \fmini\bulf{7} &
  \fmini\bulf{8} & \mini\bulf{9} & \bulf{10} & \mini\ldots\,
\\
{\small not realizable}$\,:$\rule[-3.2mm]{0mm}{4mm} &
  \multicolumn{11}{l|}{\small all proper fractions}
\\
\hline
\hline
\multicolumn{12}{|l|}{$\Sigi^+=[\mini{1\over 3}\mini,\mini{1\over
  2})\,,\,\ssp\lm=\lM=3\::$}
\\
{\small realizable}$\,:$ & \,\bulf{1\over 3} & \bulf{2\over 5} &
  \bulf{3\over 7} & \bulf{4\over 9} & \bulf{5\over 11} &
  \bulf{6\over 13} & \cirf{7\over 15} & \cirf{8\over 17} &
  \dof{9\over 19} & \nof{10\over 21} &
\\
{\small not realizable}$\,:$\rule[-3.2mm]{0mm}{4mm} &
  \hspace{.5mm}\nof{6\over 17} & \hspace{-.2mm}\dof{4\over 11} &
  \nof{7\over 19} & \nof{8\over 21} & \hspace{.7mm}\nof{5\over 13} &
  \hspace{.8mm}\nof{7\over 17} & \hspace{.7mm}\nof{8\over 19}
  & & & &
\\
 & \multicolumn{11}{l|}{\hspace{.8mm}\small and all even-denominator fractions}
\\
\hline
\multicolumn{12}{|l|}{$\Sigma_2^+=[\mini{1\over
  5}\mini,\mini{1\over 4})\,,\,\ssp\lm=\lM=5\::$}
\\
{\small realizable}$\,:$ & \,\bulf{1\over 5} & \bulf{2\over 9} &
  \hspace{-.3mm}\cirf{\mimini{3\over 13}} & \nof{4\over 17} &
  \hspace{.6mm}\nof{5\over 21} & & & & & &
\\
{\small not realizable}$\,:$\rule[-3.2mm]{0mm}{4mm} &
  \hspace{.4mm}\nof{4\over 19} &
  \multicolumn{10}{l|}{\small and all even-denominator
  fractions}
\\
\hline
\multicolumn{12}{|l|}{$\Sigma_3^+=[\mini{1\over
  7}\mini,\mini{1\over 6})\,,\,\ssp\lm=\lM=7\::$}
\\
{\small realizable}$\,:$ & \,\cirf{1\over 7} & \nof{2\over 13} &
  \nof{3\over 19} & & & & & & & &
\\
{\small not realizable}$\,:$\rule[-3.2mm]{0mm}{4mm} &
  \multicolumn{11}{l|}{\small all even-denominator fractions}
\\
\hline
\hline
\end{tabular}
\end{center}
\vspace{3mm}
\end{table}

We conclude this section by discussing Table~\ref{sGT}.1 which
summarizes the Hall fractions $\sH$ (with $\,\dH\leq 21$) in the
windows $\mini\Sigp^+\mini$ that can or cannot be realized by
elements in $\Hp^+$, with $\,p=0,1,2$, and $3\mini$.

A first inspection of Table~\ref{sGT}.1 reveals an impressive
agreement between the Hall fractions predicted by \Lmini\ CQHLs
and the experimentally observed values in the windows $\Sigp^+,\
p=1,2$, and $3\mini$. Note that CQHLs with higer dimensions
and/or higher values of $\,\lM$ are associated with less stable
QH fluids which is in accordance with our stability principle
advocated at the beginning of this section. In the windows,
$\Sigp^+,\ p=1,2$, and $3\mini$, there is only one Hall fraction,
$4/11$, for which there are some experimental indications
(however, only very weak ones!) that cannot be realized by an
\Lmini\ CQHL.

Concerning the results for the window $\Sigma_0^+$ we make three
remarks. First, it is satisfying to see that the ``standard''
integer QH fluids of the non-interacting electron approximation (see
example~{\bf (a)} at the end of Sect.\,\ref{sUC}) are naturally
included in our scheme and that they have a {\em unique} status.
They correspond to the {\em \Lmini\,}\ CQHLs in the window
$\Sigma_0^+$. We note that, contrary to the other CQHLs appearing in
Table~\ref{sGT}.1, these integer CQHLs are {\em composite}.

Second, the result that, in $\Sigma_0^+$, no proper Hall fraction
can be realized by an \Lmini\ ($\lM\!=\! 1$) CQHL leaves essentially
only two ways open for realizing (in the scaling limit) a fractional
QH fluid with $\,\sH\!>\! 1\mini$: (i) as a composite system of
independent, \Lmini\ electon- and/or hole-rich subfluids with
partial Hall fractions $\,\sH^\prime\!\leq\! 1$; see~(\ref{sHeh})
and~(\ref{sumsH}). Physically, e.g., the natural idea of adding
fully filled Landau levels to a fractional fluid with
$\,\sH^\prime\!<\! 1\,$ belongs to this situation; (ii) as an
indecomposable system described by a {\em non\mini}-\Lmini\ CQHL
(or a direct sum of such ones); see Sect.\,\ref{sLD}\mini.

Third, since the inverse shift maps $\mini\Sp^{-1}$ are relating the
CQHLs in the windows $\mini\Sigp^+,\ p\!\geq\! 1$, to the ones in
$\Sigma_0^+$, the results in Table~\ref{sGT}.1 are reminiscent of
Jain's construction~\cite{Jain} where interacting electron systems
with $\,\sH\!\in\!\Sigp^+\,$ are related to non-interacting electron
systems at the integers $\,N=\sH/(1\!-\!2\mini p\,\sH)$.
\rule[-4mm]{0mm}{5mm}

Given the discussion above, two questions emerge. First, what can
we say about the CQHL\mini-class $\mini\Hi^-$, and thus, by
Thm.\,\ref{sGT}.7, about all sets $\mini\Hp^-$ with $\,p\geq
1\mini$? Second, given some experimental evidence for the Hall
fraction $4/11$, which cannot be realized by an \Lmini\ CQHL, we
wish to get a fuller perspective on the assumption of
$L$-minimality. Hence the question: how can we go beyond the
classification of \Lmini\ CQHLs?

It turns out that already the first question, not to mention the
second one, addresses a truly formidable task of great complexity!
Sect.\,\ref{sMS} provides a partial answer to the first question by
classifying all ``maximally symmetric''\mmini, \Lmini\ CQHLs, which
represent the most natural generalizations of the CQHLs appearing in
Thm.\,\ref{sGT}.8. For low dimensions ($N\!\leq\! 4$),
Sect.\,\ref{sLD} gives the complete answer to the first question
and makes the first manageable step in the direction of answering
the second question.


\newpage
\begin{flushleft}
\section{Classification of Maximally Symmetric CQHLs}
\label{sMS}
\end{flushleft}

Maximally symmetric CQHLs correspond to the most natural
generalizations of the ``elementary'' $A\mini$- (or $su(N)$-) fluids
that appeared in  Thm.\,\ref{sGT}.8 of the last section, and which
have been shown to encompass the Laughlin fluids as well as the
``basic'' Jain fluids. Before we can give a precise definition of
the class of maximally symmetric CQHLs, we need to investigate a
general geometrical feature of CQHLs, namely their ``Witt
sublattices''. We use technical language, and then translate our
definitions into  explicit statements at the level of the data pairs
$(K,\Qv)$ associated with CQHLs; for the definition of these pairs,
see the beginning of Sect.\,\ref{sBI}\mini.

Let \QHL\, be a CQHL. Then the {\em Witt sublattice}, $\mini\Gw
\subset\G$, is defined to be the sublattice of $\mini\G$ generated by
all vectors of length squared $1$ and $2$. The general theory of
integral euclidean lattices~\cite{CSS,CSbook} tells us that $\mini\Gw$
is of the form,

\be
\Gw \:=\: \G_A \oplus \G_D \oplus \G_E \oplus I_l\ ,
\label{Witt}
\ee

\noi
where $I_l$ denotes the (self-dual) unit euclidean lattice in
$\mini l\mini$ dimensions, and $\mini\G_A,\,\G_D$, and $\mini\G_E$ are direct
sums of root lattices of the simple Lie algebras $\mini
A_{m-1}=su(m), \,D_{m+2} =so(2m+4),\ m=2,3,\ldots$, and $E_m,\
m=6,7,8$, respectively. The subscripts in the symbols $\mini
A_n,\,D_n$, and $E_n$ indicate the {\em ranks} of these algebras. We
note that all the root lattices of these Lie algebras are generated
by vectors of only one length, namely of length squared $2$. (In the
mathematical literature, the $A\mini$-, $D\mini$-, and $E\mini$-Lie
algebras are called simply-laced.)

Denoting by $\OO$ the orthogonal complement of $\mini\Gw$ in $\mini\G$,
whose dimension satisfies $\,\dim \OO = N-\dim\Gw \geq 1$, the
sublattice $\mini\Gw\oplus\OO$ is then called the {\em Kneser
shape\mini} of $\mini\G$, and one has the following embeddings of
lattices:

\be
\Gw\oplus\OO \:\subseteq\: \G \:\subseteq\: \Gs \:\subseteq\:
\Gws\oplus\OO^\ast\ ,
\label{Kneser}
\ee

\noi
where ``$\,{}^\ast\,$'' denotes the dual of a lattice, as explained
in Sect.\,\ref{sUC}\mini.

It can be shown~\cite{FT} that, for {\em indecomposable\mini} CQHLs
\QHL, $\Gw$ does {\em not\,} contain any $I_l$ and $\mini\G_{E_8}$
sublattices. In the following we will concentrate on indecomposable
CQHLs, or, correspondingly, on ``elementary'' chiral QH fluids.
\rule[-4mm]{0mm}{5mm}

{\bf Theorem~\ref{sMS}.1.} {\em Let \QHL\, be an indecomposable
CQHL with $\,\sH<2\,$. Then $\Q$ is\mini} orthogonal {\em to $\mini\Gw$,
i.e., $\Q\in\!\OO^\ast$, and $\,\Gw\subseteq\G_0\,$ where $\mini\G_0$ is
the neutral sublattice of \QHL. Moreover, if $\,\Gw\neq \emptyset$,
all the inclusions in~(\ref{Kneser}) are} proper.
\rule[-4mm]{0mm}{5mm}

For a proof of this theorem and more details on the constructions
above\,--\,which constitute the basis of the complete
classification program of (general) CQHLs\,--,
see~\cite[Sect.\,6]{FT}.

Thm.\,\ref{sMS}.1 has an interesting corollary concerning the {\em
symmetry properties} of the chiral QH fluid corresponding to \QHL.
Note that to every point in $\mini\G$ there corresponds a vertex operator
of the algebra of edge currents. Let $\mini\GG$ denote the Lie
algebra\,--\, a direct sum of simple algebras $\mini A_n,
\,D_n$, and $E_{6,7}$\,--\,whose root lattice is given by the Witt
sublattice $\mini\Gw$ of \QHL. It is not hard to show~\cite{FZ,FT}
that the algebra generated by the vertex operators corresponding to
the Witt sublattice, $\Gw$, of $\mini\G$ and the neutral
$u(1)$-currents is the enveloping algebra of the Kac-Moody current
algebra $\mini\GGh$  at level 1 (denoted $\mini\GGh_1$).

The (infinite dimensional) {\em symmetry algebra} $\mini\GGh_1$
canonically contains the (finite dimensional) Lie algebra $\mini\GG$
that can be associated with global symmetry generators. Thus, the
Lie group $G$ corresponding to $\mini\GG$ is the {\em group of global
symmetries\mini} of the QH fluid. This implies that,
given $m\mini$ electrons\,--\,fermionic quasi-particles with
charge $1$ and labelled by, say, $\,\vq_1, \ldots,
\vq_m\in\!\G\subset\Gs$\,--, they transform under particular unitary
irreducible representations (irreps.) of $\mini G$. These unitary
irreps.\ are specified as follows. Let

\be
\vq_{\mini i} \:=\: \vq_{\mini i,W} + \vq_{\mini i}^\prime, \hwh
\vq_{\mini i,W}\in\Gws, \hah \vq_{\mini i}^\prime
\in\OO^\ast\ ,
\label{weights}
\ee

\noi
be the decomposition, according to~(\ref{Kneser}), of the $i\mini$th
electron's label, $i=1,\ldots,m$. Then we may write $\,\vq_{\mini
i,W}=\vom_i+ {\bf r}_i\mini$, with $\,{\bf r}_i\in\!\Gw$ a {\em root
vector}, and with $\,\vom_i \in\!\Gws\,$ an {\em elementary
weight\mini}, i.e., a smallest length representative of the cosets
(or ``congruence classes'' in Lie group terminology) in the quotient
$\mini\Gws/\Gw$ (see, e.g., \cite{Sl}). Furthermore, by the general
representation theory of Lie and Kac-Moody algebras (see, e.g.,
\cite{GO}), the elementary weight $\vom_i$ determines uniquely a
unitary irrep., $\pi_{\vom_i}$, of $\mini G$ according to which the
one-electron state labelled by $\vq_{\mini i}$ transforms,
$i=1,\ldots,m$.

{}From the general results about lattices given in~\cite{CSS} (see
also~\cite{FT}), it follows that all the elementary weights
$\,\vom_i\in\!\Gw^\ast\,$ which can appear in~(\ref{weights}) are
such that the corresponding irreps.\ of $\mini\GG$ {\em can be
extended\,} to unitary highest-weight representations of $\mini\GGh$
at level $\mini 1$. For a discussion of the latter point, see,
e.g.,~\cite[Subsect.\,3.4]{GO}. We will call these elementary
weights ``{\em admissible\mini}'' weights, and the ones that can
occur for the simple algebras $\mini A_n,\,D_n$, and $E_{6,7}$ are
given explicitly in Appendix~A.

One can show~\cite{FT} that if $\,\dim\OO=1\,$ and $\,\G_0=\Gw\,$
then all one-electron states transform under the same unitary irrep.\
$\pi_{\vom}$ of $\mini G$, i.e., $\vq_{\mini 1,W}\equiv\cdots
\equiv\vq_{\mini m,W}\equiv \vom$ mod $\mini\Gw$.
\rule[-4mm]{0mm}{5mm}

The preceding general remarks motivate the following definition of
{\em maximally symmetric CQHLs}.
\rule[-4mm]{0mm}{5mm}

{\bf Definition.} {\em A (proper) CQHL \QHL\, is called\,} maximally
symmetric {\em if it satisfies $\,\dim \OO=1\,$ and $\,\G_0=\Gw\,$,
i.e., the neutral sublattice of \mini\QHL\, and its Witt sublattice
coincide. Furthermore, denoting by $\mini\GG$ the Lie algebra
associated with the root lattice $\mini\Gw$, the one-electron states
described by \mini\QHL\, are required to transform under a unitary
irrep.\ of $\,\GG$ which can be extended to a unitary
highest-weight representation of $\,\GGh_1\mini$.}
\rule[-4mm]{0mm}{5mm}

Maximally symmetric CQHLs \QHL\, are specified by the following data

\be
(\, L\:|\; {}^{\vom} \Gw \,)\ ,
\label{dataLoG}
\ee

\noi
where $L$ is an odd, positive integer, $\,\Gw = \G_A \oplus
\G_D \oplus \G_{E\neq 8}\,$ is the Witt sublattice of \QHL, and
$\,\vom\in\!\Gws/\Gw\,$ is an admissible weight labelling an irrep.\ of
the Lie algebra $\mini\GG$ associated with $\mini\Gw$. The possible weights
$\vom$
are further restricted by the value of $L$, namely
$\,<\!\vom,\vom\!>\:< L\,$ (see~\ref{Lgrom} below).

We note that if the Witt sublattice is a direct sum of simple root
lattices, $\Gw=\Gwi{1}\oplus\cdots\oplus\Gwi{k},\ k\geq 2$, then
the associated Lie algebra $\mini\GG$ is semi-simple with decomposition
$\,\GG=\GG_1\oplus\cdots\oplus\GG_k$, and correspondingly the
admissible weight reads $\,\vom=\vom_1 +\cdots+\vom_k\,$
where $\,\vom_i\in\!\Gwi{i}^\ast/\Gwi{i},\ i=1,\ldots,k$. In order to
get an {\em indecomposable\mini} lattice, every projection $\vom_i$
must represent a non-trivial coset in $\mini\Gwi{i}^\ast/\Gwi{i}$.
This can also be shown to be sufficient; see~\cite{FKT}. In the
sequel, we always assume that admissible weights $\vom$ fulfill this
requirement. Hence, {\em all\,} the maximally symmetric CQHLs given
in this paper are {\em indecomposable\mini}.

Equivalently to~(\ref{dataLoG}), we can also specify maximally
symmetric CQHLs \QHL\, by their corresponding data pair $(K,\Qv)$,
once a basis has been chosen in $\mini\G$; see the beginning of
Sect.\,\ref{sBI}\mini. Relative to a suitable ``normal'' basis
$\{\vq,\ve{1},\ldots,\ve{{N-1}}\}$ of $\mini\G$, \QHL\, is specified by

\be
K\:=\,
\left.\left(\renewcommand{\arraystretch}{1.7}
\begin{tabular}{c|c}
$L$ & $\omv\rule[-3.5mm]{0mm}{5mm} $  \\
\hline
$\omv^T$ & $\,C(\Gw)\rule[-3.5mm]{0mm}{5mm} $
\end{tabular} \right)\  \right\}\mbox{\scriptsize N}
\ ,\hah
\Qv \:=\:
(\mini\underbrace{1,0,\ldots,0}_{\mbox{\scriptsize N}}\,) \ ,
\label{dataKQ}
\ee

\noi
where $\,L=\; <\!\vq\,,\vq\!>\,$ is the same odd integer as
in~(\ref{dataLoG}), $C(\Gw)$ is the Gram matrix of the basis
$\{\ve{1},\ldots,\ve{{N-1}}\}$ of $\mini\Gw$\,--\,in the normal basis
chosen here, it equals the {\em Cartan matrix} of the Lie algebra
$\mini\GG$ associated with $\mini\Gw$\,--, and, finally,
$\,\omv=(\omega_1,\ldots,\omega_{N-1})\,$ is the vector of the dual
components of $\vom$ which are given by $\,\omega_j
=\:<\!\vom,\ve{j}\!>\mini,\ j=1,\ldots,N-1$. According to the
decomposition~(\ref{Kneser}), the basis vector $\vq$ can be written
as $\,\vq ={\sH}^{-1} \Q + \vom \mini$.

If $\mini\Gw$ is a direct sum of simple root lattices then
$\,C(\Gw)=C(\Gwi{1})\oplus\cdots\oplus C(\Gwi{k})\,$ is a
block-diagonal matrix, and $\,\omv=\omvi{1}+\cdots+\omvi{k} \mini$.
An example of data pairs~(\ref{dataLoG}) and~(\ref{dataKQ}) has been
given by~(\ref{Hp+}) and~(\ref{promKQ}), respectively. The explicit
forms of the Cartan matrices for the simple algebras $\mini A_n,
\,D_n$, and $E_{6,7}$, and of the dual vectors $\omv$ for the
admissible weights $\vom$ are given in Appendix~A.

We denote by $\D(\Gw)$ the discriminant of the Witt
sublattice, $\Gw$, of $\mini\G$, i.e., $\,\D(\Gw) := \det C(\Gw)
=| \Gws/\Gw\mini |$. {}From~(\ref{dataKQ}), it immediately follows that
for maximally symmetric CQHLs

\bea
\D\:=\: \det K & = & \D(\Gw)\:[\mini L-\,\omv\cdot
C(\Gw)^{-1}\omv^T\mini] \nonumber \\
 & = & \D(\Gw)\:[\mini L\,- <\!\vom,\vom\!>\mini ]\ssp
(\mini>0\mini)\ ,
\label{Lgrom}
\eea

\vav

\be
\sH \:=\; <\!\Q\,,\Q\!> \;=\: \Qv\cdot\Ki\Qv^T \:=\:
{\D(\Gw) \over \D} \:>\: {1\over L}\ .
\label{sHms}
\ee

\noi
These two equations are basic for proving the following theorem.
\rule[-4mm]{0mm}{5mm}

{\bf Theorem~\ref{sMS}.2.} {\em The symbol of a maximally symmetric
CQHL \,\QHL\, specified by~(\ref{dataLoG}) or, equivalently,
by~(\ref{dataKQ}), takes the form}

\be
\QHLsymbol{1\mini+\,\ranks\Gw\mini=\mini N}{\sH={1\over
L\,-<\!\vom,\vom\!>}}{g\mini=\mini \D(\Gw)/\hws}
{\hspace{-17.8mm}\lb\mini=\mini\hws/n_H} \hsp .
\label{symbms}
\ee

\noi
{\em where $\hw$ is the order of the elementary weight $\vom$ in
$\mini\Gws/\Gw\mini$. Furthermore, for the
relative-angular-momentum invariants $\lM$ and $\lm$, the equalities
$\,\lm=\lM=L\,$ hold.}
\rule[-4mm]{0mm}{5mm}

If $\mini\Gw$ is a direct sum of simple root lattices,
$\Gw=\Gwi{1} \oplus\cdots\oplus\Gwi{k},\ k\geq 2$, and $\,\vom=
\vom_1 +\cdots+\vom_k$, as above, then the following
identities hold:

\bea
\rank\Gw &=& \sum_{i=1}^k \rank\Gwi{i}\ , \nonumber \\
<\!\vom,\vom\!> &=& \sum_{i=1}^k <\!\vom_i,\vom_i\!>\ , \nonumber \\
\D(\Gw) \:=\: \det C(\Gw) &=& \prod_{i=1}^k \det C(\Gwi{i})\ ,
\nonumber
\eea

\vav

\bea
\hw &=& \lcm(\mini h_{\vom_1},\ldots,h_{\vom_{\mimini k}})\ ,
\label{dsum}
\eea

\noi
where the {\em least common multiple\mini} (lcm) of two
integers $\mini a\mini$ and $\mini b\mini$ is defined by
$\,\lcm(a,b)\!:= a\mini b/\gcd(a,b)$, and similarly for more than
two integers.

For the simple Lie algebras $\mini A_{m-1}=su(m), \,D_{m+2}
=so(2m+4),\ m=2,3,\ldots\,$, and $E_{6,7}$, all the ranks and
determinants of their Cartan matrices, as well as all the lengths
squared and orders of their admissible weights are
collected in Appendix~A.

\vspace{5mm}

\noi
{\bf Classification.} Exploiting the results of Thm.\,\ref{sMS}.2
and the identities in~(\ref{dsum}), it is possible to list all
maximally symmetric CQHLs which have a fixed value of $L$ and
whose Hall fractions $\sH$ ($>1/L$; see~(\ref{sHms})) belong to a
given interval. In Appendix~B, {\em all\,} maximally symmetric CQHLs
with $\,\lm=\lM=L=3\,$ and $\,\sH<1\,$ are listed. They are
organized in four infinite one-para\-meter, one infinite
two-para\-meter, and six finite series of CQHLs. For a physically
relevant subset of Hall fractions ($\dH\!<\!21$ and odd), the
resulting CQHLs are indicated in Fig.\,\ref{sI}.2, and a detailed
discussion is given presently.

Before entering this discussion, however, we state the most powerful
implication of these results. Recalling the discussion about the
shift maps in the second part of Sect.\,\ref{sGT}, we obtain the
following {\em classification result\mini}:
\rule[-4mm]{0mm}{5mm}

{\em All \mini\Lmini, maximally symmetric CQHLs are classified by
combining the series~{\bf (B1)--(B11)} given in Appendix\,B with
Thm.\,\ref{sGT}.7 of Sect.\,\ref{sGT}\mini.}
\rule[-4mm]{0mm}{5mm}

Here is a summary of the results given in Appendix~B\,--\,the
classification of \Lmini, maximally symmetric CQHLs with $\,1/\lM
=1/3 < \sH<1\mini$:

In the window $\,\Sigi^+=[\mini 1/3\mini,1/2)$, we find the infinite
series~{\bf (B1)} of CQHLs with Hall fractions $\,\sH=N/(2\mini
N\!+\!1),\ N=1,2,\ldots\,$, converging towards $1/2\mini$. This
``basic'' $A$- (or $su(N)$-) series needs no further explanation
since it coincides with the set $\Hp^+$ of Thm.~\ref{sGT}.8 which
has been discussed in detail at the end of the previous section.

In the ``complementary'' window $\Sigi^-=[\mini 1/2\mini,1)$, the
classification leads to new, physically interesting perspectives.

\begin{table}[tb]
\vspace{8mm}
\begin{center}
\parbox{147mm}{{\bf Table~\ref{sMS}.1. }{\em Symbols of\,} all
{\em\Lmini\ ($\mini\lM\!=\!3$), maximally symmetric CQHLs with
$\,\sH\in\! [\mini 1/2\mini, 2/3) \subset \Sigi^-$. Notations are as
in Fig.\,\ref{sI}.1, with the addition that ``$\:$\tbulf{\mbox{}}''
indicates a Hall fraction that has been observed in
two-layer/component systems. \rule[-4.8mm]{0mm}{5mm}}}
\renewcommand{\arraystretch}{1.8}\setlength{\tabcolsep}{2.7mm}
\setlength{\doublerulesep}{.7mm}
\begin{tabular}{|*{8}{r}|}
\hline
\hline
\multicolumn{8}{|l|}{{\small In }$\:[\mini{1\over 2},\mini{3\over
  5})\mini:\mbox{}\!^a$}  \\
\multicolumn{2}{|r}{\ptbulf{n-p}{\sub{3,4,\ldots}{({1\over
  2})}^2_2}} &
\bulf{\sub{4}{({6\over 11})}^1_1}\hspace*{3.6mm} &
\bulf{\sub{5}{({5\over 9})}^1_1} &
\multicolumn{2}{r}{\bulf{\sub{5}{({4\over 7})}^2_1}
  $\mmini,\,\sub{6}{({4\over 7})}^1_1$}\hspace*{2mm} &
\dof{\sub{6}{({10\over 17})}^1_1} &
\rule[-5.3mm]{0mm}{6mm}
\\
\hline
\multicolumn{8}{|l|}{{\small In }$\:[\mini{3\over 5},\mini{2\over
   3})\mini:\mbox{}\!^b$}  \\
\multicolumn{3}{|r}{\pbulf{B-p}{\sub{5}{({3\over 5})}^3_1}
  $\mmini,\,\sub{6}{({3\over 5})}^2_1
   \,,\,\sub{7}{({3\over 5})}^1_1
   \,,\,\sub{7}{({3\over 5})}^2_2$} &
\nof{\sub{8}{({14\over 23})}^1_1} &
\hspace*{.9mm}\bulf{\sub{9}{({8\over 13})}^2_1} &
\nof{\sub{6}{({12\over 19})}^1_1} &
\hspace*{.9mm}\cirf{\sub{7}{({7\over 11})}^1_1} &
\hspace*{-.7mm}\nof{\sub{7}{({15\over 23})}^1_1}\hspace*{1mm}
\\
\multicolumn{7}{|l}{{\small and }\nof{\sub{{n\mini +
  1}}{({2\mini n\over 3\mini n+2})}^g_\lb}\hspace*{2.9mm}{\small
  with $\;n=9,10,\ldots$}}\rule[-5.3mm]{0mm}{6mm} &
\\
\hline
\hline
\end{tabular}
\parbox{147mm}{
$^a\:${\footnotesize see {\bf (B2)}, {\bf (B3)}, {\bf (B4)}, and
  {\bf (B7)} of Appendix~B}\rule[0mm]{0mm}{5.6mm}

$^b\:${\footnotesize see {\bf (B3)}, {\bf (B4)}, and {\bf (B8)} of
  Appendix~B}}
\end{center}
\vspace{3mm}
\end{table}

First, in Table~\ref{sMS}.1, we collect the symbols, as defined
in~(\ref{symb}) of all \Lmini, maximally symmetric CQHLs with Hall
fractions $\sH$ in the subinterval $[\mini 1/2\mini,2/3)$. There are
infinitely many such lattices with Hall fractions accumulating at
$\mini 2/3\mini$.

{}From the first row in Table~\ref{sMS}.1, we conclude that, in the
subinterval [\mini 1/2\mini, 3/5), {\em no\mini} other fractions than
$1/2,\,6/11,\,5/9,\,4/7,$ and $10/17$ are realized by \Lmini,
maximally symmetric CQHLs! This result leads to the following
significant observation:

Taking also into account the classification of generic (not
necessarily maximally symmetric), low-dimensional CQHLs given in the
next section (see Table~C.2 in Appendix~C where a
three-dimensional, generic, \Lmini\ CQHL with $\,\sH=7/13\,$ is
given), we conclude that, for single-layer/component QH fluids with
$\,\sH=N/(2N\!-\!1)$, where $\,N=8,9,\ldots\,$, the ``{\em
charge-conjugation (or particle-hole symmetry) picture\,}'' provides
the only ``natural'' theoretical description. This picture
corresponds to the non-chiral decomposition $\,\Sigi^-\ni\sH
=1-\sH^\prime$, with $\,\sH^\prime<1/2\mini$; see~\cite{HH,JG} and
Appendix~E.

In general, for the ``second main series'' of Hall fractions,
$\,\sH=N/(2N\!-\!1),\ N=2,3,\ldots\,$, the charge-conjugation
picture amounts to a description in terms of the following
{\em charge-conjugated $A$- (or $su(N)-$) QH lattices}. These QH
lattices are {\em composites} of two CQHLs of {\em opposite}
chirality meaning that they describe QH fluids which consist of
electron- {\em and\,} hole-rich subfluids; see~(\ref{sHeh}). More
specifically, writing $\,\sH=1-\sH^\prime$, the charge-conjugated
$A$-QH lattices are composites of the standard CQHL for the integer
QH effect at $\,\sH=1\,$ (see example~{\bf (a)} at the end of
Sect.\,\ref{sUC}), and  an {\em \Lmini}\ ($\lM=3$) CQHL
corresponding to an elementary $A$-fluid with
$\,\sH^\prime=N/(2N\!+\!1) < 1/2\mini$. Note that, given the
uniqueness result (Thm.\,\ref{sGT}.8 of the previous section) for
the elementary $A$- (or $su(N)-$) fluids in $\Sigi^+$, the
``charge-conjugated $A$-fluids'' with $\,\sH=1-N/(2N\!+\!1)
=(N\!+\!1)/(2N\!+\!1)\,$ acquire a correspondingly {\em unique\mini}
status among all the QH fluids in $\Sigi^-$ that are proposed by the
charge-conjugation picture. Furthermore, it is shown in point~{\bf
(a)} of Appendix~E that the charge-conjugated $A$-fluids at
$\,\sH=N/(2N\!-\!1)\,$ coincide with the ``hierarchy
fluids''~\cite{HH,JG} at these fractions.

Contrary to the situation for the higher-denominator ($\dH\!\geq\!
15$) fractions of the second main series, we emphasize that, for
the fractions $\,\sH=N/(2N\!-\!1)$, with $\,N=2,3,\ldots,7$,
Tables~\ref{sMS}.1,~\ref{sMS}.2, and~C.2 show that there {\em
are\mini} strictly chiral alternatives to the charge-conjugated
$A$-fluids; (see also the discussion of the ``E-series''
in~\cite[Subsect.\,7.4]{FT}). Correspondingly, it is one of our
basic contentions in this paper that, {\em in $\Sigi^-$, the
charge-conjugation picture should not be applied without further
thought. For many fractions, there are chiral alternatives; see
Fig.\,\ref{sI}.2\,}! Actually, as will be discussed in
Sect.\,\ref{sDis}, the QH physics at many of the fractions
$\,\sH\in\!\Sigi^-\,$ turns out to be {\em very complex\,}!

It should be emphasized that non-chiral, composite QH fluids are
expected to exhibit a clear experimental signal distinguishing them
from purely chiral fluids. In {\em non-chiral\,} fluids it should be
possible to observe excitations of {\em both\mini} chiralities at
the edge of the samples, while this is, in principle, impossible
in chiral fluids. Hence, the experiments reported in~\cite{Ash},
which do {\em not\,} find edge excitations of both chiralities
at $\,\sH=2/3\,$ in the samples considered, are most interesting,
and further experimentation in this direction would clearly help to
deepen the understanding of the QH effect!

\begin{table}[tb]
\vspace{4mm}
\begin{center}
\parbox{145mm}{{\bf Table~\ref{sMS}.2. }{\em Symbols of\,} all
{\em\Lmini\ ($\mini\lM\!=\!3$), maximally symmetric CQHLs with
$\,\sH\in\! [\mini 2/3\mini, 1) \subset \Sigi^-$, and low
dimensions, $N\!\leq\! 6\mini$. Notations are as in
Tables\,\ref{sGT}.1 and~\ref{sMS}.1\mini.$\mbox{}\mini^a$
\rule[-4.8mm]{0mm}{5mm}}}
\renewcommand{\arraystretch}{1.8}\setlength{\tabcolsep}{5.5mm}
\setlength{\doublerulesep}{.7mm} \begin{tabular}{|*{7}{c}|}
\hline
\hline
\multicolumn{3}{|c}{\pbulf{B,n-p}{\sub{4,5,6}{({2\over 3})}^4_1}
  $\mmini,\,\sub{6}{({2\over 3})}^3_1$} &
\nof{\sub{5,6}{({3\over 4})}^2_2} &
\nof{\sub{6}{({10\over 13})}^1_1} &
\bulf{\sub{6}{({4\over 5})}^4_1} &
\tcirf{\sub{6}{({6\over 7})}^3_1}\rule[-4.6mm]{0mm}{5mm}
\\
\hline
\hline
\end{tabular}
\parbox{145mm}{
$^a\:${\footnotesize see {\bf (B5)}, {\bf (B6)}, {\bf (B9)}, and
  {\bf (B11)} in Appendix~B}\rule[0mm]{0mm}{5.6mm}}
\end{center}
\vspace{3mm}
\end{table}

Next, we remark that discussions and tables analogous to those for
the subinterval $[\mini 1/2\mini, 2/3)$, can be repeated for all
subintervals $\,[(n\!-\!1)/n,\, n/(n\!+\!1)\mini)\subset\Sigi^-,\
n=3,4,\ldots\ $. In each of these subintervals there is an infinite
number of \Lmini, maximally symmetric CQHLs with Hall fractions
accumulating at $n/(n\!+\!1)$.

Rather than repeating the discussions, we summarize in
Table~\ref{sMS}.2 the most relevant results for the remaining
interval $[\mini 2/3\mini, 1)$. For this interval, we present
all $L$-mi\-ni\-mal, maximally symmetric CQHLs of low dimension, say,
$N\!<\! 7\mini$. This restriction is motivated by our heuristic
stability principle (``the smaller $N$ and $\mini\lM\mini$, the more
stable the corresponding QH fluid'').

{}From Tables~\ref{sMS}.1 and~\ref{sMS}.2, and from our heuristic
stability principle, we are led to {\em predict\,} the existence
of chiral QH fluids at Hall fractions $10/17$, $10/13$, and
$12/19\mini$. Taking the symmetry structures of the corresponding
maximally symmetric CQHLs into account, the fraction $10/17$ is
clearly predicted to be the most likely, next candidate to be
observed in single-layer systems! By~{\bf (B4)}, the
one-electron states of the corresponding QH fluid are transforming
under the fundamental representations of $SU(2)\!\times\!SU(5)$.
Note also that, in the charge-conjugation picture, $10/17$ would
be ``conjugated'' to $7/17$ at which fraction there is, however,
{\em neither\mini} an \Lmini, maximally symmetric {\em nor\mini} a
generic, low-dimensional (see next section) CQHL! This conclusion
is interesting, since there are some tentative experimental results
suggesting the formation of a QH fluid at the fraction $10/17$
(see~\cite{GS}), and there is no indication of a QH fluid at the
``conjugated'' fraction $\,\sH=7/17\mini$.

Furthermore, comparing the data of Table~\ref{sGT}.1 to those of
Tables~\ref{sMS}.1 and~\ref{sMS}.2, one immediately notices a
striking {\em qualitative difference\mini} between the
``complementary'' windows $\mini\Sigp^+$ and $\mini\Sigp^-,\
p=1,2,\ldots\ $. By Thm.\,\ref{sGT}.8, we have that if a Hall
fraction in the windows $\mini\Sigp^+$ is realized by an \Lmini,
maximally symmetric CQHL then it is {\em unique}. On the other hand,
in the windows $\Sigp^-$, one often finds {\em several structurally
different\,} lattices realizing a given fraction. The CQHLs having
the same Hall fractions are typically embedded into one another.
This will be explained in more detail in Sect.\,\ref{sDis} when we
discuss the possibility of ``structural phase transitions'' in QH
fluids.

The status of {\em even-denominator\,} Hall fractions will be
discussed in Sect.\,\ref{sDis} when the classification of generic,
low-dimensional CQHLs that we present in the next section is
available.

In conclusion, we note that, except for the single fraction
$4/11$, {\em all\,} experimentally observed Hall fractions given in
Fig.\,\ref{sI}.1 can be realized by either an \Lmini, maximally
symmetric CQHL or a charge-conjugated $A$-QH lattice. All these
CQHLs are of reasonably low dimension $N$; as a matter of fact, we
have $\,N\!\leq\! 9$, except for $8/11$ where the lowest-dimensional
\Lmini, maximally symmetric CQHL has $\,N\!=\! 11\mini$.

However, before jumping to conclusions about the role of maximal
symmetry in the classification of physically relevant CQHLs, we
need to find a way of going at least one step beyond the
classification of maximally symmetric CQHLs, and see how the
resulting data compare with experimental results. Such a step
will be carried out in the following section by addressing the
classification problem of generic CQHLs in low dimensions
($N\!\leq\! 4$). Just to mention two results: we shall find, e.g.,
at $\,\sH\!=\!8/11$, a non-maximally symmetric CQHL in four
dimensions which is \Lmini\ and exhibits an $SU(2)$-sym\-me\-try;
see Table~C.4 in Appendix~C. Clearly, in describing the QH fluid
forming at $\mini 8/11$, this CQHL competes with the
$11$-dimensional, maximally symmetric one mentioned above.
Furthermore, the ``simplest'' non-\Lmini\ CQHL forms in dimension
$\,N\!=\!2\,$ just at the ``missing'' fraction
$\,\sH\!=\!4/11\mini$; see Table~C.1 in Appendix~C. It coincides
with the proposal in the ``hierarchy schemes''~\cite{HH,JG}; see
Appendix~E.


\newpage
\begin{flushleft}
\section{Classification of Low-Dimensional CQHLs}
\label{sLD}
\end{flushleft}

In this section, we venture a step beyond the classification of
maximally symmetric CQHLs presented in the last section. We
provide systematic classification results for low-dimensional CQHLs
that are neither necessarily \Lmini\ nor necessarily maximally
symmetric. This allows us to get a better understanding of the role
played by these two properties in the classification of physically
relevant CQHLs. In the second part of this section, we use
our results and the phenomenological data summarized in
Fig.\,\ref{sI}.1 to argue that the assumption of $L$-minimality for
physically relevant CQHLs is experimentally corroborated. The
maximally symmetric CQHLs turn out to be most relevant in the
windows $\mini\Sigp^+$ where they are unique in the sense of
Thm.\,\ref{sGT}.8. In the ``complementary'' windows $\mini\Sigp^-$,
they are typically competing with generic, low-dimensional, \Lmini\
CQHLs. The latter ones, however, often exhibit a form of ``partial''
symmetry, and are in most cases contained as QH sublattices (see
Sect.~\ref{sDis}) in maximally symmetric CQHLs of higher dimensions.

\vspace{5mm}

\noi
{\bf Classification.} We start by stating the precise classification
results and then sketch their derivation. We have nconstructed the
following sets of indecomposable, low-dimensional CQHLs\,--\,and,
correspondingly, of possible ``elementary'' chiral QH fluids:
\rule[-4mm]{0mm}{5mm}

\begin{flushleft}
\renewcommand{\arraystretch}{1.6}
\setlength{\tabcolsep}{2mm}
\begin{tabular}{rl}
\raisebox{2.9mm}{{\bf (N1)}} &
\parbox{130mm}{all one-dimensional CQHLs, (they correspond to the
Laughlin fluids as described at the end of Sect.\,\ref{sUC});}
\rule[-8.7mm]{0mm}{10mm}
\\
\raisebox{2.9mm}{{\bf (N2)}} &
\parbox{130mm}{all indecomposable CQHLs in dimension $\,N=2$,
(e.g., for $\,3\leq\lm\leq\lM\leq 7$, there are $42$ such
lattices);}
\rule[-8.7mm]{0mm}{10mm}
\\
\raisebox{2.9mm}{\hspace{1mm}{\bf (N3.1)}} &
\parbox{130mm}{all indecomposable CQHLs in dimension $\,N=3$, with
$\,\lm=\lM=3$, ($19$ CQHLs);}
\rule[-8.7mm]{0mm}{10mm}
\\
\raisebox{2.9mm}{{\bf (N3.2)}} &
\parbox{130mm}{all indecomposable CQHLs in dimension $\,N=3$, with
$\,3\leq\lm\leq\lM=5$, and $\,\sH\leq 3$, ($191$ CQHLs);}
\rule[-8.7mm]{0mm}{10mm}
\\
\raisebox{2.9mm}{{\bf (N4)}} &
\parbox{130mm}{all indecomposable CQHLs in dimension $\,N=4$, with
$\,\lm=\lM=3$, and $\,\sH\leq 1$, ($26$ CQHLs).}
\rule[-4mm]{0mm}{5mm}
\end{tabular}
\end{flushleft}

The explicit data characterizing the CQHLs of the sets~{\bf
(N1)}--{\bf (N4)} are summarized in Tables~C.1--C.4 of Appendix~C.

We recall that, by definition, $\,\lm=\Lm\,$ and $\,\lM=\LM\,$ for
indecomposable CQHLs; see point~{\bf (5)} in Sect.\,\ref{sBI}\mini.
Moreover, given the sets~{\bf (N1)}--{\bf (N4)}, it is
straightforward combinatorics to construct all primitive (see
point~{\bf (7)} in Sect.\,\ref{sUC}) CQHLs with bounds on
$N$ and $\mini\lM$ as above. We note that this construction has to be
carried out in order to obtain the classification of all
low-dimensional ($N\!\leq\!4$), \Lmini\ CQHLs in the windows
$\mini\Sigp$, with $\,p\geq 2$, by application of the shift maps
$\Sp$ of Sect.\,\ref{sGT}\mini.

Next, we turn to a brief sketch of the construction of the above
sets of CQHLs.

For each of the sets~{\bf (N2)}--{\bf (N3.2)}, the construction
is carried out in three steps: (i) One classifies all
indecomposable, integral, euclidean lattices $\mini\G$ with discriminants
$\D$ bounded by $\lM^N\mini$; see~(\ref{Dbound}). (ii) In the dual,
$\Gs$, of each lattice one carries out an exhaustive search for
odd, primitive vectors ($\Q\mini$-vectors). All $\Q\mini$-vectors
which belong to the same orbit under the action of the corresponding
lattice automorphism group are identified\,--\,since they give rise
to equivalent CQHLs; see~(\ref{equiv}). (iii) One has to calculate,
for each resulting CQHL \QHL, the value of $\mini\lM$ and retains
only those CQHLs satisfying the respective bounds on $\lM\mini$.

We remark that since the first step presents a highly non-trivial,
unsolved mathematical problem when $\D$ and $N$ are getting
large, the program above is bound to work only in low dimensions.
Actually, we have only been able to carry it out in two and three
dimensions! Specifically, the indecomposable, integral, euclidean
lattices with $\,N=2\,$ have been classified by Gauss;
see, e.g.,~\cite[especially Chapter~15]{CSbook}. In three
dimensions, the very detailed discussion of ``reduced forms'' for
the corresponding lattices by Dickson~\cite[especially the tables
in Chapter~11]{Dick} make a computer implementation for
classifying all such lattices with, say, $\,\D\leq 5^3=125\mini$,
straightforward. Further interesting mathematical considerations
related to this fist step can be found in~\cite{Sieg}.

The second step is easily realized for two-dimensional lattices.
Again in three dimensions, the work in~\cite{Dick} is most helpful,
since it provides precise algorithms for determining the
automorphism group of a given lattice. Given these algorithms
and the bounds in~(\ref{Qbound}), it is straightforward to find a
computer implementation of a search routine for orbits of
$\mini\Q\mini$-vectors.

The third step is tedious but computationally  straightforward. The
main work is to find all charge\mini-1 vectors in $\mini\G$ which then
have to be combined to form all possible symmetric bases needed in
order to calculate  $\lM\mini$; see~(\ref{Lmax}).

For a better organization of the CQHLs in~{\bf (N3.2)}, it is
convenient to introduce another relative-angular-momentum
invariant: Similarly to~(\ref{Lmax}), we denote by $\oBQ$ the set
of all ordered, symmetric bases of $\mini\G,\ \{\vq_1,\vq_2,\vq_3\}$,
i.e., $\,<\!\Q\,,\vq_i\!>\:=\!1$, for $\,i=1,2,3$, and
$\,<\!\vq_1,\vq_1\!> \;\leq\; <\!\vq_2,\vq_2\!> \;\leq\;
<\!\vq_3,\vq_3\!>$. Then one can show that, for all lattices
considered in~{\bf (N3.2)}, the following invariant is well-defined:

\be
\Lii \::=\: \min_{\stackrel{\scriptstyle \{\vq_1,\,\vq_2,\,\vq_3
\}\,\in\,\oBQs}{\scriptstyle<\vq_1,\,\vq_1>\mini=
\mini\lm\mini,\: <\vq_3,\,\vq_3>\mini=\mini\lM}}
<\!\vq_2,\vq_2\!> \ ,
\label{L2}
\ee

\noi
and its possible values are $3$ and $5$. The set~{\bf (N3.2)}
can be split into three subsets characterized by
$\,[\mini\lm,\Lii,\lM] = [3,3,5],\ [3,5,5]$, and $[5,5,5]$,
respectively. The corresponding compilations of CQHLs are summarized
in Table~C.3 of Appendix~C. Clearly, the subset with invariants
$[5,5,5]$ contains all the (indecomposable) images under the shift
map $\Si$ of the CQHLs listed in set~{\bf (N3.1)}; the corresponding
inverse images are indicated in Table~C.3\mini.

In order to obtain the set~{\bf (N4)} we have applied the
following procedure. Making use of the special form the data pairs
$(K,\Qv)$ characterizing these CQHLs take in suitable symmetric
bases (see~(C.3) in Appendix~C), the positivity of $\mini K\mini$
implies that all six coefficients, $a_1,\mini a_2,\ldots,\mini c$,
necessarily have an absolute value which is strictly less than
three. Based on this observation a simple computer routine can be
used to generate the data pairs $(K,\Qv)$ (relative to symmetric
bases) of all CQHLs which belong to the set~{\bf (N4)}. Identifying
all the data pairs which are related by a mere change of basis in an
underlying CQHL (see~(\ref{equiv})) and checking for
indecomposability, one obtains the result summarized in
Table~C.4\mini. Actually, the indecomposability of lattices with
discriminant $\,\D\leq 25\,$ could be checked by comparison with the
classification results given in~\cite{CSS}. The lattices with
discriminants $\D$ exceeding $25$ had to be considered case by case.

This completes the description of our procedures for obtaining the
sets~{\bf (N1)}--{\bf (N4)}. Next, we shall see what these results
imply with respect to the role played by $L$-minimality and maximal
symmetry in the classification of physically relevant CQHLs.

\vspace{5mm}

\noi
{\bf $L$-Minimality and Maximal Symmetry vs.~Experiment.} We
first recall that an \Lmini\ CQHL with $\,\sH\in\!\Sigp= [\mini
1/(2\mini p\!+\!1),1/(2\mini p\!-\!1)\mini)\,$ is characterized by
its primitivity (see point~{\bf (7)} in Sect.\,\ref{sUC}) and the
equalities $\,\Lm=\lm=\LM=\lM=2\mini p\!+\!1,\ p=1,2,\ldots\ $.
Given the explicit data in Appendices~B and~C, we can ask the
question: Which Hall fractions $\mini\sH$, e.g, in the window
$\Sigi$ are ``strongly non-\Lmini''? Here, {\em strongly
non-\Lmini\,} means that these fractions {\em can\mini} be realized
by a non-\Lmini\ (indecomposable or composite) CQHL with
$\,N\!\leq\!3$, but {\em neither\,} by a low-dimensional ($N\!\leq\!
4$), \Lmini\ CQHL, {\em nor} by a maximally symmetric one of {\em
arbitrary\mini} dimension. Besides this ``strong'' form of
non-$L$-minimality we may also define a ``weaker'' form. Let us call
a Hall fraction {\em weakly non-\Lmini\,} if it {\em can} be
realized by a non-\Lmini\ CQHL with $\,N\!\leq\! 3$, and if there is
also a maximally symmetric, \Lmini\ realization, however, {\em
only\mini} in higher dimensions, say, with $\,N\!\geq\!
10\mini$.\,--\,Recall the phenomenological discussion at the end of
the last section, where $\,N\!\simeq\! 10\,$ has been argued to
provide an approximate, heuristic upper bound on the dimension of
maximally symmetric CQHLs which are physically relevant.

\begin{table}[htb]
\vspace{4mm}
\begin{center}
\parbox{147mm}{{\bf Table~\ref{sLD}.1. }{\em Strongly and
weakly non-\Lmini\ Hall fractions $\mini\sH\mini$ in the window
$\,\Sigi=[\mini 1/3\mini,1)$. Notations are explained in the
text.\rule[-4.8mm]{0mm}{5mm}}}
\renewcommand{\arraystretch}{1.3}\setlength{\tabcolsep}{1.1mm}
\setlength{\doublerulesep}{.7mm}
\begin{tabular}{|*{7}{c}|}
\hline
\hline
$\ba\mbox{\dof{4\over 11}}\rule{0mm}{7mm} \\
  {[3,5]} \\ {[5,5,5]}\ea$  &
$\ba\mbox{\nof{7\over 19}}\rule{0mm}{7mm} \\
  {[3,5,5]} \\ {[5,5,5]}\ea$  &
$\ba\mbox{\nof{3\over 8}}\rule{0mm}{7mm} \\
  {[3,5,5]} \\ \, \ea$  &
$\ba\mbox{\nof{5\over 13}}\rule{0mm}{7mm} \\
  {[3,5,5]} \\ \,  \ea$  &
$\ba\mbox{\nof{7\over 17}}\rule{0mm}{7mm} \\
  {[3,3,5]} \\ {[5,5,5]}\ea$  &
$\ba\mbox{\cirf{8\over 15}}={1\over 3}+{1\over 5}\rule{0mm}{7mm} \\
  {[3]\oplus[5]} \\ {[5,5]\oplus[5]}\ea$  &
$\ba\mbox{\nof{11\over 19}}\rule{0mm}{7mm}\hspace*{1.7mm} \\
  {[5,5,5]}\hspace*{1.7mm} \\ \,  \ea$
\\
$\ba\mbox{\nof{13\over 21}}={1\over 3}+{2\over 7}\rule{0mm}{8mm}
  \\ {[3]\oplus[5,5]}\rule[-4.6mm]{0mm}{5mm} \ea$  &
$\ldots$ & & & & &
\\
\hline
$\ba\mbox{\nof{9\over 11}}\rule{0mm}{7mm} \\ {[3,5,5]} \\
(N\geq 17)\rule[-4.2mm]{0mm}{5mm}\ea$  &
\multicolumn{2}{c}{$\ba\mbox{\nof{13\over 15}}={1\over 3}+
  {1\over 3}+{1\over 5}\rule{0mm}{7mm} \\ {[3]\oplus[3]\oplus[5]} \\
(N\geq 25)\rule[-4.2mm]{0mm}{5mm} \ea$}  &
$\ba\mbox{\nof{17\over 19}}\rule{0mm}{7mm} \\ {[3,5,5]} \\
  (N\geq 33)\rule[-4.2mm]{0mm}{5mm} \ea$  &
$\ldots$ & &
\\
\hline
\hline
\end{tabular}
\end{center}
\vspace{3mm}
\end{table}

A compilation of strongly and weakly non-\Lmini\ Hall
fractions is given in Table~\ref{sLD}.1\mini. The non-\Lmini\
(indecomposable or composite) CQHLs realizing these fractions are
indicated by the values of their invariants $\mini\lm$, $\Lii$, and
$\mini\lM$, respectively, and the corresponding explicit data pairs
$(K,\Qv)$ can be found in Tables~C.1 and C.3 of Appendix~C. In
Table~\ref{sLD}.1, the dimensions in which maximally symmetric
lattices exist for the  weakly non-\Lmini\ fractions are indicated
in brackets. All other notations are as in Table~\ref{sGT}.1\mini.

Upon closer inspection, Table~\ref{sLD}.1 is most revealing. The
``simplest'' strongly non-\Lmini\ situations are encountered at
$\,\sH=4/11$ and $8/15\mini$. For both fractions, there is a
{\em two\mini}-dimensional $[3,5]$-CQHL with invariants
$\,\lb\!=\!g\!=\!1\mini$. It is indecomposable in the first, and
composite in the second case. As a matter of fact, we note that the
latter situation provides one of the ``simplest'' examples of a
{\em composite chiral\,} QH fluid, namely a composite of two basic
Laughlin fluids. Clearly, at $\,\sH=8/15$, the description in the
charge-conjugation picture, $\,8/15=1-7/15\,$ (where the $7/15$
hole-subfluid is described by the unique \Lmini\ CQHL $(3\,|\,^1\mA
A_{6})$ in dimension $N\!=\!7$; see the discussion in
Sect.\,\ref{sMS}), competes with the above non-\Lmini\ solution.
Applying the results of Appendix~E, the above $[3,5]$-CQHL at
$\,\sH=4/11\,$ can be seen to correspond to the QH fluids predicted
by the Haldane-Halperin (HH)~\cite{HH} and Jain-Goldman
(JG)~\cite{JG} hierarchy schemes at ``level'' two and three,
respectively.

Experimentally, there seems to be only very weak support for a QH
fluid at $\,\sH=4/11\,$ (see~\cite[and Ref.\,12 therein]{Hu}), and
some first indications of the Hall effect at $\mini 8/15$ have only
been found recently in very high quality samples~\cite{main,Kang}.
Apparently, the formation of QH fluids at these two fractions is a
very delicate matter!

More surprisingly, there is a persistent absence of experimental
indications of the QH effect at the non-\Lmini\ fractions $\,7/19$,
$\,5/13\mini$(!), $\,7/17\mini$(!), $11/19$, $13/21$,
$\,9/11\mini$(!), $\,13/15\mini$(!), and $\mini 17/19\mini$. The
fractions marked with ``(!)'' are well separated from experimentally
strong fractions nearby and thus, a priori, they are expected to be
experimentally observable! This should be further confronted with the
fact that none of the fractions in $\Sigi$ which are realizable by
\Lmini\ CQHLs with $\,N\!\leq\! 3\,$ is lacking experimental
observation!\,--\,We note that, in the two hierarchy schemes, fluids
at {\em low(!)} ``levels'' are predicted at all these fractions.
In the HH picture, there are, at all fractions above, fluids at
``level'' $3$, with the exception of $11/19$ and $13/21$ where
fluids form at ``level'' $5$. In the JG scheme, the corresponding
fluids are found at ``level'' $2$, except for the last three
fractions where they form at ``level'' $3$, $4$, and $5$,
respectively. {}From the point of view of QH lattices, all ``hierarchy
fluids'' predicted at the fractions above are {\em
non-euclidean\mini} with the exception of those at $\mini 7/17$ and
$\mini 7/19$; see Appendix~E. In these two cases, they coincide
with our non-\Lmini\ proposals with $\,\lm=3\,$ and $\,\lM=5$,
respectively, listed in Table~\ref{sLD}.1\mini.

Recalling the heuristic stability principle of Sect.\,\ref{sGT},
the observations above lead to the following
\rule[-4mm]{0mm}{5mm}

{\bf Strong Stability Principle.} {\em The most stable chiral QH
fluids are described by \Lmini\ CQHLs, and the smaller the lattice
dimension $N$, the greater the stability of the corresponding
fluid.} \rule[-4mm]{0mm}{5mm}

This heuristic stability principle, with the prominence of \Lmini\
CQHLs implied by it, is rather pleasing in the light of
Thm.\,\ref{sGT}.7 which states that all sets, $\Hp\mini$, of \Lmini,
primitive CQHLs in the windows $\mini\Sigp,\ p=2,3,\ldots\,$, stand
in one-to-one correspondence with $\Hi$ in $\Sigi$.

Furthermore, given the stability principle above and the result
of Thm.\,\ref{sGT}.8, it would appear to be justified to claim that
there is now a firm understanding of the ``structural organization''
of QH fluids in the windows $\mini\Sigp^+,\ p=0,1,2,\ldots\
$.\,--\,We note that, in particular, at the Hall fractions
$\,\sH=N/(2\mini p N\!+\!1),\ N=1,2,\ldots\,$, which belong to the
windows $\mini\Sigp^+$, the HH-hierarchy picture~\cite{HH}, the
JG-picture~\cite{JG} and our ``\Lmini\ CQHL picture'' are
equivalent! For details, see Appendix~E.

Combining the two preceding remarks, we conclude that the
challenging ground for deepening the understanding of the QH effect
lies in the ``complementary'' windows $\mini\Sigp^-,\
p=1,2,\ldots\,$, and, in particular, in the ``fundamental domain''
$\,\Sigi^-= [\mini 1/2\mini,1)\mini$! In this window, room is found
for an interesting competition between three classes of \Lmini\
CQHLs; namely, (i) the generic, low-dimensional ($N\!\leq\!4$) CQHLs
with no symmetry restrictions on their structure, (ii) the class of
maximally symmetric CQHLs of fairly low dimensions (typically
$N\lsim 9$), and (iii) the (non-chiral) charge-conjugated $A$-QH
lattices discussed in Sect.\,\ref{sMS}\mini. This competition and
its consequences, such as the prediction of possible ``structural
phase transitions'', appears to be missed in the hierarchy schemes.
It is one of the main issues we address in our final section.


\newpage
\begin{flushleft}
\section{Summary and Physical Implications of the Classification
Results}
\label{sDis}
\end{flushleft}

In this final section, the key insights and conclusions of the
previous sections are summarized and completed. In particular, the
status of the two main restrictions assumed in our classification,
{\em chirality\mini} and {\em $L$-minimality\mini}, is discussed in
detail. Several new experiments that could help to further deepen
the understanding of the QH effect, in particular, of the
``structural organization'' of QH fluids, are proposed.

\vspace{5mm}

\noi
{\bf Stability Principles.} Based on the physical meaning of the
CQHL invariants $N$ (the number of channels in the corresponding QH
fluid; see~{\bf(A2)} in Sect.\,\ref{sUC}) and $\mini\lM$ (the
smallest relative angular momentum of a pair of a certain type of
electrons that are excited above the QH fluid's ground state;
see~(\ref{lmax})), we have motivated, in Sect.\,\ref{sGT}, the
heuristic stability principle that {\em the smaller the invariants
$N$ and $\mini\lM$, the more stable the corresponding QH fluid}.

For a sharpening of this stability principle, the introduction of
the notion of $L\mini$-minimality has proven to be
effective.\,--\,{\em $L\mini$-minimality\mini} says that all the
minimal relative angular momenta between any two identical types of
electrons excited above a QH fluid's ground state are the {\em same}
(``homogeneity''), and that, furthermore, $\lM$ assumes the {\em
smallest possible} value (``minimality'') consistent with the value
of the Hall fraction $\sH$; see below~(\ref{Sig}).\,--\,A detailed
confrontation of our classification results (summarized in
Appendices~B and~C and discussed in Sects.\,\ref{sMS} and~\ref{sLD})
with the experimental data summarized in Fig.\,\ref{sI}.1 then leads
to the following strong stability principle: {\em The most stable
chiral QH fluids are described by \Lmini\  CQHLs, and the smaller
the lattice dimension $N$, the greater the corresponding fluid's
stability.}

Furthermore, the presently available experimental data on
single-layer systems suggest the respective values {\em $10$ and
$\mini 7$ as heuristic upper bounds for the invariants $N$ and
$\mini\lM$ of physically relevant CQHLs}, (see also the discussion
preceding Thm.~\ref{sGT}.2). This observation is most powerful in
combination with Thm.\,\ref{sGT}.1 which states that the set of
CQHLs satisfying such bounds is {\em finite\mini}.

We continue this subsection with two compilations of Hall fractions
where experimental indications for a QH fluid would, in the first
case, strengthen the conclusions above, and, in the second case,
would pose new interesting questions about the physics underlying
the QH effect. For a partial summary of the subsequent results, see
Fig.\,\ref{sI}.2 in Sect.\,\ref{sI}\mini.

{\bf (a)} {\em New fractions at which QH fluids can be expected
to form.} Given the above stability principles, there are basically
two ways to predict new Hall fractions at which one could expect the
formation of QH fluids in single-layer systems from the data given
in Appendices~B and~C.

First, we shall argue for new fractions in the window
$\,\Sigi=[\mini 1/3\mini,1)$. There, candidates are fractions that
can be realized by ``simple'' maximally symmetric CQHLs where
``simple'' means \Lmini, low-dimensional, and the Witt sublattice
(which encodes the symmetry properties of the fluid;
see~(\ref{Witt})) is either simple or semi-simple but with at most
{\em two} summands. The most obvious such candidates are the three
fractions $10/13$, $10/17$, and $12/19$ of Table~\ref{sMS}.1, and
the next ``member'' in the basic $A\mini$- (or $su(N)$-) series
(see~{\bf (B1)}), namely, $10/21\mini$! The first three fractions
are realized by CQHLs in six dimensions, the latter by one in ten
dimensions. All four lattices are indecomposable and have level
$\,l\!=\!\lb\mini g\!=\!1\,$ which means that, by Thm.\,\ref{sGT}.5,
a charge-statistics relation holds for them. In addition to these
fractions, further candidates in the window $\mini\Sigi$ can be
inferred form Table~C.4 containing all indecomposable, \Lmini\ CQHLs
in four dimensions. Here, two fluids with a partial $SU(2)$- and one
with a partial $SU(2)\!\times\!SU(2)$-symmetry are predicted to form
at $\,\sH=6/7$, $13/17$, and $\mini 14/19$, respectively! Moreover, a
generic fluid exhibiting no continuous symmetries might form at
$\,\sH=11/13\mini$.

Second, in the windows $\,\Sigp=[\mini 1/(2\mini p\!+\!1),1/(2\mini
p\!-\!1) \mini),\ p=2,3,\ldots\,$, new QH fluids are predicted by
acting with the shift maps $\mini{\cal S}_{p-1}$ on the CQHLs
corresponding to well-established fluids with $\,\sH\in\!\Sigi$;
see Sect.\,\ref{sGT}, in particular transformation
property~(\ref{shift}). The most immediate fluids whose shift map
images might be considered are the ones belonging to the
$A\mini$-series with Hall fractions $\,\sH={N/(2\mini N\!+\!1)}$.
This leads to predictions of QH fluids at, e.g., $\mini 2/13$,
$4/17$, and $\mini 5/21\mini$! We note that, from a QH lattice point
of view, our results at the fractions $\,\sH={N/(2\mini N\!+\!1)}$
and at their shifted images coincide with the proposals given in
both the Haldane-Halperin~\cite{HH} and the Jain-Goldman~\cite{JG}
hierarchy schemes; see Appendix~E. However, at most of the other
fractions, the pictures can differ significantly, as we explain in
detail in the remaining part of this section.

{\bf (b)} {\em ``Missing'' Hall fractions.} Our considerations,
here, are not only based on the two sets of classification results
summarized in Appendices~B (\Lmini, maximally symmetric CQHLs) and~C
(all indecomposable CQHLs with $\,N\leq 3\,(4)$ and $\mini\lM\leq
5\,(3)$), but also on the investigation of the {\em composite\mini}
CQHLs that can be built from the ones listed there, provided their
invariants $N$ and $\mini\lM$ satisfy the respective bounds. For
brevity, we restrict attention to odd-denominator fractions in the
window $\Sigi\mini$.\,--\,A general discussion of the status of
even-denominator fractions will be given below.

The strongest statement we can make about ``missing'' fractions in
$\Sigi$ is the following: The data mentioned above provide {\em no}
CQHLs at the fractions $\,6/17$, \,\cirf{9/17}, $8/19$, $10/19$,
$13/19$, $8/21$, $11/21,\ldots\,$, and hence, {\em no chiral\,} QH
fluids are expected to form at these fractions\mini!\,--\,When
listing fractions in this section, the dots ``$\ldots$'' are always
indicating further fractions with $\dH>21$, and the experimental
status of the fractions in single-layer systems is indicated as in
Fig.\,\ref{sI}.1\mini.\,--\,In other words, finding an experimental
signal at one of these fractions forces us either to go beyond our
classification results or to reconsider some of our basic
assumptions. E.g., the implications for the status of the
chirality assumption which follow form the experimental data at
$\,\sH=9/17 \,$--\,and, for that matter, would also result from
signals at $10/19$ and $\mini 11/21$\,--\,are discussed in the next
subsection.

By reversing the line of arguments that lead to the strong
stability principle in Sect.\,\ref{sLD}, we can make further
non-trivial predictions of ``missing'' fractions. Namely, assuming
(i) $L\mini$-minimality to be a necessary property of stable QH
fluids, and (ii) that our data is exhaustive (which means, in
particular, that generic \Lmini\ CQHLs with $\,N\geq 5\,$ are
physically irrelevant), then {\em no} stable chiral QH fluid can
form at the fractions \,\dof{4/11}, $\,5/13$, \,\cirf{8/15},
$\,7/17$, $\,7/19$, $11/19$, $13/21,\ldots\,$! These fractions have
been called strongly non-\Lmini\ in Sect.\,\ref{sLD}; see
Table~\ref{sLD}.1\mini. We note that a detailed analysis of the
implications resulting from the experimental indications at
$\mini 4/11$ and $\mini 8/15$ can also be found there. (The fraction
$8/15$ finds a natural explanation in the charge-conjugation
picture, as discussed presently, and the weak experimental data at
$4/11\mini$ might indeed indicate the {\em only\mini} QH fluid
corresponding to a non-\Lmini\ CQHL which, in this case, would be
two-dimensional.) Assuming, in addition, a heuristic upper bound on
the dimension $N$ of CQHLs that can be realized physically, say
$\,N\!\leq\! 10$, as mentioned above, then further  ``missing''
fractions are predicted to be $\,9/11\,(17)$, $\,13/15\,(25)$,
$\,17/19\,(33)$, as well as $\mini 11/17\,(23)$, $14/17\,(20)$,
$16/17\,(18)$, $15/19\,(15)$, $16/19\,(19)$, $18/19\,(20)$,
$16/21\,(19)$, $17/21\,(17)$,  $19/21\,(37),\ldots\,$! The first
three fractions in this list have been called weakly non-\Lmini\ and
appeared in Table~\ref{sLD}.1\mini. All fractions are listed together
with the dimension in which the lowest-dimensional maximally
symmetric, \Lmini\ CQHL can be found realizing that Hall fraction.

Given these predictions, it would certainly be most interesting to
carry out further experimental investigations in the regions around
the indicated ``missing'' Hall fractions! The status of some of
these fractions in the hierarchy schemes has been discussed towards
the end of Sect.\,\ref{sLD}\mini.

\vspace{5mm}

\noi
{\bf Composite CQHLs and Charge-Conjugation.} What can we infer
from experiment about the necessity to consider composite {\em
chiral\,} QH lattices in the description of single-layer QH fluids?
The answer is, there is {\em no} experimental data in
Fig.\,\ref{sI}.1 conveying need for composite CQHLs, {\em except\,}
possibly at $\,\sH=2\mini N/(2\mini N\!+\!1)\,$  where direct sums
of two identical (indecomposable) CQHLs from the basic
$A\mini$-series should not be ruled out, a priori; see the discussion
below, in the subsection about ``structural phase transitions''. To
substantiate this claim, let us list, e.g., all Hall fractions
exhibited by low-dimensional ($N\!\leq\! 4$), \Lmini, composite CQHLs
in $\mini\Sigi^-=[\mini 1/2\mini,1)$: \,\bulf{2/3}$\;=1/3+1/3$,
\,\bulf{4/5}$\;=2/5+2/5$, $\,5/6=1/3+1/2$, $\,9/10=1/2+2/5$,
$\,11/15=1/3+2/5$, $\,14/15=1/3+3/5$, $\,16/21=1/3+3/7,\ldots\ $.
We note that all such composite lattices necessarily have $\,\sH\geq
2/3\mini$. The claim can be further corroborated by also inspecting
higher-dimensional, as well as non-\Lmini, composite CQHLs.

In multi-layer/component systems with nearly independent
components\,--\,e.g., with a strong suppression of tunneling
between the different layers\,--, the picture will, of course, be
different, and fractions listed above might possibly arise.

The second question is whether the experimental data in
Fig.\,\ref{sI}.1 are suggestive of QH fluids that are composites of
subfluids with {\em opposite} chiralities? For single-layer systems,
the commonly accepted charge-conjugation (or particle-hole symmetry)
picture~\cite{HH,JG} assumes this to be so. Actually, in this picture, the Hall
physics at the fractions $\,\sH\in\!\Sigi^- = [\mini 1/2\mini,1) \,$
is assumed to be the ``charge-conjugated'' mirror image,
$\,\sH=1\!-\!\sH^\prime$, of the one at the corresponding fractions
$\,\sH^\prime\in\! (0\mini ,1/2]\mini$. In particular, at two
``conjugated fractions'' $(\sH,\sH^\prime)$, the likelihoods of
formation and the stability properties of the corresponding QH
fluids are expected to be approximately the same~\cite{JG}. Although
this picture is contained in our general framework presented in
Sect.\,\ref{sUC} (see~(\ref{sHeh}) and Appendix~E), we argue that it
is not, {\em in general}, in accordance with the experimental data
available so far.

Let us see, more precisely, what the experimental evidence for or
against the charge-conjugation picture is in single-layer systems.
A first look at Fig.\,\ref{sI}.1 shows that there are $\mini 11$
pairs of conjugated fractions $(\sH,\sH^\prime)$ where, at both
fractions, QH fluids of similar stability have been established, and
which thus are consistent with the charge-conjugation picture. These
$\mini 11$ pairs, however, have to be confronted with $\mini
10\,$(!) pairs of conjugated fractions $(\sH,\sH^\prime)$ where
either only one member is observed or the stability status of the
two members is markedly different. Taking a closer look at the
experimental data, one realizes that $\mini 8\mini$ of the $\mini
11$ pairs supporting charge-conjugation are of the form $(N/(2\mini
N\!+\!1),\,(N\!+\!1)/(2\mini N\!+\!1))$, i.e., they are relating
fractions of the basic $A\mini$-series with ones belonging to the
``second main experimental series''.

As we have discussed at the end of Sect.\,\ref{sMS}, it is natural
and, in some cases, necessary to take the charge-conjugation picture
into account when discussing the  QH physics at the fractions of the
second main series, $\,\sH=N/(2\mini N\!-\!1),\ N=2,3,\ldots\ $. The
particular {\em non-chiral, composite} QH lattices associated with
these fractions in the charge-conjugation picture have been called
{\em charge-conjugated $A\mini$-QH lattices}. They have a {\em
unique\mini} status among all charge-conjugated QH lattices in
$\mini\Sigi^-$; see \mbox{Sect.\,\ref{sMS}\mini.}

We note, however, that for the first six members ($2/3$ through
$\mini 7/13$) of the second main series, there are also {\em strictly
chiral, \Lmini\ alternatives}; a fact that is rather interesting, in
the light of the results reported in~\cite{Ash}. In the experiments
reported there, one has been looking for the signature of a
charge-conjugation QH fluid at $\,\sH=2/3\ (=1\!-\!1/3$), namely,
the existence of edge excitations of {\em both\mini} chiralities; see
Sect.\,\ref{sUC}\mini. But no evidence was found for this signature,
a result that would be consistent with the proposal of a {\em
strictly chiral\,} fluid at that fraction. Further physically
interesting implications of {\em chiral\,} QH lattices are discussed
below, in the subsection about ``structural phase transitions''.

There is another important observation to be made: In the realm of
CQHLs, there are only {\em non-\Lmini\,} CQHLs at the fractions
\,\dof{4/11}, $\,5/13$, and $\mini 7/17$, while at the ``conjugated''
values \,\cirf{7/11}, \,\bulf{8/13}, and \,\dof{10/17} there are {\em
\Lmini\,} (maximally symmetric) CQHLs of dimension $\mini 7$, $9$,
and $\mini 6$, respectively! Given the fact that the first three
fractions are experimentally only very weakly indicated or
unobserved, while the latter three are clearly observed or indicated,
we favour the chiral explanations for the latter three fractions
over the ones of the charge-conjugation picture.

In conclusion, we are tempted to claim that, for single-layer
systems, the presently available experimental data do not
support the charge-conjugation picture \mbox{{\em in general\mini}.}
Since this claim may appear to remain doubtful, further experiments
of the type reported in~\cite{Ash} would be most welcome!

\vspace{5mm}

\noi
{\bf Status of Even-Denominator Hall Fluids.} First, we  emphasize
that in the framework adopted in the present work, the description
of QH fluids at fractions with even denominators $\dH$ is {\em
not\,} an impossibility. This is satisfying since, experimentally,
even-denominator QH fluids are well-established at
$\,\sH=1/2$~\cite{1/2,Su} in {\em two-layer/component\,} systems, and
there are celebrated data at $\,\sH=5/2$~\cite{Wil5/2,5/2} observed
in {\em single-layer\mini} systems.

Second, theoretically, the most interesting {\em fact about
even-denominator CQHLs is that their charge parameters $\mini\lb$
are necessarily even\mini}; see Thm.\,\ref{sGT}.3\mini.
Phenomenologically, this translates into the prediction that, {\em
in such fluids, quasi-particles may be excited above the ground
state which have (fractional) charges} $\,e^\ast=1/(\lb\dH)\leq
1/(2\mini\dH)\,(!)$; see~(\ref{emin}). The even-$\lb$ observation
acquires further meaning when we note that all odd-denominator QH
lattices which are consistent with the above strong stability
principle and the respective phenomenological bounds on $N$ and
$\mini\lM\mini$, are characterized by $\,\lb\!=\!1\mini$! Thus, the
charge parameter $\mini\lb$ appears to play a dichotomizing role
between odd- and even-denominator QH fluids.

Third, we must ask the crucial question: Which even-denominator
fractions are predicted in our framework? To be more precise,
taking over (i) the strong stability principle, (ii) the
experimentally supported upper bounds on the invariants $N$ and
$\mini\lM\mini$, and (iii) that, phenomenologically, there is little
need for composite CQHLs, we ask: Which even-denominator Hall
fractions in $\mini\Sigi$ can be realized by \Lmini, indecomposable
CQHLs that are either maximally symmetric with $\,N\!\leq\! 10$, or
generic with $\,N\!\leq\! 4\mini$? The answer is surprisingly short!
We give the resulting fractions and indicate in round and square
brackets the dimensions of the corresponding maximally symmetric and
generic CQHLs, respectively: $\,1/2\:[2],(3,4,\ldots)$, $\,3/4
\:[4],[4]\!\supset\!su(3),(5,6,\ldots)$, $\,5/6\:(7,8,\ldots)$,
$\,5/8\:[4]\!\supset\!su(2),(9,10,\ldots)$, $\,7/8\:(9,10,\ldots)
\mini$. The generic lattices at $\,1/2$, $3/4$, and $\mini 5/8$ are
given explicitly in Tables~C.1 and~C.4 in Appendix~C, while all the
maximally symmetric ones with Hall fractions $\,(2\mini n\!-\!1)/
(2\mini n)\,$ are structurally similar. Their Witt sublattices are
given by $\,\mbox{}^1\mA A_{2(n-1)}\,\mbox{}^1\mA A_1 \,\mbox{}^1\mA
A_1\mini,\ \,\mbox{}^1\mA A_{2(n-1)}\,\mbox{}^2\mA A_3\mini,
\ldots\,$; see~{\bf(B2)} and~{\bf(B5)} in Appendix~B, and the
discussion in the next subsection. Since, for $\,n=2,3,\ldots\,$,
the Witt sublattices of the lowest-dimensional realizations are
semi-simple with {\em three\mini} summands, we do not expect these
lattices to present phenomenologically plausible proposals. This, in
turn, leaves us, {\em for the window $\mini\Sigi\mini$, with the
prediction of even-denominator QH fluids at $\,\sH=1/2$, $3/4$,
and\, $5/8\mini$}!

We recall that, as mentioned in Sect.\,\ref{sI}, there are convincing
arguments~\cite{HLRAIM} that, in a {\em single-layer\mini} QH
system, there are {\em no\mini} plateau at $\,\sH=1/2,\ 1/4,\ 3/4$,
(and other even-denominator fractions). The ground state of a QH
system at the corresponding filling factors is argued to be a
gapless Fermi liquid.

For {\em double-layer\mini} (or {\em wide-single-quantum-well\,}) QH
systems, however, the proposals made above are very natural. For
example, at $\,\sH=1/2$, we have a maximally symmetric CQHL with
symbol (see~(\ref{symb})) and data (see~(\ref{dataLoG})) given by
$\,\subi{3}{(1/2\mini)}^2_2 \fsp(3\,|\,\mbox{}^1\mA A_1\mini
\mbox{}^1\mA A_1)$. This three-dimensional example has been
discussed in Sect.\,\ref{sI}. The two $\mini A_1\!=\!su(2)\mini$
summands forming its Witt sublattice $\mini\Gw$ make it a natural
candidate for describing a QH fluid with an \SUs\ {\em and\,} an
$SU(2)_{layer}$ symmetry. Similar discussions can be repeated for
the other even-denominator QH lattices mendioned above.

\vspace{5mm}

\noi
{\bf Embeddings of CQHLs and Structural Phase Transitions.} A rather
remarkable consequence of our study of QH lattices is that, staying
in the context of {\em chiral\,} and {\em \Lmini\,} QH lattices, as
motivated above, the interval of Hall fractions $\,0<\sH\leq 1\,$
can naturally be organized into ``windows'' in a two-fold way.

First, defining the windows $\,\Sigp=[\mini 1/(2\mini p\!+\!1),
1/(2\mini p\!-\!1)\mini), \ p=1,2,\ldots\,$, the characterizing
property of \Lmini\ CQHL with $\,\sH\in\!\Sigp\,$ is that they
saturate the bound $\,1/\sH\leq \lM\,$ given in Thm.\,\ref{sGT}.2,
i.e., they have $\,\lM=2\mini p\!+\!1\mini$. We recall that, by
Thm.\,\ref{sGT}.7, {\em all the sets of \Lmini\ CQHLs with
$\,\sH\in\!\Sigp\,$ are in one-to-one correspondences with one
another.} These correspondences are realized by the {\em shift maps}
discussed in Sect.\,\ref{sGT}, and lead to the result that, when
discussing \Lmini\ CQHLs, we can restrict attention to the ``{\em
fundamental window\,}'' $\Sigi\mini$. We will make use of this
fact in the remaining part of this subsection.

Second, each window $\Sigp$ can be divided into two subwindows,
$\Sigp^+$ and $\mini\Sigp^-$, by the mid value of $\,1/(2\mini
p)$. The interesting fact behind this division is that the two
resulting subwindows exhibit very different ``{\em structural
organization\,}''. While, in the windows $\,\Sigp^+=[\mini
1/(2\mini p\!+\!1), 1/(2\mini p)\mini)$, there are {\em unique}
\Lmini\ CQHLs at the fractions $\,\sH=N/(2\mini pN\!+\!1),\
N=1,2,\ldots\,$, (see Thm.\,\ref{sGT}.8), one infers from the data
in Appendices~B and~C that, in the ``complementary'' windows
$\,\Sigp^-=[\mini 1/(2\mini p), 1/(2\mini p\!-\!1)\mini)$, typically
{\em several inequivalent\,} CQHLs can be found at a given Hall
fraction $\sH$. An interesting question then is: What is the
relationship between CQHLs which have the same Hall fraction?
Furthermore, what does this relationship imply at the level of QH
fluids? In order to answer these two questions, we introduce the
concept of QH-lattice embeddings.
\rule[-4mm]{0mm}{5mm}

{\bf Definition.} {\em A QH lattice $(\G^\prime,\Q^\prime\in\!
\G^{\prime\mini\ast})$ is embedded into another QH lattice
$(\G,\Q\in\!\Gs)$ if (i) both QH lattices exhibit the same Hall
fraction, i.e., $\sH^\prime = \:<\!\Q^\prime,\Q^\prime\!> \:=\:
<\!\Q\,,\Q\!> \:= \sH\mini$, (ii) $\G^\prime$ is a
sublattice of $\,\G$, and (iii) the two charge vectors $\Q^\prime$
and $\Q$ are compatible in the sense that all multi-electron/hole
states described by $(\G^\prime,\Q^\prime)$ remain physical states
when viewed (via the lattice embedding $\,\G^\prime\subset\G$) as
states described by $(\G,\Q)$. In particular, all the electric
charges stay the same, i.e., $<\!\Q^\prime,\vq^\prime\!> \:=\:
<\!\Q\,,\vq^\prime\!> \mini$, for all $\,\vq^\prime \in\!\G^\prime
\subset\G$.}
\rule[-4mm]{0mm}{5mm}

At the level of symbols (see~(\ref{symb})), we denote such embeddings
by

\be
\thesymbolpr\:\emb\thesymbol\ .
\label{embed}
\ee

\noi
Note that, as an immediate consequence of definition~(\ref{lmin}),
$\,\lm^{\mini\prime}\geq\lm\mini$.

Physically, a QH fluid described by the QH lattice $(\G^\prime,
\Q^\prime)$ which is embedded into another lattice \QHL\, is
characterized by a {\em restricted\,} set of possible
multi-electron/hole excitations above the ground state, as compared
to the corresponding set of the fluid associated with the lattice
\QHL. Furthermore, since the neutral sublattice
(see~(\ref{neutraldef})) of $(\G^\prime,\Q^\prime)$ is a sublattice
of the neutral sublattice of \QHL, the embedded fluid exhibits a (global)
symmetry
group $G^{\mini\prime}$ (see~(\ref{weights})) which is a {\em
subgroup} of $\mini G$, the symmetry group exhibited by the fluid
associated with \QHL. Thus, in this precise sense, {\em the embedded
fluid exhibits a more restricted symmetry than the one it embeds
into.} Put differently, {\em going from a QH fluid to an embedded
subfluid corresponds to a ``reduction or breaking of symmetries''.}
(As a mathematical aside, we remark that the study of embeddings of
maximally symmetric CQHLs into one another is equivalent to the
study of regular conformal embeddings of level-$1$ Kac-Moody
algebras and the respective branching rules. For recent results on
the latter subject, see, e.g., the references in~\cite{KMembed}.)
Experimentally, symmetry breaking might be realized in {\em
phase transitions\mini} that are driven, at a given Hall fraction, by
varying external control parameters. Hence, it is most interesting
to see at which fractions in $\mini\Sigi^-$ such ``structural'' phase
transitions can be expected within our framework.

Motivated by the observations in the first two subsections above,
we answer this question by taking into account the following
physically relevant sets of CQHLs: (i) all generic, \Lmini\ CQHLs
in low dimensions, $N\!\leq\! 4$ (see Appendix~C), (ii) all maximally
symmetric, \Lmini\ CQHLs in dimensions $\,N\!\leq\! 10$ (see
Appendix~B), and (iii) all composites of two identical lattices
belonging to the basic $A\mini$-series given in~{\bf(B1)} of
Appendix~B. The Hall fractions in $\mini\Sigi^-$ at which a
CQHL embedding, or ``chains'' of CQHL embeddings, can be
found are listed, together with the corresponding lattices, in
Table~D.1 of Appendix~D. The resulting fractions are\,
\mbox{$\,$\pbulf{B,n-p}{2/3}}, \mbox{$\,$\pbulf{B-p}{3/5}},
$\,$\bulf{4/5}, $\,$\bulf{4/7}, \mbox{$\,$\pbulf{(B-p)}{5/7}},
\mbox{$\,$\tcirf{6/7}}, $\,$\bulf{5/9}, and the even-denominator
fractions \mbox{$\,$\tbulf{1/2}} and $\,(2\mini n\!-\!1)/(2\mini
n)$, with $\,n=2,3$, and $4\mini$.

This result can actually be sharpened by taking the structure of
the involved CQHLs into account (especially, their symmetry groups).
Given that, at the fractions $\,n/(n\!+\!1)$, with $\,n=3,4,5,6$,
and $7$, already the lowest-dimensional pairs of embedded CQHLs
involve structurally complex Witt sublattices (with three summands
and dimensions $N\!\geq\! 5$), we do not expect the proposals at
these fractions to be phenomenologically very relevant. To
summarize, {\em in $\Sigi^-$, the Hall fractions at which structural
phase transitions are likely to occur are predicted to be}
\mbox{$\,$\pbulf{B,n-p}{2/3}}, \mbox{$\,$\pbulf{B-p}{3/5}},
$\,$\bulf{4/7}, \mbox{$\,$\pbulf{(B-p)}{5/7}}, $\,$\bulf{5/9}, {\em
and} \mbox{$\,$\tbulf{1/2}\mini!} Confronted with the experimental
data, we find it most remarkable that precisely at the three
fractions $\mini 2/3$, $3/5$, and $\mini 5/7$ at which there are
low-dimensional CQHL embeddings ($N\!\leq\! 4$), phase transitions
have been observed or are experimentally plausible. Observations of
phase transitions at $\,\sH=4/7$ and $\mini 5/9$ would, of course,
further support the proposed picture of structural phase
transitions. Thus, experiments are encouraged at these fractions!

One question that remains is whether {\em other types of phase
transitions} can occur {\em in the windows} $\mini\Sigp^+$ where we
have the $A\mini$-series of unique \Lmini\ CQHLs? The answer is {\em
yes\mini}! We briefly explain why. So far, we have basically ignored
the spin degrees of freedom in our discussion. However, a systematic
incorporation of {\em spin phenomena\mini} into our framework is
straightforward and has been discussed in detail in~\cite{FS2}; see
also~\cite{FT}. Basically, such an extended framework for QH fluids
with dynamical spin degrees of freedom incorporates (i) all the data
forming a QH lattice \QHL, and (ii) it additionally requires a {\em
polarization vector}, $\bfdelta\in\!\Gs$. The polarization vector
$\bfdelta$ specifies the spin-polarization of the excitations in the
system (relative to some given direction) similarly to the way the
charge vector $\Q$ specifies their electric charges; see~(\ref{cha}).
Given, e.g., a CQHL with a (neutral) $\mini A_1\!=\!su(2)\mini$
sublattice, it has been shown in~\cite[Sect.\,6]{FS2} that such a
lattice can naturally be used to describe {\em either} a QH fluid
with a spin-singlet ground state (from which \SUs-degrees of freedom
can be excited), {\em or\mini} a QH fluid with a fully polarized
ground state (from which only polarized quasi-particles can be
excited) exhibiting, however, an internal $SU(2)$-symmetry.
Datawise, the two QH fluids are only distinct by the form of their
associated polarization vectors! In~\cite[Sect.\,7]{FS2}, the most
simple examples of such fluids have been discussed. They form at the
fractions $\,\sH=2/(4\mini p\!+\!1),\ p=1,2,\ldots\,$, and are based
on the maximally symmetric, \Lmini\  CQHLs with data $(2\mini
p\!+\!1\,|\,\mbox{}^1\mA A_1)$; see~(\ref{dataLoG}). Experimentally,
we recall that the two QH fluids\,--\,one having a spin-singlet
ground state and the other a polarized ground state with an internal
symmetry\,--\,can be distinguished, in principle, by their magnetic
susceptibilities and by their quantum Hall effects for the spin
currents; see~\cite[Sect.\,7]{FS2}! In conclusion, {\em at fractions
in $\mini\Sigp^+$, we do not expect structural phase transitions;
however, spin-induced phase transitions are clearly possible!} More
details on this will be given elsewhere,~\cite{FKT}.

Finally, we ask whether one should expect to observe {\em phase
transitions at} $\,\sH\!=\!1\mini$. The unique \Lmini\ ($\lM\!=\!1$)
CQHL is the one-dimensional Laughlin lattice with $m\!=\!1$; see
example~{\bf (b)} in Sect.\,\ref{sUC}\mini. Thus, any other CQHL
realizing this fraction necessarily has to be {\em non-\Lmini\,}
($\lM\!\geq\! 3$), a fact that suggests a {\em markedly reduced
stability\mini} for the corresponding fluids, as compared to the
(\Lmini) Laughlin fluid! Moreover, by Thm.\,\ref{sGT}.4, we know
that any other indecomposable CQHL at this fraction exhibits a
charge parameter $\mini\lb$ strictly larger than $1$. By an
argument similar to the one in~(\ref{even}), this leads to the
prediction of {\em fractional\,} charges in these fluids! For the
purpose of illustration, we give the lowest-dimensional examples of
such lattices from Tables~C.1 and~C.2 in Appendix~C. Using the same
notations as in Appendix~D, one finds the following embeddings for
these non-\Lmini\ CQHLs at $\,\sH=1$

\be
\subi{2}{(1)}^4_2\fsp[3^{-1}3]
\emb
\left\{\renewcommand{\arraystretch}{1.2}
\ba
 \subi{3}{(1)}^6_2\fsp(2\mini {-1};0) \supset
   A_1\rule[-3.8mm]{0mm}{5mm}
\\
 \subi{3}{(1)}^8_2\fsp(1\mini {-1};1)
\ea\right\}
\emb
\subi{5}{(1)}^8_2\fsp(3\,|\,\mbox{}^1\mA A_1\mini \mbox{}^1\mA
A_1\mini\mbox{}^1\mA A_1\mini\mbox{}^1\mA A_1)
\emb\ldots\;.
\label{embed1}
\ee

\noi
We note that this chain of embeddings, with the corresponding
possibilities of structural phase transitions, is particularly
interesting in the light of the recent experimental data given
in~\cite{ptat1}. There, evidence for a phase transition between
different QH fluids at $\,\sH=1\,$ has been reported. The phase
transition seems to be driven by an in-plane magnetic field,
$\Bcpara$, and is observed in {\em double-layer\mini} QH systems.
Note that, in~(\ref{embed1}), e.g., the first two CQHLs, (the
lattice with symbol $\,\subi{2}{(1)}^4_2\,$ and the one with symbol
$\,\subi{3}{(1)}^6_2$), both are natural candidates for describing
double-layer QH fluids. The first one can be interpreted as
showing a discrete $\BB{Z}_{\mini 2}$ layer symmetry, while the
second one can be thought to exhibit a continuous $\,A_1=su(2)\,$
layer symmetry; see also the discussion in
\mbox{Sect.\,\ref{sI}\mini.} Furthermore, since, for all lattices
in~(\ref{embed1}), the charge parameter $\mini\lb$ equals $2$, we
would expect, as mentioned above, that quasi-particles with
fractional charge $1/2$ can be excited above the ground state of
the corresponding QH fluids. An experimental investigation of this
prediction would seem to be revealing and is encouraged!

\vspace{15mm}
\begin{flushleft}
{\Large\bf Acknowledgements}
\end{flushleft}

\noi
We are grateful to Thomas Kerler and Rudolf Morf for many useful
discussions. We thank Duncan Haldane for a helpful discussion on
hierarchy fluids and Paul Wiegmann for much needed encouragement
during times when a number of colleagues appeared to try to
discourage us. One of us (U.M.S.) thanks Yosi Avron for hospitality
at Technion (Haifa) and for many stimulating discussions in an early
phase of the present work. The research of U.M.S.~has been supported
by Onderzoeksfonds K.U.~Leuven, grant OT/92/9.


\newpage
\setcounter{secnumdepth}{0}
\begin{flushleft}
\section{Appendix A\mini: Simple Lie Algebras}
\label{AppA}
\end{flushleft}

\appendix
\setcounter{secnumdepth}{1}
\setcounter{section}{1}

\noi
The purpose of this appendix is to collect those facts about the
simple Lie algebras $\mini A_n = su(n\!-\!1), \ n=1,2,\ldots\,,\
D_n=so(2\mini n),\ n=4,5,\ldots\,,\ E_6\mini$, and $\mini E_7\mini$
which are basic for the classification of maximally symmetric CQHLs,
as discussed in Sect.\,\ref{sMS}\mini. For our explicit notations
we adopt the conventions of Ref.~\cite{Sl}\,--\,they are followed,
in particular, for the numbering of the simple roots of the
algebras above, and we note that this numbering differs from the
one chosen in~\cite{GO}. Furthermore, for notational simplicity,
we often only write the symbol $\mini\GG$ denoting a simple Lie
algebra when we are actually referring to the associated root
lattice $\mini\G_{\!\GG}$.

As stated in the text, the {\em ranks} of the Lie algebras
$\mini A_n,\ D_n\mini$, and $\mini E_n\mini$, and correspondingly of
their associated root lattices are given by the index $\mini n\mini$
in their symbols.

Further data about these algebras, which we generally denoted by
$\mini\GG$, are given as follows: First, we specify the {\em
Cartan matrices}, $C(\GG)$, which characterize the associated root
lattices, $\G_{\!\GG}$, and we give the corresponding {\em
discriminants}, $\,\D(\GG)=\det C(\GG)$. Second, we provide the {\em
admissible weights}, $\vom$, in the dual lattices,
$\G^\ast_{\!\GG}$, by stating explicitly their dual-component
vectors, $\omv$, the so-called Dynkin labels. Moreover, the lengths
squared, $<\!\vom,\vom\!>$, and the orders, $\hw$, of these weights
in $\mini\G^\ast_{\!\GG}/\G_{\!\GG}$ are listed.
\rule[-4mm]{0mm}{5mm}

\noi
$\bullet\ $ For $\,A_{m-1}=su(m),\ m=2,3,\ldots\,$, we have relative
to a basis of simple roots $\{\ve{1},\ldots,\ve{{m-1}}\}$:

\bea
C(A_{m-1}) & = & \left.\left(\renewcommand{\arraystretch}{.8}
\begin{array}{cccccccc}
2 & -1 & 0 & \cdot & \cdot & \cdot & 0 & 0 \\
-1 & 2 & -1 & 0 & \cdot & \cdot & 0 & 0 \\
0 & -1 & 2 & -1 & 0 & \cdot & 0 & 0 \\
\cdot & \cdot & \cdot & \cdot & \cdot & \cdot & \cdot & \cdot \\
0 & 0 & 0 & \cdot & 0 & -1 & 2 & -1 \\
0 & 0 & 0 & \cdot & \cdot & 0 & -1 & 2
\end{array} \right)\  \right\} m-1 \ , \hwh \nonumber \\
\det C(A_{m-1}) & = & m\ .
\eea

The admissible weights, $\vom_t,\ t=1,\ldots,m-1$, correspond to
the unitary irreducible representations (irreps.)\mmini of
$\,su(m)\,$ with ``$m$-alities'' $t\mini$ and dimensions
${m\cdot(m-1)\cdot\,\cdots\, \cdot(m-t+1)\over 1\cdot 2\cdot\,\cdots
\,\cdot t}$. They are given by the dual-component vectors
$\,\omvi{t}=(<\!\vom_t,\ve{1}\!>, \ldots,<\!\vom_t,\ve{{m-1}}\!>)$
which read explicitly

\bea
\omvi{t} & = &
(\mini\underbrace{0,\ldots,0,1,0,\ldots,0}_{\mbox{\scriptsize
m-1}}\,) \ ,\ \mbox{ with $1$ in the $\mini t\mini$th position}\ .
\eea

\noi
Moreover, their lengths squared and orders are given by

\be
<\vom_t,\vom_t> \;=\: {t\,(m-t)\over m}\ , \hah
h_{\vom_{\mimini\scriptstyle t}} \:=\: {m\over \gcd(m,t)}\ .
\ee

\noi
We note that, from the point of view of characterizing CQHLs, the
elementary weights $\vom_t$ and $\vom_{m-t}$ are equivalent; see the
equivalence relation~(\ref{equiv}).
\rule[-4mm]{0mm}{5mm}

\noi
$\bullet\ $ For $\,D_n=so(2\mini n),\ n=4,5,\ldots\,$, we have:

\bea
C(D_n) & = & \left.\left(\renewcommand{\arraystretch}{.8}
\begin{array}{ccccccccc}
2 & -1 & 0 & \cdot & \cdot & \cdot & 0 & 0 & 0 \\
-1 & 2 & -1 & 0 & \cdot & \cdot & 0 & 0 & 0 \\
0 & -1 & 2 & -1 & 0 & \cdot & 0 & 0 & 0 \\
\cdot & \cdot & \cdot & \cdot & \cdot & \cdot & \cdot & \cdot & \cdot \\
0 & 0 & 0 & \cdot & 0 & -1 & 2 & -1 & -1 \\
0 & 0 & 0 & \cdot & \cdot & 0 & -1 & 2 & 0 \\
0 & 0 & 0 & \cdot & \cdot & 0 & -1 & 0 & 2
\end{array} \right)\ \right\} n \ , \hwh \nonumber \\
\det C(D_n) & = & 4\ .
\eea

There are three admissible weights, $\vom_v,\,\vom_{\mimini s}$, and
$\vom_{\bar{s}}$, corresponding to the $2\mini n$-di\-men\-sional
vector, the $2^{\mini n-1}$-dimensional spinor, and the conjugate
spinor irrep.\ of $so(2\mini n)$, respectively. The corresponding
$n$-di\-men\-sional dual-component vectors read

\bea
\omvi{v} & = & (1,0,\ldots,0)\ , \nonumber \\
\omvi{s} & = & (0,\ldots,0,1)\ , \ha \nonumber \\
\omvi{{\bar{s}}} & = & (0,\ldots,0,1,0)\ .
\label{vss}
\eea

\vspace{-4.5mm}\noi
Furthermore,

\be
<\vom_v,\vom_v> \;=\: 1\ , \hah h_{\vom_{\scriptstyle v}} \:=\: 2\ ,
\ee

\vav

\be
<\vom_{\mimini s},\vom_{\mimini s}> \;=\: {n\over 4} \:=\;
<\vom_{\bar{s}},\vom_{\bar{s}}> \ , \hah
\mbox{$h_{\vom_{\scriptstyle\mmini s}} \:=\:
h_{\vom_{\scriptstyle\mimini\bar{s}}} \:=\: \left\{\ba 4 \\ 2\ea
\right.$ if $\,n\,$ is $\left\{\ba \mbox{odd} \\ \mbox{even}\ea
\right.$}\ .
\ee

\noi
For the labelling of CQHLs, $\vom_{\mimini s}$ and $\vom_{\bar{s}}$ are
equivalent by~(\ref{equiv}). Moreover, for $D_4$, all three admissible
weights in~(\ref{vss}) are equivalent (the so-called ``triality'' of
$so(8)$).
\rule[-4mm]{0mm}{5mm}

\noi
$\bullet\ $ For $\mini E_6$, we have:

\bea
C(E_6) & = & \left(\renewcommand{\arraystretch}{.8}
\begin{array}{cccccc}
2 & -1 & 0 & 0 & 0 & 0 \\
-1 & 2 & -1 & 0 & 0 & 0 \\
0 & -1 & 2 & -1 & 0 & -1 \\
0 & 0 & -1 & 2 & -1 & 0 \\
0 & 0 & 0 & -1 & 2 & 0 \\
0 & 0 & -1 & 0 & 0 & 2
\end{array} \right) \ , \hwh \nonumber \\
\det C(E_6) & = & 3\ .
\eea

There are two admissible weights, $\vom_{\mimini f}$ and
$\vom_{\mimini\bar{f}}$, corresponding to the $27$-di\-men\-sional
fundamental, and to its contragredient irrep.\ of $E_6$,
respectively. The corresponding dual-component vectors read

\bea
\omvi{f} & = & (1,0,0,0,0,0)\ , \nonumber \\
\omvi{{\bar{f}}} & = & (0,0,0,0,1,0)\ ,
\eea

\vspace{-4.5mm}\noi
Furthermore,

\be
<\vom_{\mimini f},\vom_{\mimini f}> \;=\: {4\over 3} \:=\;
<\vom_{\mimini\bar{f}},\vom_{\mimini\bar{f}}> \ , \hah
h_{\vom_{\mmini f}} \:=\: h_{\vom_{\mmini\bar{f}}} \:=\: 3\ .
\ee

\noi
For the labelling of CQHLs, these two elementary weights are equivalent.
\rule[-4mm]{0mm}{5mm}

\noi
$\bullet\ $ Finally, for $\mini E_7$, we have:

\bea
C(E_7) & = & \left(\renewcommand{\arraystretch}{.8}
\begin{array}{ccccccc}
2 & -1 & 0 & 0 & 0 & 0 & 0 \\
-1 & 2 & -1 & 0 & 0 & 0 & 0 \\
0 & -1 & 2 & -1 & 0 & 0 & -1 \\
0 & 0 & -1 & 2 & -1 & 0 & 0 \\
0 & 0 & 0 & -1 & 2 & -1 & 0 \\
0 & 0 & 0 & 0 & -1 & 2 & 0 \\
0 & 0 & -1 & 0 & 0 & 0 & 2
\end{array} \right) \ , \hwh \nonumber \\
\det C(E_7) & = & 2\ .
\eea

There is one admissible weight, $\vom_{\mimini f}$,
corresponding to the $56$-di\-men\-sional fundamental irrep.\ of
$E_7$, with

\bea
\omvi{f} & = & (0,0,0,0,0,1,0)\ ,
\eea

\vav

\be
<\vom_{\mimini f},\vom_{\mimini f}> \;=\: {3\over 2}\ , \hah
h_{\vom_{\mmini f}} \:=\: 2\ .
\ee


\newpage
\setcounter{secnumdepth}{0}
\begin{flushleft}
\section{Appendix B\mini: Maximally Symmetric CQHLs}
\label{AppB}
\end{flushleft}

\appendix
\setcounter{secnumdepth}{1}
\setcounter{section}{2}

\noi
In this appendix, all maximally symmetric CQHLs with $\,\lm=\lM=
L=3$, and $\,\sH<1\,$ are listed. The compilation has been obtained
by systematically exploiting Thm.\,\ref{sMS}.2 in Sect.\,\ref{sMS}
and the identities~(\ref{dsum}). The data is organized in
11~series~{\bf (B1)--(B11)}, and for each series the following
format is chosen:

First, the symbols of the CQHLs, $\subs{N}{(\ndH)}^g_\lambda\mini$,
are given; see~(\ref{symb}). They are followed by the characterizing
data of maximally symmetric CQHLs, $(\mini
L\:|\;\mbox{}^{\vom}\Gw\mini)$; see~(\ref{dataLoG}). Actually, since
we are considering exclusively CQHLs with $\,L=3\,$ in this
appendix, the quantity $\mini L\mini$ is omitted from the notation
and only the data $\mini\mbox{}^{\vom}\Gw\mini$ is stated
explicitly. If the Witt lattice is composite, $\Gw=\Gwi{1} \oplus
\cdots \oplus\Gwi{k},\ k\geq 2$, and the elementary weight reads
correspondingly $\,\vom=\vom_1+\cdots+ \vom_k\mini$, then we write
$\,\mbox{}^{\vom}\Gw = \mbox{}^{\vom_1}\Gwi{1} \cdots
\mbox{}^{\vom_k}\Gwi{k}\mini$. As in Appendix~A, the root lattices
$\mini\Gwi{i}\mini$ are denoted by the symbols of the associated
(simple) Lie algebras $\mini A_n,\ D_n\mini$, and $\mini E_n\mini$,
respectively. Furthermore, the notation for the elementary weights
$\,\vom_t,\,\vom_v,\,\vom_{\mimini s}\mini$, and $\vom_{\mimini f}$
which are all given explicitly in Appendix~A is simplified by only
writing the indexing letters $\, t,\, v,\, s$, and $f$,
respectively. Finally, we adopt the convention of writing $\,a\,|\,
b\,$ and $\,a\mA\not|\ b\:$ if $\,a\,$ divides, respectively, does
not divide $\,b\mini$.

Second, for each series, explicit examples of Hall fractions which
can be realized by a CQHL of that series are given together with
indications of their experimental status, typically in single-layer
systems. For the corresponding notations, see Fig.\,\ref{sI}.1 in
Sect.\,\ref{sI} and Table~\ref{sMS}.1 in Sect.\,\ref{sMS}\mini.

\begin{table}[htbp]
\vspace{16mm}
\begin{flushleft}
\parbox{145mm}{{\bf Table~B.1. }{\em All maximally symmetric
CQHLs with $\,L=3\,$ and $\,\sH<1$.\rule[-4.8mm]{0mm}{5mm}}}
\renewcommand{\arraystretch}{1.8}
\setlength{\tabcolsep}{2.1mm}
\setlength{\doublerulesep}{.7mm}
\begin{tabular}{lr*{11}{l}}
\hline
\hline
 & $\subs{N}{(\ndH)}^g_\lambda\mini$ &
  \multicolumn{11}{l}{$\mbox{}^{\vom}\Gw\,$,\hspace{6.1mm}
  Parameters and Examples\rule[-4.6mm]{0mm}{6mm}}
\\
\hline
{\bf (B1)}\rule[0mm]{0mm}{9.4mm} &
  $\displaystyle \QHLsymbol{N}{N\over 2\mini N+1}{1}{1}$ &
  \multicolumn{11}{l}{$\mbox{}^1\mA A_{N-1}\,,\hsp
  N=1,2,\ldots\ $:}
\\
 & & \bulf{1\over 3} & \bulf{2\over 5} & \,\bulf{3\over 7} &
  \bulf{4\over 9} & \bulf{5\over 11} & \bulf{6\over 13} &
  \cirf{7\over 15} & \cirf{8\over 17} & \dof{9\over 19} &
  \mA\nof{10\over 21} & \mA\ldots\rule[-4.6mm]{0mm}{5mm}
\\ \hline
{\bf (B2)}\rule[0mm]{0mm}{9.4mm} &
  $\displaystyle \QHLsymbol{N}{1\over 2}{2}{2}$ &
  \multicolumn{11}{l}{$\mbox{}^v\mmini D_{N-1}\,,\hsp
  N=3,4,\ldots\ $:}
\\
 & & \multicolumn{11}{l}{[Remark\mini: $\:\mbox{}^v\mmini D_2 \simeq
  \mbox{}^1\mA A_1\,\mbox{}^1\mA A_1\mini,\ \,\mbox{}^v\mmini D_3
  \simeq \mbox{}^2\mA A_3\mini$, \,and $\:\mbox{}^v\mmini D_4
  \simeq\mbox{}^s\mmini D_4\mini$.]}
\\
 & & \multicolumn{11}{l}{\tbulf{1\over 2}\rule[-4.6mm]{0mm}{5mm}}
\\ \hline
\end{tabular}
\end{flushleft}
\end{table}

\newpage

\samepage{
\begin{flushleft}
\parbox{145mm}{{\bf Table~B.1. }{\em (Continued).}
\rule[-4.8mm]{0mm}{5mm}}
\renewcommand{\arraystretch}{1.8}
\setlength{\tabcolsep}{2.1mm}
\begin{tabular}{lr*{11}{l}}
\hline
{\bf (B3)}\rule[0mm]{0mm}{9.4mm} &
  $\displaystyle \QHLsymbol{N}{N\over N+4}{g}{\lb}$ &
  \multicolumn{11}{l}{$\mbox{}^2\mA A_{N-1}\,$,\hsp
  with $\,g=1\:(2)\,$ and $\,\lb=1\:(1\mbox{ or }2)\,$ if $\mini
  N\mini$ is \,odd}
\\
 & & \multicolumn{11}{l}{\hspace{17.4mm}(even, \,and
  $\,4\!\not|\:N\,\mbox{ or }\,4\,|\mini N\,$)\mini;
  $\:N=5,6,\ldots\ $:}
\\
 & & \multicolumn{11}{l}{[Remark\mini: $\:\mbox{}^2\mA A_4 \simeq$
  ``$\,\mbox{}^f\! E_4\mini$''.]}
\\
 & & \bulf{5\over 9} & \bulf{3\over 5} & \cirf{7\over 11} &
  \bulf{2\over 3} & \cirf{9\over 13} & \,\bulf{5\over 7} &
  \nof{11\over 15} & \mini\ldots\rule[-4.6mm]{0mm}{5mm} & & &
\\ \hline
{\bf (B4)}\rule[0mm]{0mm}{9.6mm} &
  \multicolumn{5}{l}{$\displaystyle\QHLsymbol{n_1+\mini
  n_2-1}{n_1\mini n_2/g\lb \over (n_1\mini n_2+n_1+n_2)/
  g\lb}{g}{\lb}$} &
  \multicolumn{7}{l}{\hspace{-3.7mm}$\mbox{}^1\mA A_{n_1-1}\,
  \mbox{}^1\mA A_{n_2-1}\,,\hsp n_1=g\mini r_1,\ n_2=g\mini
  r_2\mini$,}
\\
 & & \multicolumn{11}{l}{\hspace{17.4mm}with $\,g=\gcd(n_1,n_2)
  \mini$, \,and $\,\lb=\gcd(r_1+r_2,g)\mini$;}
\\
 & & \multicolumn{11}{l}{\hspace{17.4mm}$N=n_1+n_2-1=4,5,
  \ldots\,$, \,and $\,2\leq n_1\leq n_2\ $:}
\\
 & & \multicolumn{2}{l}{$n_1=2\:$:} & \bulf{6\over 11} &
  \bulf{4\over 7} & \dof{10\over 17} & \,\bulf{3\over 5} &
  \nof{14\over 23} & \bulf{8\over 13} & \nof{18\over 29} &
  \mini\ldots &
\\
 & & \multicolumn{2}{l}{$n_1=3\:$:} & \bulf{3\over 5} &
  \nof{12\over 19} & \mini\nof{15\over 23} & \,\bulf{2\over 3} &
  \nof{21\over 31} & \mini\ldots\rule[-4.6mm]{0mm}{5mm} & & &
\\
 & & $\mini\cdots$\rule[-4.6mm]{0mm}{5mm} &\multicolumn{9}{l}{ }
\\ \hline
{\bf (B5)}\rule[0mm]{0mm}{9.4mm} &
  $\displaystyle \QHLsymbol{N}{n\over n+1}{g}{\lb}$ &
  \multicolumn{11}{l}{$\mbox{}^1\mA A_{n-1}\, \mbox{}^v\mmini
  D_{N-n}\,$,\hsp with $\,g=2\:(4)\,$ and $\,\lb=2\:(1)\,$ if
  $\mini N\mini$ is}
\\
 & & \multicolumn{11}{l}{\hspace{17.4mm}odd (even)\mini;
  $\:N=4,5,\ldots\,$, \,and $\,2\leq n\leq N-2\ $:}
\\
 & & \multicolumn{11}{l}{[For $\:\mbox{}^v\mmini D_2\mini,\
  \,\mbox{}^v\mmini D_3\mini$, \,and $\:\mbox{}^v\mmini
  D_4\mini$, \,see {\bf (B2)}.]}
\\
 & & \bulf{2\over 3} & \nof{3\over 4} & \bulf{4\over 5} &
  \nof{5\over 6} & \mA\tcirf{6\over 7} & \,\nof{7\over 8} &
  \mini\ldots\rule[-4.6mm]{0mm}{5mm} & & & &
\\ \hline
{\bf (B6)}\rule[0mm]{0mm}{9.4mm} &
  $\displaystyle \QHLsymbol{N}{N\over 9}{g}{1}$ &
  \multicolumn{11}{l}{$\mbox{}^3\mA A_{N-1}\,$,\hsp with
  $\,g=\gcd(N,3)\mini;\ \:N=6,7$, \,and $\mini 8\ $:}
\\
 & & \bulf{2\over 3} & \nof{7\over 9} &
  \nof{8\over 9}\rule[-4.6mm]{0mm}{5mm} & \multicolumn{8}{l}{ }
\\ \hline
{\bf (B7)}\rule[0mm]{0mm}{9.4mm} &
  $\displaystyle \QHLsymbol{N}{4\over 13-N}{g}{1}$ &
  \multicolumn{11}{l}{$\mbox{}^s\mmini D_{N-1}\,$,\hsp with
  $\,g=2\:(1)\,$ if $\mini N\mini$ is \,odd (even)\mini;}
\\
 & & \multicolumn{11}{l}{\hspace{17.4mm}$\:N=6,7$, \,and
  $\mini 8\  $:}
\\
 & & \multicolumn{11}{l}{[Remark\mini: $\:\mbox{}^s\mmini D_5
  \simeq$  ``$\,\mbox{}^f\! E_5\mini$''.]}
\\
 & & \bulf{4\over 7} & \bulf{2\over 3} &
  \bulf{4\over 5}\rule[-4.6mm]{0mm}{5mm} & \multicolumn{8}{l}{ }
\\ \hline
\end{tabular}
\end{flushleft}
}

\newpage

\samepage{
\begin{flushleft}
\parbox{145mm}{{\bf Table~B.1. }{\em (Continued).}
\rule[-4.8mm]{0mm}{5mm}}
\renewcommand{\arraystretch}{1.8}
\setlength{\tabcolsep}{2.1mm}
\setlength{\doublerulesep}{.7mm}
\begin{tabular}{lr*{11}{l}}
\hline
{\bf (B8)}\rule[0mm]{0mm}{9.4mm} &
  \bulf{\displaystyle \QHLsymbol{7}{3\over 5}{1}{1}} &
  \multicolumn{11}{l}{$\mbox{}^f\! E_6$\rule[-6mm]{0mm}{5mm}}
\\
 & \bulf{\displaystyle \QHLsymbol{8}{2\over 3}{1}{1}} &
  \multicolumn{11}{l}{$\mbox{}^f\! E_7$\rule[-6.8mm]{0mm}{5mm}}
\\ \hline
{\bf (B9)}\rule[0mm]{0mm}{9.4mm} &
  $\displaystyle \QHLsymbol{N}{2\mini N-2\over N+7}{g}{1}$ &
  \multicolumn{11}{l}{$\mbox{}^1\mA A_1\,\mbox{}^2\mA A_{N-2}\,$,
  \hsp with $\,g=2\:(1)\,$ if $\mini N\mini$ is \,odd (even)\mini;}
\\
 & & \multicolumn{11}{l}{\hspace{21.7mm}$N=6,7$, \,and $\mini 8\
$:} \\
 & & \nof{10\over 13} & \mA\tcirf{6\over 7} &
  \nof{14\over 15}\rule[-4.6mm]{0mm}{5mm} & \multicolumn{8}{l}{ }
\\ \hline
{\bf (B10)}\rule[0mm]{0mm}{9.4mm} &
  \bulf{\displaystyle \QHLsymbol{7}{4\over 5}{2}{1}} &
  \multicolumn{11}{l}{$\mbox{}^1\mA A_1\,\mbox{}^s\mmini D_5\,$,
  \hsp [Remark\mini: $\:\mbox{}^s\mmini D_5 \simeq$
  ``$\,\mbox{}^f\! E_5\mini$''.]\rule[-6mm]{0mm}{5mm}}
\\
 & \tcirf{\displaystyle \QHLsymbol{8}{6\over 7}{1}{1}} &
  \multicolumn{11}{l}{$\mbox{}^1\mA A_1\,\mbox{}^f\!
  E_6$\rule[-6mm]{0mm}{5mm}}
\\
 & \nof{\displaystyle \QHLsymbol{7}{15\over 17}{1}{1}} &
  \multicolumn{11}{l}{$\mbox{}^1\mA A_2\,\mbox{}^2\mA A_4\,$,
  \hsp [Remark\mini: $\:\mbox{}^2\mA A_4 \simeq$
  ``$\,\mbox{}^f\! E_4\mini$''.]\rule[-6mm]{0mm}{5mm}}
\\
 & \nof{\displaystyle \QHLsymbol{8}{12\over 13}{1}{1}} &
  \multicolumn{11}{l}{$\mbox{}^1\mA A_2\,\mbox{}^s\mmini
  D_5$\rule[-6mm]{0mm}{5mm}}
\\
 & \nof{\displaystyle \QHLsymbol{8}{20\over 21}{1}{1}} &
  \multicolumn{11}{l}{$\mbox{}^1\mA A_3\,
  \mbox{}^2\mA A_4$\rule[-6.8mm]{0mm}{5mm}}
\\ \hline
{\bf (B11)}\rule[0mm]{0mm}{9.4mm} &
  $\displaystyle \QHLsymbol{N}{6\mini N-18\over 5\mini N-9}{g}{1}$ &
  \multicolumn{11}{l}{$\mbox{}^1\mA A_1\,\mbox{}^1\mA A_2\,
  \mbox{}^1\mA A_{N-4}\,$,\hsp with $\,g=3,\mini 2,\mini 1\mini$,
  for $\,N=6,7$, \,and $\mini 8\ $:}
\\
 & & \tcirf{6\over 7} & \nof{12\over 13} &
  \nof{30\over 31}\rule[-4.6mm]{0mm}{5mm} & \multicolumn{8}{l}{ }
\\ \hline
\hline
\end{tabular}
\end{flushleft}
}


\newpage
\setcounter{secnumdepth}{0}
\begin{flushleft}
\section{Appendix C\mini: Low-Dimensional, Indecomposable CQHLs}
\label{AppC}
\end{flushleft}

\appendix
\setcounter{secnumdepth}{1}
\setcounter{section}{3}

\noi
The purpose of this appendix is to summarize the classification of
all indecomposable CQHLs in two and three dimensions with
relative-angular-momentum invariant $\,\lM\leq 5$, and of all such
lattices in four dimensions with $\,\lM=3\mini$. We recall that, by
definition (see~(\ref{lmax})), we have $\,\lM=\LM\,$ for indecomposable
CQHLs.

In Tables~C.1, C.2, and~C.4, the CQHLs are organized according to
increasing values of their Hall fractions $\sH$, and for each CQHL,
the symbol $\subs{N}{(\ndH)}^g_\lambda\mini$ is given together with
indications of the experimental status of the corresponding Hall
fraction. For the latter indications, notations are as in
Appendix~B. The symbols are followed by the explicit data
$(K,\Qv)$ which characterize the CQHLs completely; see the
beginning of Sect.\,\ref{sBI}\mini. For a succinct presentation of
the data $(K,\Qv)$, we choose symmetric bases in the
corresponding CQHLs (see~(\ref{Lmin})), and adopt the following
notations:

\bea
N=2\ : & [\mini\lm^{\hspace{5.5mm}\textstyle a\,}\lM\mini]\ ,  &
\fh K\:=\left( \begin{array}{cc}
\lm & a \\  a & \lM
\end{array} \right) \hah \nonumber \\
 & & \hspace{11.1mm}\Qv \:=\: (1,1) \ ;  \\
\nonumber \\
N=3\ : & (\mini a_1\mini a_2\,;b\,)\ , &  \fh
K\:=\left( \begin{array}{ccc}
3 & a_1 & a_2 \\  a_1 & 3 & b \\ a_2 & b & 3
\end{array} \right) \hah \nonumber \\
 & & \hspace{11.1mm}\Qv \:=\: (1,1,1)\ ;  \\
\nonumber \\
N=4\ : & \,\ (\mini a_1\mini a_2\,a_3\,;b_1\mini b_2\,;c\,)\ , &
\fh K\:=\left( \begin{array}{cccc}
3 & a_1 & a_2 & a_3 \\  a_1 & 3 & b_1 & b_2 \\ a_2 & b_1 & 3 & c \\
a_3 & b_2 & c & 3 \end{array} \right) \hah \nonumber \\
 & & \hspace{11.1mm}\Qv \:=\: (1,1,1,1)\ .
\eea

\noi
Furthermore, in Tables~C.1--\mini 4, we indicate as remarks the
corresponding Witt sublattices and/or preimages under the shift maps
when they exist. We note that, in Tables~C.1 and~C.2, the Witt
sublattices of the CQHLs with $\,\sH\!\geq\! 2\,$ are not fully
included in their neutral sublattices, i.e., some of the associated
symmetry generators have a non-vanishing electric charge.

In Table~C.3, the symbols of a physically relevant subset of
all three-dimensional, indecomposable CQHLs with $\,\lM=5\,$ are
provided. They are organized according to the values of their
relative-angular-momentum invariants $[\mini\lm,\Lii,\lM\mini]$;
see~(\ref{L2}). The symbols are followed by triples $(\mini a_1\mini
a_2\,;b\,)$ which have the same meaning as in~(C.2) above with the
only change that the diagonal elements of $\mini K\mini$ are not
$\,3\,3\,3\,$ but given, from left to right, by
$\,\lm\:\Lii\:\lM\mini$, as specified at the beginning of each
sublist. Morover, in the sublist with invariants $[5,5,5]$, all
those inverse images under the shift map ${\cal S}_1$ are indicated
which belong to Table~C.2 with invariants $[3,3,3]$;
see~(\ref{shift}). Since the invariants $N,\ g$, and $\lb$ do not
change under the shift maps, they are suppressed in the labelling of
the inverse images. Finally, only CQHLs with $\,\lb\mini\dH\leq 22\,$
are listed. For the physical interpretation of $\lb\mini\dH$ as the
smallest possible (fractional) charge of quasi-particle excitations
in the corresponding QH fluids, see~(\ref{emin}).

\samepage{
\vspace{14mm}
\begin{center}
\parbox{147mm}{{\bf Table~C.1. }{\em All indecomposable CQHLs
with $\,N=2\,$ and $\,3\leq\lm\leq\lM\leq 5$.
\rule[-4.8mm]{0mm}{5mm}}}
\renewcommand{\arraystretch}{1.6}
\setlength{\tabcolsep}{2.1mm}
\setlength{\doublerulesep}{.7mm}
\begin{tabular}{rrl}
\hline
\hline
 & $\subs{N}{(\ndH)}^g_\lambda\mini$ &
  $[\mini\lm^{\hspace{5.5mm}\textstyle a\mini}\lM\mini]$
  \hspace{10mm}Remarks\rule[-4.5mm]{0mm}{6mm}\hspace{29.6mm}
\\
\hline
$0<\sH<{1\over 5}\ :$ & none, &
  by~(\ref{sHbound})\rule[-3.9mm]{0mm}{5mm}
\\
\hline
$\Sigma_2^+,\:{1\over 5}\leq\sH<{1\over 4}\ :$  &
  \bulf{\sub{2}{({2\over 9})}^1_1}  &  $[5^4 5]
  \ssp=\:(\mini 5\,|\,\mbox{}^1\mA A_1) \:=\:
  {\cal S}_2(\,\subi{1}{(1)}^1_1 \oplus
  \subi{1}{(1)}^1_1\,)$\rule[-3.9mm]{0mm}{5mm}
\\
\hline
$\Sigma_2^-,\:{1\over 4}\leq\sH<{1\over 3}\ :$  &
  \nof{\sub{2}{({1\over 4})}^2_2}  &  $[5^3 5]
  \ssp=\: {\cal S}_1(\,\sub{2}{({1\over 2})}^2_2\,)$
\\
 & \bulf{\sub{2}{({2\over 7})}^3_1}  &  $[5^2 5]
  \ssp=\: {\cal S}_1(\,\sub{1}{({1\over 3})}^1_1
  \oplus \sub{1}{({1\over 3})}^1_1\,)$\rule[-3.9mm]{0mm}{5mm}
\\
\hline
$\Sigma_1^+,\:{1\over 3}\leq\sH<{1\over 2}\ :$  &
  \bulf{\sub{2}{({1\over 3})}^4_2}  &  $[5^1 5]
  \ssp=\: {\cal S}_1(\,\subi{2}{(1)}^4_2\,)$
\\
 & \dof{\sub{2}{({4\over 11})}^1_1}  &  $[3^2 5]
  \ssp=\: {\cal S}_1(\,\subi{1}{(1)}^1_1 \oplus
  \sub{1}{({1\over 3})}^1_1\,)$
\\
 & \bulf{\sub{2}{({2\over 5})}^1_1}  &  $[3^2 3]
  \ssp=\:(\mini 3\,|\,\mbox{}^1\mA A_1) \:=\:
  {\cal S}_1(\,\subi{1}{(1)}^1_1 \oplus \subi{1}{(1)}^1_1\,)$
\\
 & \bulf{\sub{2}{({3\over 7})}^2_1}  &
  $[3^1 5]$\rule[-3.9mm]{0mm}{5mm}
\\
\hline
$\Sigma_1^-,\:{1\over 2}\leq\sH<1\ :$  &
  \tbulf{\sub{2}{({1\over 2})}^2_2}  &  $[3^1 3]$
\\
 & \tbulf{\sub{2}{({1\over 2})}^6_2}  &  $[5^{-1} 5]$
\\
 & \pbulf{B,n-p}{\sub{2}{({2\over 3})}^7_1}  &  $[5^{-2} 5]$
\\
 & \pbulf{(B-p)}{\sub{2}{({5\over 7})}^2_1}  &
  $[3^{-1} 5]$\rule[-3.9mm]{0mm}{5mm}
\\
\hline
$\Sigma_0^+,\:1\leq\sH<\infty\ :$  &
  \bulf{\subi{2}{(1)}^4_2}  &  $[3^{-1} 3]$
\\
 & \bulf{\subi{2}{(1)}^8_2}  &  $[5^{-3} 5]$
\\
 & \nof{\sub{2}{({12\over 11})}^1_1}  &  $[3^{-2} 5]$
\\
 & \bulf{\subi{2}{(2)}^5_1}  &  $[3^{-2} 3]\ssp\supset\:A_1$
\\
 & \bulf{\subi{2}{(2)}^9_1}  &  $[5^{-4} 5]
  \ssp\supset\:A_1$\rule[-3.9mm]{0mm}{5mm}
\\
\hline\hline
\end{tabular}
\end{center}
}

\newpage

\samepage{
\begin{flushleft}
\parbox{145mm}{{\bf Table~C.2. }{\em All indecomposable CQHLs
with $\,N=3\,$ and $\,\lm=\lM=3$.\rule[-4.8mm]{0mm}{5mm}}}
\renewcommand{\arraystretch}{1.6}
\setlength{\tabcolsep}{2.1mm}
\setlength{\doublerulesep}{.7mm}
\begin{tabular}{rrl}
\hline
\hline
 & $\subs{N}{(\ndH)}^g_\lambda\mini$ &
  $(\mini a_1\mini a_2\,;b\,)$
  \hspace{10mm}Remarks\rule[-4.5mm]{0mm}{6mm} \\
\hline
$0<\sH<{1\over 3}\ :$ & none, &
  by~(\ref{sHbound})\rule[-3.9mm]{0mm}{5mm}
\\
\hline
$\Sigma_1^+,\:{1\over 3}\leq\sH<{1\over 2}\ :$  &
  \bulf{\sub{3}{({3\over 7})}^1_1}  &  $(2\mini 2\,;2)
  \ssp=\:(\mini 3\,|\,\mbox{}^1\mA A_2) \:=\:
  {\cal S}_1(\,\subi{1}{(1)}^1_1 \oplus \subi{1}{(1)}^1_1
  \oplus \subi{1}{(1)}^1_1\,)$\rule[-3.9mm]{0mm}{5mm}
\\
\hline
$\Sigma_1^-,\:{1\over 2}\leq\sH<1\ :$  &
  \tbulf{\sub{3}{({1\over 2})}^2_2}  &  $(2\mini 1\,;2)
  \ssp=\:(\mini 3\,|\,\mbox{}^1\mA A_1\,\mbox{}^1\mA A_1)$
\\
 & \bulf{\sub{3}{({7\over 13})}^1_1}  &  $(2\mini 1\,;1)
  \ssp\supset\: A_1$
\\
 & \pbulf{B-p}{\sub{3}{({3\over 5})}^4_1}  &  $(1\mini 1\,;1)$
\\
 & \pbulf{B,n-p}{\sub{3}{({2\over 3})}^4_1}  &  $(2\mini 0\,;1)
  \ssp\supset\: A_1$
\\
 & \pbulf{(B-p)}{\sub{3}{({5\over 7})}^3_1}  &
  $(1\mini 0\,;1)$\rule[-3.9mm]{0mm}{5mm}
\\
\hline
$\Sigma_0^+,\:1\leq\sH<2\ :$  &
  \bulf{\subi{3}{(1)}^6_2}  &  $(2\,{-1}\,;0)
  \hspace{7.3mm}\supset\: A_1$
\\
 & \bulf{\subi{3}{(1)}^8_2}  &  $(1\,{-1}\,;1)$
\\
 & \nof{\sub{3}{({23\over 21})}^1_1}  &  $(1\,{-1}\,;0)$
\\
 & \nof{\sub{3}{({15\over 13})}^1_1}  &  $(2\,{-1}\,;-1)
  \ssp\supset\: A_1$
\\
 & \bulf{\sub{3}{({7\over 5})}^4_1}  &  $(1\,{-1}\,;-1)$
\\
 & \nof{\sub{3}{({13\over 7})}^3_1}  &  $(0\,{-1}\,;-1)$
\rule[-7mm]{0mm}{5mm}
\\
$2\leq\sH<3\ :$ & \bulf{\subi{3}{(2)}^8_1}  &  $(2\,{-2}\,;-1)
  \ssp\supset\:A_1\,A_1$
\\
 & \bulf{\subi{3}{(2)}^{12}_1}  &  $(1\,{-2}\,;0)
  \hspace{7.3mm}\supset\:A_1$
\\
 & \nof{\sub{3}{({31\over 13})}^1_1}  &  $(1\,{-2}\,;-1)
  \ssp\supset\:A_1$
\\
 & \cirf{\sub{3}{({19\over 7})}^1_1}  &  $(2\,{-2}\,;-2)
  \ssp\supset\:A_2$
\rule[-7mm]{0mm}{5mm}
\\
$3\leq\sH<\infty\ :$ & \bulf{\subi{3}{(3)}^{16}_1}  &
  $({-1}\,{-1}\,;-1)$
\\
 & \nof{\sub{3}{({11\over 3})}^2_2}  &  $(0\,{-2}\,;-1)
  \ssp\supset\:A_1$
\\
 & \nof{\sub{3}{({9\over 2})}^2_2}  &  $(1\,{-2}\,;-2)
  \ssp\supset\:A_1\,A_1$\rule[-3.9mm]{0mm}{5mm}
\\
\hline\hline
\end{tabular}
\end{flushleft}
}

\newpage

\samepage{
\begin{flushleft}
\parbox{154mm}{{\bf Table~C.3. }{\em Symbols of indecomposable
CQHLs with $\,N=3\mini,\ 3\leq\lm\leq\lM=5$, and $\,\sH<1$. The dots
``$\,\ldots$'' indicate omitted fractions with $\,\lb\mini\dH>22$.
\rule[-4.8mm]{0mm}{5mm}}}
\renewcommand{\arraystretch}{1.6}
\setlength{\tabcolsep}{3.4mm}
\setlength{\doublerulesep}{.7mm}
\begin{tabular}{*{4}{l}}
\hline
\hline
\multicolumn{4}{l}{$[\mini\lm\mini,\Lii\mini,\lM\mini]\:=\:
  [\mini 3,3,5\mini]\ $:}
\\
$\sub{3}{({7\over 17})}^1_1\tsp(2\, 2\mini;2)$  &
  $\sub{3}{({4\over 9})}^2_1\tsp(2\, 1\mini;2)$  &
  $\sub{3}{({1\over 2})}^6_2\tsp(1\, 2\mini;2)$  &
  $\sub{3}{({6\over 11})}^2_1\tsp(2\, 0\mini;1)$
\\
$\sub{3}{({5\over 9})}^4_1\tsp(1\, 1\mini;1)$  &
  $\sub{3}{({4\over 7})}^4_1\tsp(1\, 0\mini;2)$  &
  $\sub{3}{({2\over 3})}^{10}_1\tsp(0\, 1\mini;2)$  &
  $\sub{3}{({9\over 13})}^3_1\tsp(0\, 1\mini;1)$
\\
$\sub{3}{({8\over 11})}^2_1\tsp(2\, 0\mini;-1)$  &
  $\sub{3}{({3\over 4})}^4_2\tsp(1\, 1\mini;-1)$  &
  $\sub{3}{({4\over 5})}^6_1\tsp(0\, 2\mini;-1)$  &
  $\ \ldots$
\\
\multicolumn{4}{l}{(in total $17$ CQHLs)}\rule[-3.9mm]{0mm}{5mm}
\\
\hline
\multicolumn{4}{l}{$[\mini\lm\mini,\Lii\mini,\lM\mini]\:=\:
  [\mini 3,5,5\mini]\ $:}
\\
$\sub{3}{({7\over 19})}^1_1\tsp(3\, 2\mini;3)$  &
  $\sub{3}{({3\over 8})}^2_2\tsp(2\, 2\mini;3)$  &
  $\sub{3}{({5\over 13})}^3_1\tsp(2\, 2\mini;2)$  &
  $\sub{3}{({2\over 5})}^4_2\tsp(2\, 2\mini;1)$
\\
$\sub{3}{({3\over 7})}^5_1\tsp(2\, 1\mini;3)$  &
  $\sub{3}{({5\over 11})}^4_1\tsp(1\, 1\mini;3)$  &
  $\sub{3}{({9\over 19})}^3_1\tsp(1\, 1\mini;2)$  &
  $\sub{3}{({1\over 2})}^8_4\tsp(1\, 1\mini;1)$
\\
$\sub{3}{({7\over 13})}^5_1\tsp(1\, 1\mini;0)$  &
  $\sub{3}{({3\over 5})}^{12}_1\tsp(1\, 1\mini;-1)$  &
  $\sub{3}{({8\over 13})}^4_1\tsp(2\, 0\mini;-1)$  &
  $\sub{3}{({5\over 7})}^7_1\tsp(1\, 1\mini;-2)$
\\
$\sub{3}{({11\over 15})}^4_1\tsp(1\, {-1}\mini;1)$  &
  $\sub{3}{({13\over 17})}^3_1\tsp(2\, {-1}\mini;-1)$  &
  $\sub{3}{({9\over 11})}^4_1\tsp({-1}\, {-1}\mini;3)$  &
  $\sub{3}{({7\over 8})}^4_2\tsp(1\, {-1}\mini;-1)$
\\
$\sub{3}{({17\over 19})}^3_1\tsp({-1}\, {-1}\mini;2)$  &
$\ \ldots$ & &
\\
\multicolumn{4}{l}{(in total $34$ CQHLs)}\rule[-3.9mm]{0mm}{5mm}
\\
\hline
\multicolumn{4}{l}{$[\mini\lm\mini,\Lii\mini,\lM\mini]\:=\:
  [\mini 5,5,5\mini]\ $:}
\\
$\sub{3}{({3\over 13})}^1_1\:=\:\Si(\,{3\over 7}\,)$  &
  $\sub{3}{({1\over 4})}^2_2\:=\:\Si(\,{1\over 2}\,)$  &
  $\sub{3}{({3\over 11})}^4_1\:=\:\Si(\,{3\over 5}\,)$  &
  $\sub{3}{({2\over 7})}^4_1\:=\:\Si(\,{2\over 3}\,)$
\\
$\sub{3}{({5\over 17})}^3_1\:=\:\Si(\,{5\over 7}\,)$  &
  $\sub{3}{({1\over 3})}^6_2\:=\:\Si(\mini 1\mini)$  &
  $\sub{3}{({1\over 3})}^8_2\:=\:\Si(\mini 1\mini)$  &
  $\sub{3}{({1\over 3})}^9_3\tsp(2\, 2\mini;2)$
\\
$\sub{3}{({4\over 11})}^4_2\tsp(2\, 2\mini;1)$  &
  $\sub{3}{({7\over 19})}^4_1\:=\:\Si(\,{7\over 5}\,)$  &
  $\sub{3}{({2\over 5})}^8_1\:=\:\Si(\mini 2\mini)$  &
  $\sub{3}{({2\over 5})}^{12}_1\:=\:\Si(\mini 2\mini)$
\\
$\sub{3}{({7\over 17})}^5_1\tsp(2\, 0\mini;2)$  &
  $\sub{3}{({3\over 7})}^{16}_1\:=\:\Si(\mini 3\mini)$  &
  $\sub{3}{({1\over 2})}^{10}_2\tsp(4\, 0\mini;-1)$  &
  $\sub{3}{({1\over 2})}^{16}_2\tsp(3\, 1\mini;-1)$
\\
$\sub{3}{({1\over 2})}^{18}_2\tsp(2\, 2\mini;-1)$  &
  $\sub{3}{({5\over 9})}^{12}_1\tsp(1\, 1\mini;-1)$ &
  $\sub{3}{({11\over 19})}^4_1\tsp(3\, {-1}\mini;-1)$  &
  $\sub{3}{({7\over 11})}^9_1\tsp(2\, {-1}\mini;-1)$
\\
$\sub{3}{({2\over 3})}^{12}_1\tsp(4\, {-2}\mini;-1)$  &
  $\sub{3}{({2\over 3})}^{20}_1\tsp(3\, 0\mini;-2)$  &
  $\sub{3}{({2\over 3})}^{24}_1\tsp(2\, 1\mini;-2)$  &
  $\sub{3}{({9\over 13})}^7_1\tsp(1\, 1\mini;-2)$
\\
$\sub{3}{({5\over 7})}^8_2\tsp(1\, {-1}\mini;-1)$  &
  $\ \ldots$ & &
\\
\multicolumn{4}{l}{(in total $48$ CQHLs)}\rule[-3.9mm]{0mm}{5mm}
\\
\hline\hline
\end{tabular}
\end{flushleft}
}

\newpage

\samepage{
\begin{flushleft}
\parbox{154mm}{{\bf Table~C.4. }{\em All indecomposable CQHLs
with $\,N=4,\ \lm=\lM=3$, and $\,\sH<1$.
\rule[-4.8mm]{0mm}{5mm}}}
\renewcommand{\arraystretch}{1.6}
\setlength{\tabcolsep}{2.1mm}
\setlength{\doublerulesep}{.7mm}
\begin{tabular}{rrl}
\hline
\hline
 & $\subs{N}{(\ndH)}^g_\lambda\mini$ &
  $(\mini a_1\mini a_2\,a_3\,;b_1\mini b_2\,;c\,)$
  \hspace{10mm}Remarks\rule[-4.5mm]{0mm}{6mm}\hspace{29.4mm}
\\
\hline
$0<\sH<{1\over 3}\ :$ & none, &
  by~(\ref{sHbound})\rule[-3.9mm]{0mm}{5mm}
\\
\hline
$\Sigma_1^+,\:{1\over 3}\leq\sH<{1\over 2}\ :$  &
  \bulf{\sub{4}{({4\over 9})}^1_1}  &
  $(2\mini 2\mini 2\,;2\mini 2\,;2)
  \ssp=\:(\mini 3\,|\,\mbox{}^1\mA A_3)$\rule[-3.9mm]{0mm}{5mm}
\\
\hline
$\Sigma_1^-,\:{1\over 2}\leq\sH<{2\over 3}\ :$  &
  \tbulf{\sub{4}{({1\over 2})}^2_2}  &
  $(2\mini 2\mini 1\,;2\mini 2\,;2)
  \ssp=\:(\mini 3\,|\,\mbox{}^2\mA A_3)$
\\
 & \bulf{\sub{4}{({6\over 11})}^1_1}  &
  $(2\mini 2\mini 1\,;2\mini 1\,;2)
  \ssp=\:(\mini 3\,|\,\mbox{}^1\mA A_1\,\mbox{}^1\mA A_2)$
\\
 & \bulf{\sub{4}{({5\over 9})}^2_1}  &
  $(2\mini 2\mini 1\,;2\mini 1\,;1) \ssp\supset\:A_2$
\\
 & \bulf{\sub{4}{({4\over 7})}^3_1}  &
  $(2\mini 1\mini 1\,;1\mini 1\,;2) \ssp\supset\:A_1\,A_1$
\\
 & \pbulf{B-p}{\sub{4}{({3\over 5})}^4_1}  &
  $(2\mini 1\mini 1\,;2\mini 1\,;1) \ssp\supset\:A_1\,A_1$
\\
 & \nof{\sub{4}{({5\over 8})}^2_2}  &
  $(2\mini 1\mini 1\,;1\mini 1\,;1) \ssp\supset\:A_1$
\rule[-7mm]{0mm}{5mm}
\\
${2\over 3}\leq\sH<1\ :$ &
  \pbulf{B,n-p}{\sub{4}{({2\over 3})}^4_1}  &
  $(2\mini 1\mini 0\,;2\mini 1\,;2) \ssp=\:(\mini 3\,|\,\mbox{}^1\mA
  A_1\,\mbox{}^1\mA A_1\,\mbox{}^1\mA A_1)$
\\
 & \pbulf{B,n-p}{\sub{4}{({2\over 3})}^5_1}  &
  $(2\mini 2\mini 0\,;2\mini 1\,;1) \ssp\supset\:A_2$
\\
 & \pbulf{B,n-p}{\sub{4}{({2\over 3})}^8_2}  &
  $(1\mini 1\mini 1\,;1\mini 1\,;1)$
\\
 & \pbulf{(B-p)}{\sub{4}{({5\over 7})}^4_1}  &
  $(2\mini 1\mini 0\,;1\mini 1\,;1) \ssp\supset\:A_1$
\\
 & \bulf{\sub{4}{({8\over 11})}^3_1}  &
  $(2\mini 1\mini 1\,;1\mini 1\,;0) \ssp\supset\:A_1$
\\
 & \nof{\sub{4}{({14\over 19})}^1_1}  &
  $(2\mini 1\mini 0\,;2\mini 0\,;1) \ssp\supset\:A_1\,A_1$
\\
 & \nof{\sub{4}{({3\over 4})}^2_2}  &
  $(2\mini 2\mini 0\,;2\mini 0\,;1) \ssp\supset\:A_2$
\\
 & \nof{\sub{4}{({3\over 4})}^6_2}  &
  $(1\mini 1\mini 0\,;1\mini 1\,;1)$
\\
 & \nof{\sub{4}{({13\over 17})}^2_1}  &
  $(2\mini 1\mini 0\,;1\mini 0\,;1) \ssp\supset\:A_1$
\\
 & \bulf{\sub{4}{({4\over 5})}^9_1}  &
  $(1\mini 1\mini 0\,;0\mini 1\,;1)$
\\
 & \nof{\sub{4}{({26\over 31})}^1_1}  &
  $(2\mini 0\mini 0\,;1\mini 0\,;1) \ssp\supset\:A_1$
\\
 & \nof{\sub{4}{({11\over 13})}^4_1}  &
  $(1\mini 1\mini 0\,;1\mini 0\,;1)$
\\
 & \tcirf{\sub{4}{({6\over 7})}^4_1}  &
  $(2\mini 0\mini 0\,;1\mini 1\,;1) \ssp\supset\:A_1$
\\
 & \nof{\sub{4}{({10\over 11})}^5_1}  &
  $(1\mini 0\mini 0\,;1\mini 0\,;1)$\rule[-3.9mm]{0mm}{5mm}
\\
\hline\hline
\end{tabular}
\end{flushleft}
}


\newpage
\setcounter{secnumdepth}{0}
\begin{flushleft}
\section{Appendix D\mini: Embeddings of $L-\mini$Minimal CQHLs}
\label{AppD}
\end{flushleft}

\appendix
\setcounter{secnumdepth}{1}
\setcounter{section}{4}

\noi
In this appendix, embeddings (see~(\ref{embed})) of \Lmini\ CQHLs with
Hall fractions in the window $\,\sH\in\!\Sigi^-=[\mini
1/2\mini,1)\,$ are listed. More precisely, in accordance with the
results presented in Sect.\,\ref{sDis}, we are taking the following
sets of CQHLs into account: (i) all generic, \Lmini\ CQHLs in low
dimensions, $N\!\leq\! 4$ (see Appendix~C), (ii) all maximally
symmetric, \Lmini\ CQHLs in dimensions $\,N\!\leq\! 10$ (see
Appendix~B), and (iii) all composites of two identical lattices
belonging to the prominent $A$-series given by~{\bf(B1)} in
Appendix~B. In Table~D.1, CQHLs are specified by their symbols,
$\subs{N}{(\ndH)}^g_\lambda\mini$, and the explicit data
characterizing their structure. These data are given in the
conventions chosen in Appendices~B and~C, respectively.

In order to simplify notation in the subsequent table, we note
that, at the fractions $\,\sH=n/(n\!+\!1),\ n=1,2,\ldots\,$, there
are infinite ``chains'' of embeddings,

\bea
&&\hspace*{-8mm}\sub{n+2}{(\mbox{$n\over
n+1$})}^g_\lb\fsp\mbox{}^1\mA A_{n-1}\, \mbox{}^1\mA
A_1\,\mbox{}^1\mA A_1 \emb \sub{n+3}{(\mbox{$n\over
n+1$})}^g_\lb\fsp\mbox{}^1\mA A_{n-1}\, \mbox{}^2\mA A_3
\emb\rule[-3.3mm]{0mm}{4mm} \nonumber \\ &&\emb
\sub{n+4}{(\mbox{$n\over n+1$})}^g_\lb\fsp\mbox{}^1\mA A_{n-1}\,
\mbox{}^v\mmini D_4
\emb \ldots \emb
\sub{N}{(\mbox{$n\over n+1$})}^g_\lb\fsp\mbox{}^1\mA A_{n-1}\,
\mbox{}^v\mmini D_{N-n} \emb \ldots\ .  \nonumber \\
&&
\eea

\noi
In the following table, the respective next members of these
chains of embeddings are understood when we write the dots
``$\ldots$''\mini.

\begin{table}[htbp]
\vspace{10mm}
\begin{flushleft}
\parbox{153mm}{{\bf Table~D.1. }{\em All embeddings of \Lmini\
CQHLs that have $\,\sH\in\!\Sigi^-\,$ and belong to the heuristic
classes mentioned above.\rule[-4.8mm]{0mm}{5mm}}}
\renewcommand{\arraystretch}{1.8}
\setlength{\tabcolsep}{0mm}
\setlength{\doublerulesep}{.7mm}
\begin{tabular}{rl}
\hline
\hline
\tbulf{1\over 2}\fssp &
$\sub{2}{({1\over 2})}^2_2\fsp[3^1 3]\rule[-4.6mm]{0mm}{5mm}
 \emb
 \sub{3}{({1\over 2})}^2_2\fsp\mbox{}^1\mA A_1\,\mbox{}^1\mA A_1
 \emb
 \sub{4}{({1\over 2})}^2_2\fsp\mbox{}^2\mA A_3
 \emb
 \sub{5}{({1\over 2})}^2_2\fsp\mbox{}^v\mmini D_4
 \emb \ldots\hspace*{12.6mm}$
\\
\hline
\bulf{5\over 9}\fssp &
$\sub{4}{({5\over 9})}^2_1\fsp(2\mini 2\mini 1\,;2\mini 1\,;1)
   \supset A_2\rule[-4.6mm]{0mm}{5mm}
 \emb
 \sub{5}{({5\over 9})}^1_1\fsp\mbox{}^2\mA A_4$
\\
\hline
\bulf{4\over 7}\fssp &
$\sub{4}{({4\over 7})}^3_1\fsp(2\mini 1\mini 1\,;1\mini 1\,;2)
   \supset A_1\mini A_1\rule[-4.6mm]{0mm}{5mm}
 \emb
 \sub{5}{({4\over 7})}^2_1\fsp\mbox{}^1\mA A_1\,\mbox{}^1\mA A_3
 \emb
 \sub{6}{({4\over 7})}^1_1\fsp\mbox{}^s\mmini
   D_5$
\\
\hline
\pbulf{B-p}{3\over 5}\fssp &
$\sub{3}{({3\over 5})}^4_1\fsp(1\mini 1\,;1)
 \emb
 \sub{4}{({3\over 5})}^4_1\fsp(2\mini 1\mini 1\,;2\mini 1\,;1)
   \supset A_1\mini A_1\rule[-4.6mm]{0mm}{5mm}
 \emb$
\\
&
$\hspace{20.5mm}\left.\renewcommand{\arraystretch}{1.2}
 \barr
   \emb
   \sub{5}{({3\over 5})}^3_1\fsp\mbox{}^1\mA A_2\,\mbox{}^1\mA A_2
   \emb
   \sub{6}{({3\over 5})}^2_1\fsp\mbox{}^2\mA
     A_5\rule[-3.8mm]{0mm}{5mm}
 \\
   \sub{7}{({3\over 5})}^2_2\fsp\mbox{}^1\mA A_1\,\mbox{}^1\mA
     A_5
 \earr\right\}
 \emb
 \sub{7}{({3\over 5})}^1_1\fsp\mbox{}^f\!E_6\rule[-9mm]{0mm}{10mm}$
\\
\hline
\end{tabular}
\end{flushleft}
\end{table}

\newpage

\samepage{
\begin{flushleft}
\parbox{154mm}{{\bf Table~D.1. }{\em (Continued).}
\rule[-4.8mm]{0mm}{5mm}}
\renewcommand{\arraystretch}{1.8}
\setlength{\tabcolsep}{0mm}
\setlength{\doublerulesep}{.7mm}
\begin{tabular}{rl}
\hline
\pbulf{B,n-p}{2\over 3}\fssp &
$\sub{1}{({1\over 3})}^1_1\fsp\![3] \oplus \sub{1}{({1\over
   3})}^1_1\fsp\![3]\rule[-4.6mm]{0mm}{5mm}
 \emb
 \sub{3}{({2\over 3})}^4_1\fsp(2\mini 0\,;1)\supset A_1
 \emb$
\\
 &
$\hspace*{-2.3mm}\emb
 \left\{\renewcommand{\arraystretch}{1.2}
 \ba
   \sub{4}{({2\over 3})}^5_1\fsp(2\mini 2\mini 0\,;2\mini 1\,;1)
     \supset A_2\rule[-3.8mm]{0mm}{5mm}
 \\
   \sub{4}{({2\over 3})}^4_1\fsp\mbox{}^1\mA A_1\,\mbox{}^1\mA A_1
     \mbox{}^1\mA A_1
 \ea\right\}
 \mmini\emb
 \sub{5}{({2\over 3})}^4_1\fsp\mbox{}^2\mA A_3\,
   \mbox{}^1\mA A_1\rule[-9.2mm]{0mm}{10mm}
 \emb$
\\
 &
$\left.\renewcommand{\arraystretch}{1.2}
 \ba
   \left.\renewcommand{\arraystretch}{1.2}
   \barr
     \hspace*{-6.7mm}\emb \sub{6}{({2\over 3})}^4_1\fsp\mbox{}^1\mA
       A_1\mini\mbox{}^v\mmini D_4\fsp\rule[-3.8mm]{0mm}{5mm}
   \\
     \sub{6}{({2\over 3})}^3_1\fsp\mbox{}^3\mA
       A_5\fsp\rule[-3.8mm]{0mm}{5mm}
   \\
     \sub{7}{({2\over 3})}^4_2\fsp\mbox{}^1\mA A_3\,
     \mbox{}^1\mA A_3\fsp\rule[-3.8mm]{0mm}{5mm}
   \earr\right\}
   \left.\renewcommand{\arraystretch}{1.2}
   \bal
     \hspace*{-9.5mm}\emb \sub{7}{({2\over 3})}^4_1\fsp\mbox{}^1\mA
       A_1\mini\mbox{}^v\mmini D_5\rule[-3.8mm]{0mm}{5mm}
     \emb
     \sub{8}{({2\over 3})}^4_1\fsp\mbox{}^1\mA A_1\mini
       \mbox{}^v\mmini D_6
     \emb \ldots\hspace*{-17.7mm}
   \\
     \hspace*{-2.7mm}\emb \sub{7}{({2\over 3})}^2_1\fsp
       \mbox{}^s\mmini D_6\rule[-3.8mm]{0mm}{5mm}
   \\
     \hspace*{-9.5mm}\emb \sub{8}{({2\over 3})}^2_2\fsp
       \mbox{}^2\mA A_7\rule[-3.8mm]{0mm}{5mm}
   \eal\right.
 \\
   \hspace{12.1mm}\sub{8}{({2\over 3})}^3_3\fsp\mbox{}^1\mA A_2\,
     \mbox{}^1\mA A_5
 \ea\right\}
 \mmini\emb
 \sub{8}{({2\over 3})}^1_1\fsp\!\mbox{}^f\!
   E_7\rule[-17.6mm]{0mm}{19mm}$
\\
\hline
\pbulf{(B-p)}{5\over 7}\fssp &
$\sub{3}{({5\over 7})}^3_1\fsp(1\mini 0\,;1)\rule[-4.6mm]{0mm}{5mm}
 \emb
 \sub{4}{({5\over 7})}^4_1\fsp(2\mini 1\mini 0\,;1\mini 1\,;1)
   \supset A_1$
\\
\hline
\nof{3\over 4}\fssp &
$\hspace*{-2.2mm}\left.\renewcommand{\arraystretch}{1.2}
 \barr
   \sub{4}{({3\over 4})}^2_2\fsp(2\mini 2\mini 0\,;2\mini 0\,;1)
     \mmini\supset\mmini A_2\hspace*{-1mm}\rule[-3.8mm]{0mm}{5mm}
   \\
   \sub{4}{({3\over 4})}^6_2\fsp(1\mini 1\mini 0\,;1\mini
     1\,;1)
 \earr\right\}
 \!\emb
 \sub{5}{({3\over 4})}^2_2\fsp\mbox{}^1\mA A_2\,\mbox{}^1\mA
   A_1\mini \mbox{}^1\mA A_1\rule[-9mm]{0mm}{19.9mm}
 \mimini\emb
 \sub{6}{({3\over 4})}^2_2\fsp\mbox{}^1\mA A_2\, \mbox{}^2\mA A_3
 \emb\mimini\ldots$
\\
\hline
\bulf{4\over 5}\fssp &
$\hspace*{-2.2mm}\left.\renewcommand{\arraystretch}{1.2}
 \barr
   \sub{2}{({2\over 5})}^1_1\fsp\mbox{}^1\mA A_1 \oplus
   \sub{2}{({2\over 5})}^1_1\fsp\mbox{}^1\mA
     A_1\rule[-3.8mm]{0mm}{5mm}
 \\
   \sub{4}{({4\over 5})}^9_1\fsp(1\mini 1\mini 0\,;0\mini 1\,;1)
 \earr\right\}
 \emb
 \sub{6}{({4\over 5})}^4_1\fsp\mbox{}^1\mA A_3\,\mbox{}^1\mA A_1\,
   \mbox{}^1\mA A_1
 \emb\rule{0mm}{11mm}$
\\
&
$\hspace*{42.6mm}\emb\rule[-9mm]{0mm}{19.5mm}
 \left\{\renewcommand{\arraystretch}{1.2}
 \bal
   \sub{7}{({4\over 5})}^4_1\fsp\mbox{}^1\mA A_3\,\mbox{}^2\mA A_3
   \emb\ldots\rule[-3.8mm]{0mm}{5mm}
 \\
   \sub{7}{({4\over 5})}^2_1\fsp\mbox{}^1\mA A_1\mini
     \mbox{}^s\mmini D_5
   \emb
   \sub{7}{({4\over 5})}^1_1\fsp\mbox{}^s\mmini D_7
 \eal\right.$
\\
\hline
\nof{5\over 6}\fssp &
$\sub{7}{({5\over 6})}^2_2\fsp\mbox{}^1\mA A_4\,\mbox{}^1\mA
   A_1\,\mbox{}^1\mA A_1\rule[-4.6mm]{0mm}{5mm}
 \emb
 \sub{8}{({5\over 6})}^2_2\fsp\mbox{}^1\mA A_4\,\mbox{}^2\mA A_3
 \emb \ldots$
\\
\hline
\tcirf{6\over 7}\fssp &
$\hspace*{-2.2mm}\left.\renewcommand{\arraystretch}{1.2}
 \barr
   \sub{4}{({6\over 7})}^4_1\fsp(2\mini 0\mini 0\,;1\mini 1\,;1)
     \mmini\supset\mmini A_1
   \emb
   \sub{6}{({6\over 7})}^3_1\fsp\mbox{}^1\mA A_1\,\mbox{}^1\mA
     A_2\,\mbox{}^1\mA A_2\rule[-3.8mm]{0mm}{5mm}
   \\
   \sub{3}{({3\over 7})}^1_1\fsp\mbox{}^1\mA A_2 \oplus
   \sub{3}{({3\over 7})}^1_1\fsp\mbox{}^1\mA A_2
 \earr\right\}
 \emb\rule{0mm}{11mm}$
\\
&
$\hspace*{39.5mm}\left.\renewcommand{\arraystretch}{1.2}
 \barr
   \emb
   \sub{7}{({6\over 7})}^2_1\fsp\mbox{}^1\mA A_1\,\mbox{}^2\mA
   A_5\hspace*{1.7mm}\rule[-11.8mm]{0mm}{5mm}
   \\
   \sub{8}{({6\over 7})}^4_1\fsp\mbox{}^1\mA A_5\,\mbox{}^1\mA
   A_1\,\mbox{}^1\mA A_1
   \emb
   \sub{9}{({6\over 7})}^4_1\fsp\mbox{}^1\mA A_5\,\mbox{}^2\mA A_3
   \emb\ldots\hspace*{-49.4mm}
 \earr\right\}
 \emb
 \sub{8}{({6\over 7})}^1_1\fsp\mbox{}^1\mA A_1\,\mbox{}^f\!
   E_6\rule[-13.1mm]{0mm}{28.3mm}$
\\
\hline
\nof{7\over 8}\fssp &
$\sub{9}{({7\over 8})}^2_2\fsp\mbox{}^1\mA A_6\,\mbox{}^1\mA
   A_1\,\mbox{}^1\mA A_1\rule[-4.6mm]{0mm}{5mm}
 \emb
 \sub{10}{({7\over 8})}^2_2\fsp\mbox{}^1\mA A_6\,\mbox{}^2\mA A_3
 \emb \ldots$
\\
\hline
\hline
\end{tabular}
\end{flushleft}
}


\newpage
\setcounter{secnumdepth}{0}
\begin{flushleft}
\section{Appendix E\mini: Hierarchy QH~Lattices}
\label{AppE}
\end{flushleft}

\appendix
\setcounter{secnumdepth}{1}
\setcounter{section}{5}

\noi
In this appendix, we collect some basic facts about the
description of the Haldane-Halperin~\cite{HH} and the
Jain-Goldman~\cite{JG} hierarchy fluids in terms of QH lattices,
\QHL. First, we follow the ideas presented by Read in~\cite{Read}.

The Gram matrix $\mini K\mini$ (see (\ref{Gram})) which
characterizes the integral lattice $\mini\G\mini$ associated with a
hierarchy fluid with Hall conductivity $\,\sH=\nH/\dH$, where
$\dH$ is {\em odd}, can be read off from the ``continued fraction
expansion'' of $\sH\mini$. Let

\be
\sH \:=\: {\nH\over\dH} \:=\: {1\over m-{1\over a_1-{1\over
a_2-{1\over \ddots\,-{1\over a_{N-1}}}}}}\ ,
\ee

\noi
where $\,m\,$ is an {\em odd}, positive integer, and $\,a_1,
\ldots,\mini a_{N-1}\,$ are {\em even} integers of either sign. Then
the associated Gram matrix, $K$, is given by

\bea
(K_{ij}) & = & \left.\left(\renewcommand{\arraystretch}{.8}
\begin{array}{cccccccc}
m & -1 & 0 & \cdot & \cdot & \cdot & \cdot & 0 \\
-1 & a_1 & -1 & 0 & \cdot & \cdot & \cdot & 0 \\
0 & -1 & a_2 & -1 & 0 & \cdot & \cdot & 0 \\
\cdot & \cdot & \cdot & \cdot & \cdot & \cdot & \cdot & \cdot \\
0 & \cdot & \cdot & \cdot & \cdot & 0 & -1 & a_{N-1}
\end{array} \right)\  \right\} N \ ,
\label{Khier}
\eea

\noi
which we abbreviate by the symbol

\be
[\,m\,;a_1\mini,\ldots,a_{N-1}\,]\ .
\label{Khierab}
\ee

\noi
We note that the signs of the $1$'s in~(\ref{Khier}) can be
changed by suitable equivalence transformations~(\ref{equiv}). The
choice of all the negative signs in~(\ref{Khier}) is our
convention. Moreover, we remark that, {\em from a QH lattice point
of view, the two hierarchy schemes of Haldane-Halperin~\cite{HH}
and Jain-Goldman~\cite{JG} are equivalent\mini}; see~(\ref{equiv})
and also the examples below. For this reason, we simply talk about
``hierarchy QH lattices''.

In the dual basis associated with~(\ref{Khier}), the
integer-valued linear functional (or charge vector) $\Q$ is given
by

\be
\Qv \:=\:(\mini\underbrace{1,0,\ldots,0}_{\mbox{\scriptsize N}}\,)
\ ;
\label{Qhier}
\ee

\noi

see the beginning of Sect.\,\ref{sBI}.

With the help of Kramer's rule~(\ref{Kram}), one easily verifies
that

\be
\sH \:=\; <\!\Q\,,\Q\!> \;=\: \Qv\cdot\Ki \Qv^T\ .
\ee

{}From Eqs.\,(\ref{Khier}) and~(\ref{Qhier}), it is clear that the
charge vector $\Q$ is {\em primitive\mini} and {\em odd}, as
defined in~(\ref{prim}) and~(\ref{odd}), respectively.

We note that, in general, the integral lattice $\mini\G\mini$
specified by~(\ref{Khier}) is {\em not\,} euclidean. In order for
it to be {\em euclidean}, the Gram matrix $\mini K\mini$
in~(\ref{Khier}) has to be positive-definite. One can show that
$K$ is positive-definite if and only if all the coefficients $a_i,\
i=1,\ldots,N\!-\!1$, are {\em positive}. In this situation, the
hierarchy QH lattice \QHL\, is a CQHL, as defined in
Sect.\,\ref{sUC}\mini. In particular, it satisfies
assumption~{\bf (A5)} there.
\rule[-4mm]{0mm}{5mm}

In the remaining part of this appendix, we comment on the status of
assumption~{\bf (A5)} for the {\em non-euclidean} hierarchy
fluids.\,--\,We recall that all (euclidean {\em and\,}
non-euclidean) hierarchy QH lattices satisfy assumptions~{\bf
(A1--4)} of Sect.\,\ref{sUC}\mini.

We exemplify the situation of non-euclidean hierarchy QH lattices
by discussing in some detail the two physically important series
of hierarchy fluids with $\,\sH=N/(2\mini N\!-\!1)$, and $\,
N/(4\mini N\!-\!1),\ N\geq 2\mini$.
\rule[-4mm]{0mm}{5mm}

{\bf (a)} $\;\sH=N/(2\mini N\!-\!1)\,$: By~(\ref{Khier})
and~(\ref{Khierab}), the Gram matrices $K$ of these hierarchy fluids
are given by

\be
K \:=\: [\,1\,;\mini\underbrace{-2,\ldots,-2\,}_{\mbox{\scriptsize
N-1}}\,]\ ,
\label{Khier2}
\ee

\noi
and the charge vectors $\Q$ are given by~(\ref{Qhier}). In order to
make the lattice structures behind~(\ref{Khier2}) more explicit, we
apply equivalence transformations~(\ref{equiv}), with $S$ given by

\bea
S & = & \left.\left(\renewcommand{\arraystretch}{.8}
\begin{array}{cccccccc}
1 & -1 & 0 & \cdot & \cdot & \cdot & \cdot & 0 \\
-1 & 2 & -1 & 0 & \cdot & \cdot & \cdot & \cdot \\
0 & 0 & 1 & 0 & \cdot & \cdot & \cdot & \cdot \\
\cdot & \cdot & 0 & -1 & 0 & \cdot & \cdot & \cdot \\
\cdot & \cdot & \cdot & 0 & 1 & 0 & \cdot & \cdot \\
\cdot & \cdot & \cdot & \cdot & \cdot & \cdot & \cdot & \cdot \\
0 & \cdot & \cdot & \cdot & \cdot & \cdot & 0 & \pm 1
\end{array} \right)\  \right\} N \ .
\eea

\noi
We find

\ben
K^\prime \:=\: [\, 1\,] \:\oplus\:
(-1)\cdot[\,3\,;\mini\underbrace{2\mini,\ldots,2\,}_{\mbox{\scriptsize
N-2}}\,]\ ,
\een

\vav

\be
\Qv^\prime \:=\: 1 \:+\;
(\mini\underbrace{-1,\mini 0,\ldots,\mini 0}_{\mbox{\scriptsize
N-1}}\,) \label{Qhier2} \ .
\ee

\noi
The interpretation of~(\ref{Qhier2}) is that, from a QH lattice
point of view, the hierarchy fluids at $\,\sH = N/(2\mini N\!-\!1)
= 1-(N\!-\!1)/[2\mini(N\!-\!1)+1]\,$ are indeed the ``charge
conjugates'' of the ``elementary''
$(N\!-\!1)/[2\mini(N\!-\!1)+1]\mini$-fluids exhibiting
$\,su(N\!-\!1)$-current algebras at level 1; see example~{\bf (c)}
at the end of Sect.\,\ref{sBI}\mini.

We note that from~(\ref{Qhier2}) it is clear that these
non-euclidean hierarchy QH lattices satisfy assumption~{\bf (A5)}
of Sect.\,\ref{sUC}.
\rule[-4mm]{0mm}{5mm}

{\bf (b)} $\;\sH=N/(4\mini N\!-\!1)\,$: By~(\ref{Khier})
and~(\ref{Khierab}), the Gram matrices $K$ of these hierarchy fluids
read

\be
K \:=\: [\,3\,;\mini\underbrace{-2,\ldots,-2\,}_{\mbox{\scriptsize
N-1}}\,]\ ,
\label{Khier4}
\ee

\noi
and the charge vectors $\Q$ are given by~(\ref{Qhier}). Again, in
order to make the composite nature of the lattices described
by~(\ref{Khier2}) explicit, we apply equivalence
transformations~(\ref{equiv}), with $S$ given by

\bea
S & = & \left.\left(\renewcommand{\arraystretch}{.8}
\begin{array}{cccccccc}
2\mini N\!-\!1 & -1 & \cdot & \cdot & \cdot & \cdot & \cdot & -1 \\
2\mini N\!-\!2 & -1 & \cdot & \cdot & \cdot & \cdot & \cdot & -1 \\
-(2\mini N\!-\!4) & 0 & 1 & \cdot & \cdot & \cdot & \cdot & 1 \\
2\mini N\!-\!6 & 0 & 0 & -1 & \cdot & \cdot & \cdot & -1 \\
-(2\mini N\!-\!8) & 0 & \cdot & 0 & 1 & \cdot & \cdot & 1 \\
\cdot & \cdot & \cdot & \cdot & \cdot & \cdot & \cdot & \cdot \\
\pm 2 & 0 & \cdot & \cdot & \cdot & \cdot & 0 & \mp 1
\end{array} \right)\  \right\} N \ .
\eea

\noi
This results in

\ben
K^\prime \:=\: [\, 4\mini N\!-\!1\,] \:\oplus\:
(-1)\cdot\Bigl(\,\underbrace{[\,1\,] \:\oplus\, \cdots\, \oplus\:
[\,1\,]}_{\mbox{\scriptsize N-1}}\,\Bigr) \ ,
\een

\vav

\be
\Qv^\prime \:=\: (2\mini N\!-\!1) \:+\;
\underbrace{(-1) \:+\:\cdots \:+\:(-1)}_{\mbox{\scriptsize N-1}}\ .
\label{Qhier4}
\ee

\noi
At the level of Hall conductivities, the
decompositions~(\ref{Qhier4}) can be expressed as $\,\sH = N/(4\mini
N\!-\!1) = (2\mini N\!-\!1)^2/(4\mini N\!-\!1)-1-\cdots -1$, with
$N\!-\!1$ summands of $-1\mini$.

Hence, similarly to~{\bf (a)}, the lattices $\mini\G\mini$ of this
series are composed out of positive- {\em and\,} negative-definite
sublattices, $\Ge$ and $\mini\Gh$, respectively. Contrary to~{\bf
(a)}, however, it follows form~(\ref{Qhier4}) that the
restrictions of the charge vector $\Q$ to the positive- and
negative-definite components of $\mini\G$\,--\,$\Qe$ and $\Qh$,
respectively (see~(\ref{deco}) and~(\ref{sumsH})\,--\,are {\em not
separately\mini} primitive. Rather, it is only the {\em full\,}
integer-valued linear form
$\,\Q=\Qe+\Qh\in\!\Gs=\Ge^\ast\oplus\Gh^\ast\,$ which is
primitive; see~(\ref{prim}).

In physical terms, this means that, similarly to assumption~{\bf
(A5)} in Sect.\,\ref{sUC}, the dynamics of the positively and of
the negatively charged (quasi-)particle rich
subfluids\,--\,corresponding to $\mini\Ge$ and $\mini\Gh$,
respectively\,--\,are independent in the scaling limit. Contrary
to~{\bf (A5)}, however, the physics of these two subfluids are {\em
not\,} identical up to charge conjugation. (The pair
\mbox{$(\Ge,\Qe)$} is {\em not\,} a CQHL, as defined in
Sect.\,\ref{sUC}, since $\Qe$ is {\em not\,} primitive.) We note
that the fundamental charge carriers of these QH fluids, electrons
{\em and\,} holes, are described as {\em composites\mini} of the
``basic'' positively and negatively charged (quasi-)particles
described by $\mini\Ge$ and $\mini\Gh$, respectively.

In conclusion, a slightly weaker assumption than~{\bf (A5)},
accounting for the situation above, would be as follows:
\rule[-4mm]{0mm}{5mm}

{\bf (A5')} The ``basic'' charge carriers of a QH fluid are
positively and/or negatively charged (quasi-)particles. We assume
that, in the scaling limit, the dynamics of
positive-(quasi-)particle-rich subfluids of a QH fluid is {\em
independent\mini} of the dynamics of negative-(quasi-)particle-rich
subfluids. The physically fundamental charge carriers of a QH
fluid, electrons and/or holes, are {\em composites\mini} of
positive and/or negative ``basic'' (quasi-)particles, respectively,
or electrons and/or holes are {\em composites\mini} of {\em
both\mini}, positive {\em and\,} negative ``basic''
(quasi-)particles.
\rule[-4mm]{0mm}{5mm}

Adopting assumption~{\bf (A5')} instead of~{\bf (A5)}, the
classification problem of QH fluids (see Sect.\,\ref{sUC}) would be
generalized according to: {\em In the scaling limit, the
quantum-mechanical description of an (incompressible) QH fluid is
universal. It is coded into a pair of odd, integral, euclidean
lattices\,--\,$\,\Ge$ positive- and $\,\Gh$ negative-definite,
respectively\,--\, and an odd, primitive vector
$\Q\in\Gs=\Ge^\ast\oplus\Gh$.}

For the reasons stated in Sect.\,\ref{sUC}, we do not study the
resulting, slightly more general classification problem. We remark,
however, that all hierarchy fluids, which are of physical
relevance in the region $\,0\!<\!\sH\!<\!1$, have been checked to
belong to this more general classification program if they are not
already contained in the one treated in this paper.



\clearpage
\begin{thebibliography}{99}
\addcontentsline{toc}{section}{References}



\bibitem{QHE} K.~von Klitzing, G.~Dorda, and M.~Pepper \PRL {\bf
45}, 494 (1980);

D.C.~Tsui, H.L.~Stormer, and A.C.~Gossard, \PRB {\bf 48}, 1559
(1982);

for reviews, see, e.g.: R.E.~Prange and S.M.~Girvin (eds.), {\em
The Quantum Hall Effect}, Second Edition, Graduate Texts in
Contemporary Physics (Springer, New York, 1990);

T.~Chakraborty and P.~Pietil\"ainen, {\em The Fractional Quantum
Hall Effect: Properties of an Incompressible Quantum Fluid},
Springer Series in Solid State Science {\bf 85} (Springer,
Berlin, 1988);

M.~Stone (ed.), {\em Quantum Hall Effect} (World Scientific,
Singapore, 1992).


\bibitem{FK} J.~Fr\"ohlich and T.~Kerler, \NPB{\bf 354}, 369
(1991).


\bibitem{FZ} J.~Fr\"ohlich and A.~Zee, \NPB{\bf 364}, 517 (1991).


\bibitem{FS1} J.~Fr\"ohlich and U.M.~Studer, \CMP {\bf 148}, 553
(1992).


\bibitem{FS2} J.~Fr\"ohlich and U.M.~Studer, Rev.\ Mod.\ Phys.\
{\bf 65}, 733 (1993).


\bibitem{FT} J.~Fr\"ohlich and E.~Thiran, J.\ Stat.\ Phys. {\bf
76}, 209 (1994).


\bibitem{FKST} J.~Fr\"ohlich, T.~Kerler, U.M.~Studer, and
E.~Thiran, ``Structuring the set of incompressible quantum Hall
fluids'', preprint ETH-TH/95-5.


\bibitem{Ha} B.I.~Halperin, \PRB {\bf 25}, 2185 (1982).


\bibitem{Read} N.~Read, \PRL {\bf 65}, 1502 (1990).


\bibitem{edgetheo} X.G.~Wen, \PRL {\bf 64}, 2206 (1990); \PRB
{\bf 41}, 12838 (1990);

B.~Blok and X.G.~Wen, \PRB {\bf 42}, 8133 and 8145 (1990);

D.H.~Lee and X.G.~Wen \PRL {\bf 66}, 1765 (1991);

M.~Stone, \AP {\bf 207}, 38 (1991).


\bibitem{edge} M.~B\"uttiker, \PRB {\bf 38}, 9375 (1988);

C.W.J.~Beenakker, \PRL {\bf 64}, 216 (1990);

A.H.~MacDonald, {\em ibid.} {\bf 64}, 220 (1990);

A.M.~Chang, \SSC {\bf 74}, 871 (1990).


\bibitem{edgeexp} K.~von Klitzing, Physica {\bf B 184}, 1 (1993);
and references therein.


\bibitem{FGM} J.~Fr\"ohlich, R.~G\"otschmann, and P.-A.~Marchetti,
``Bosonization of Fermi systems in arbitrary dimensions in terms of
gauge forms'', preprint DFPD 94/TH/36.


\bibitem{HLRAIM} B.I.~Halperin, P.A.~Lee, and N.~Read, \PRB {\bf
47}, 7312 (1993);

B.L.~Altshuler, L.B.~Ioffe, and A.J.~Millis, \PRB {\bf 50}, 14048
(1994).


\bibitem{nus} D.C.~Tsui, Physica B {\bf 164}, 59 (1990);

H.L.~Stormer, Physica B {\bf 177}, 401 (1992).


\bibitem{Saj} T.~Sajoto, Y.W.~Suen, L.W.~Engel, M.B.~Santos, and
M.~Shayegan, \PRB {\bf 41}, 8449 (1990).


\bibitem{main} R.R.~Du, H.L.~Stormer, D.C.~Tsui, L.N.~Pfeiffer,
and K.W.~West, \PRL {\bf 70} 2944 (1993).


\bibitem{Kang} W.~Kang, H.L.~Stormer, L.N.~Pfeiffer, K.W.~Baldwin,
and K.W.~West, \PRL {\bf 71} 3850 (1993).


\bibitem{GS} V.J.~Goldman and M.~Shayegan, \SurS {\bf 229}, 10
(1990).


\bibitem{Hu} C.-R.~Hu, \IJMP B {\bf 5}, 1739 (1991); see
especially Ref.\,12 therein.


\bibitem{Cl90} R.G.~Clark, S.R.~Haynes, J.V. Branch, A.M.~Suckling,
P.A.~Wright, P.M.W.~Oswald, J.J.~Harris, and C.T.~Foxon, \SurS {\bf
229}, 25 (1990).


\bibitem{Eis90} J.P.~Eisenstein, H.L.~Stormer, L.N.~Pfeiffer, and
K.W.~West, \PRB {\bf 41}, 7910 (1990).


\bibitem{Eng} L.W.~Engel, S.W.~Hwang, T.~Sajoto, D.C.~Tsui,
and M.~Shayegan, \PRB {\bf 45}, 3418 (1992).


\bibitem{HaHPA} Halperin, B.~I., 1983, \HPA {\bf 56}, 75.


\bibitem{1/2} Y.W.~Suen, L.W.~Engel, M.B.~Santos, M.~Shayegan, and
D.C.~Tsui, \PRL {\bf 68}, 1379 (1992);

J.P.~Eisenstein, G.S.~Boebinger, L.N.~Pfeiffer, K.W.~West, and Song
He, \PRL {\bf 68}, 1383 (1992).


\bibitem{Su} Y.W.~Suen, H.C.~Manoharan, X.~Ying, M.B.~Santos and
M.~Shayegan, \SurS {\bf 305}, 13 (1994).


\bibitem{Wil5/2} R.L.~Willett, J.P.~Eisenstein, H.L.~Stormer,
D.C.~Tsui, A.C.~Gossard, and J.H.~English, \PRL {\bf 59}, 1776
(1987); the first observations of fractional QH fluids with
$\,\sH>2\,$ are given in this reference.


\bibitem{5/2} J.P.~Eisenstein, R.L.~Willett, H.L.~Stormer,
D.C.~Tsui, A.C.~Gossard, and J.H.~English, \PRL {\bf 61}, 997
(1988);

J.P.~Eisenstein, R.L.~Willett, H.L.~Stormer, L.N.~Pfeiffer, and
K.W.~West, \SurS {\bf 229}, 31 (1990).


\bibitem{HH} F.D.M.~Haldane, \PRL {\bf 51}, 605 (1983);

B.I.~Halperin, {\em ibid.} {\bf 52}, 1583 (1984).


\bibitem{JG} J.K.~Jain and V.J.~Goldman, \PRB {\bf 45}, 1255
(1992).


\bibitem{Jain} J.K.~Jain, \PRL {\bf 63}, 199 (1989); \PRB {\bf
41}, 7653 (1990); Adv.\ Phys.\ {\bf 41}, 105 (1992).


\bibitem{Sak} B.~Sakita, \PL B {\bf 315}, 124 (1993).


\bibitem{Winf} S.~Iso, D.~Karabali, and B.~Sakita, \PL B {\bf 296},
143 (1992);

A.Cappelli, C.A.~Trugenberger, and G.R.~Zemba, \NPB {\bf 396}, 465
(1993); \PL B {\bf 306}, 100 (1993);

M.~Flohr and R.~Varnhagen, J.\ Phys.\ A: {\bf 27}, 3999 (1994);

D.~Karabali, \NPB {\bf 428}, 531 (1994).


\bibitem{Luet} C.A.~L\"utken, \NPB {\bf 396}, 670 (1993); J.\
Phys.\ A: {\bf 26}, L811 (1993);

C.A.~L\"utken and G.G.~Ross, \PRB {\bf 48}, 2500 (1993).



\bibitem{L} R.B.~Laughlin, \PRL {\bf 50}, 1395 (1983); \PRB
{\bf 27}, 3383 (1983).


\bibitem{Wi} E.~Witten, \CMP {\bf 121}, 351 (1989).


\bibitem{Bala} A.P.~Balachandran, G.~Bimonte, K.S.~Gupta, and
A.~Stern, \IJMP \mbox{A {\bf 7}}, 4655 and 5855 (1992); and
references therein.



\bibitem{FSTL} J.~Fr\"ohlich, U.M.~Studer, and E.~Thiran, ``An
$ADE-O\mini$ classification of minimal incompressible quantum
Hall fluids'', in ``On Three Levels'', M.~Fannes, C.~Maes, and
A.~Verbeure, eds., (Plenum, New York, 1994); see also
cond-mat/9406009.


\bibitem{Sieg} C.L.~Siegel, {\em Lectures on the geometry of
numbers}, Springer-Verlag (Berlin, Heidelberg, 1989).


\bibitem{WigCry} P.K.~Lam and S.M.~Girvin, \PRB {\bf 30}, 473
(1984);

D.~Levesque, J.J.~Weiss, and A.H.~MacDonald, \PRB {\bf 30}, 1056
(1984).


\bibitem{WigCryexp}  B.~McCombe and A.~Nurmikko (eds.), {\em
Electronic Properties of Two-\-Dimensional Systems}
(North-\-Holland, New York, 1994).


\bibitem{FKT} J.~Fr\"ohlich, T.~Kerler, and E.~Thiran, in
preparation.


\bibitem{CSS} J.H.~Conway and N.J.A.~Sloane, Proc.\ R.\ Soc.\
Lond.\ A {\bf 418}, 17 (1988); and references therein.



\bibitem{CSbook} J.H.~Conway and N.J.A.~Sloane, {\em
Sphere-packings, lattices and groups}, Springer (New York, 1988).


\bibitem{Sl} R.~Slansky, Phys.~Reports {\bf 79}, 1 (1981); and
references therein.


\bibitem{GO} P.~Goddard and D.~Olive, \IJMP A{\bf 1}, 303 (1986).


\bibitem{Ash} R.C.~Ashoori, H.L.~Stormer, L.N.~Pfeiffer,
K.W.~Baldwin, and K.W.~West , \PRB {\bf 45}, 3894 (1992).



\bibitem{Dick} L.E.~Dickson, {\em Studies in the theory of numbers},
University of Chicago Press (Chicago, 1930), reprinted by Chelsea
Publishing Company (New York, 1957).



\bibitem{KMembed} A.N.~Schellekens and N.P.~Warner, \PRD {\bf 34},
3092 (1986);

F.A.~Bais and P.G.~Bouwknegt, \NPB{\bf 279}, 561 (1987);

V.G.~Kac and M.N.~Sanielevici, \PRD {\bf 37}, 2231 (1988).


\bibitem{ptat1} S.Q.~Murphy, J.P.~Eisenstein, G.S.~Boebinger,
L.N.~Pfeiffer, and K.W.~West, \PRL {\bf 72}, 728 (1994).

\end{thebibliography}
\end{document}